\documentclass[a4paper,twoside,12pt]{article}
\usepackage{natbib, mnemonic}
\usepackage{epsfig}
\usepackage{lscape}

\newcommand{\ignore}[1]{}
\begin{document}

\null
\vspace{2truecm}

\centerline{\bf \Large Report by the ESA--ESO Working Group} 
\centerline{\bf \Large on Extra-Solar Planets}

\vspace{30pt}
\centerline{\large 4 March 2005}

\vfill

\section*{Summary}

Various techniques are being used to search for extra-solar planetary
signatures, including accurate measurement of radial velocity and
positional (astrometric) displacements, gravitational microlensing,
and photometric transits.  Planned space experiments promise a
considerable increase in the detections and statistical knowledge
arising especially from transit and astrometric measurements over the
years 2005--15, with some hundreds of terrestrial-type planets
expected from transit measurements, and many thousands of Jupiter-mass
planets expected from astrometric measurements. 

Beyond 2015, very ambitious space (Darwin/TPF) and ground (OWL)
experiments are targeting direct detection of nearby Earth-mass
planets in the habitable zone and the measurement of their spectral
characteristics.  Beyond these, `Life Finder' (aiming to produce
confirmatory evidence of the presence of life) and `Earth Imager'
(some massive interferometric array providing resolved images of a
distant Earth) appear as distant visions. 

This report, to ESA and ESO, summarises the direction of exo-planet
research that can be expected over the next 10~years or so, identifies
the roles of the major facilities of the two organisations in the
field, and concludes with some recommendations which may assist
development of the field.

The report has been compiled by the Working Group members and
experts (page~iii) over the period June--December 2004.

\vfill

\newpage
\pagestyle{plain}
\vspace{-10pt}

\centerline{\bf Introduction \& Background}


Following an agreement to cooperate on science planning issues, the executives of the European Southern Observatory (ESO) and the European Space Agency (ESA) Science Programme and representatives of their science advisory structures have met to share information and to identify potential synergies within their future projects. The agreement arose from their joint founding membership of EIROforum (http://www.eiroforum.org) and a recognition that, as pan-European organisations, they served essentially the same scientific community.

At a meeting at ESO in Garching during September 2003, it was agreed to establish a number of working groups that would be tasked to explore these synergies in important areas of mutual interest and to make recommendations to both organisations. The chair and co-chair of each group were to be chosen by the executives but thereafter, the groups would be free to select their membership and to act independently of the sponsoring organisations. The first working group to be established was on the topic of Extra-Solar Planet research, both detection and physical study, over a period extending from now until around 2015. The group worked on its report from June until December 2004 and reported its conclusions and recommendations to a second ESA-ESO meeting, held at ESA HQ in Paris in February 2005.

\vspace{50pt}

\centerline{\bf Terms of Reference and Composition}

The goals set for the working group were to provide: 
\vspace{-10pt}
\begin{itemize}

\item {A survey of the field: this will comprise: (a) a review of the methods
used or envisaged for extra-solar planet detection and study; (b) a
survey of the associated instrumentation world-wide (operational,
planned, or proposed, on-ground and in space); (c) for each, a summary
of the potential targets, accuracy and sensitivity limits, and
scientific capabilities and limitations. }

\item{An examination of the role of ESO and ESA facilities: this will: (a) identify areas in
which current and planned ESA and ESO facilities will contribute; (b)
analyse the expected scientific returns and risks of each; (c)
identify areas of potential scientific overlap, and thus assess the
extent to which the facilities complement or compete; (d) identify
open areas which merit attention by one or both organisations (for
example, follow-up observations by ESO to maximise the return from
other major facilities); (e) conclude on the scientific case for the
very large facilities planned or proposed.  }

\end{itemize}

\newpage

\vspace{3.0cm}
The working group membership was established by the 
chair and co-chair: the report is not a result of consultation with 
the community as a whole. The experts contributed considerable information
for the report, but the conclusions and recommendations are the 
responsibility of the members.

\vspace{1.0cm}

\begin{table}[h]
\small
\begin{center}
\setlength\tabcolsep{5pt}
\begin{tabular}{llll}
Chair:&    Michael Perryman &   ESA \\
Co-Chair:& Olivier Hainaut  &   ESO \\
\noalign{\smallskip}
Members:  
&       Dainis Dravins      &  Lund \\
&       Alain L\'eger       &  IAS \\
&       Andreas Quirrenbach &  Leiden \\
&       Heike Rauer         &  DLR \\
\noalign{\smallskip}
ECF support: 
&       Florian Kerber      &  ESO--ECF \\       
&       Bob Fosbury         &  ESA--ECF \\
\noalign{\smallskip}
Experts:  
&       Fran{\c c}ois Bouchy     &  OHP Marseilles      & COROT \\
&       Fabio Favata        &  ESA                 & Eddington \\
&       Malcolm Fridlund    &  ESA                 & Darwin \\
&       Roberto Gilmozzi    &  ESO                 & OWL \\
&       Anne-Marie Lagrange &  LAOG Grenoble       & Planet Finder \\
&       Tsevi Mazeh         &  Tel Aviv            & Transits \\
&       Daniel Rouan        &  Obs de Paris-Meudon & Genie \\
&       Stephane Udry       &  Gen\`eve            & Radial velocity \\
&       Joachim Wambsganss  &  Heidelberg          & Microlensing \\
\end{tabular}
\end{center}
\end{table}

\vspace{30pt}
{\bf Catherine Cesarsky (ESO)
\hspace{3.6cm}\'Alvaro Gim\'enez Ca\~nete (ESA)\\

March 2005}

\clearpage

\parskip 9pt
\tableofcontents
\parskip 10pt

\cleardoublepage
\pagenumbering{arabic}
\setcounter{page}{1}

\section{Survey of the Field}

\subsection{Introduction}

The field of exo-planet research has exploded dramatically
since the discovery of the first such systems in 1995. Underlying this
huge interest three main themes of exo-planet research can be
identified: (a) characterising and understanding the planetary
populations in our Galaxy; (b) understanding the formation and
evolution of planetary systems (e.g., accretion, migration,
interaction, mass-radius relation, albedo, distribution, host star
properties, etc.); (c) the search for and study of biological markers
in exo-planets, with resolved imaging and the search for intelligent
life as `ultimate' and much more distant goals.  

Detection methods for extra-solar planets can be broadly classified
into those based on: (i) dynamical effects (radial velocity,
astrometry, or timing in the case of the pulsar planets); (ii)
microlensing (astrometric or photometric); (iii) photometric signals
(transits and reflected light); (iv) direct imaging from ground or
space in the optical or infrared; and (v) miscellaneous effects (such
as magnetic superflares, or radio emission). Each have their
strengths, and advances in each field will bring specific and often
complementary discovery and diagnostic capabilities. Detections are a
pre-requisite for the subsequent steps of detailed physical-chemical
characterisation demanded by the emerging discipline of
exo-planetology.

As of December 2004, 135 extra-solar planets have been discovered
from their radial velocity signature, comprising 119 systems of which
12~are double and 2~are triple. One of these planets has also been
observed to transit the parent star. Four additional confirmed
planets have been discovered through transit detections using data
from OGLE (and confirmed through radial velocity measurements),
and one, TrES-1, using a small 10-cm ground-based telescope.  One
further, seemingly reliable, planet candidate has been detected
through its microlensing signature.  The planets detected to date
(apart from those surrounding radio pulsars, which are not considered
further in this report) are primarily `massive' planets, of order
1\,$M_{\rm J}$, but extending down to perhaps 0.05\,$M_{\rm J}$
(around 15\,$M_\oplus$) for three short-period systems, although the
inclination (and hence true mass) of two of these is unknown\footnote{
  the following notation is used: $M_{\rm J}$ = Jupiter mass;
  $M_\oplus$ = Earth mass $\sim0.003\,M_{\rm J}$.}.

\begin{figure}[t]
\begin{center}
\centerline{\epsfig{file=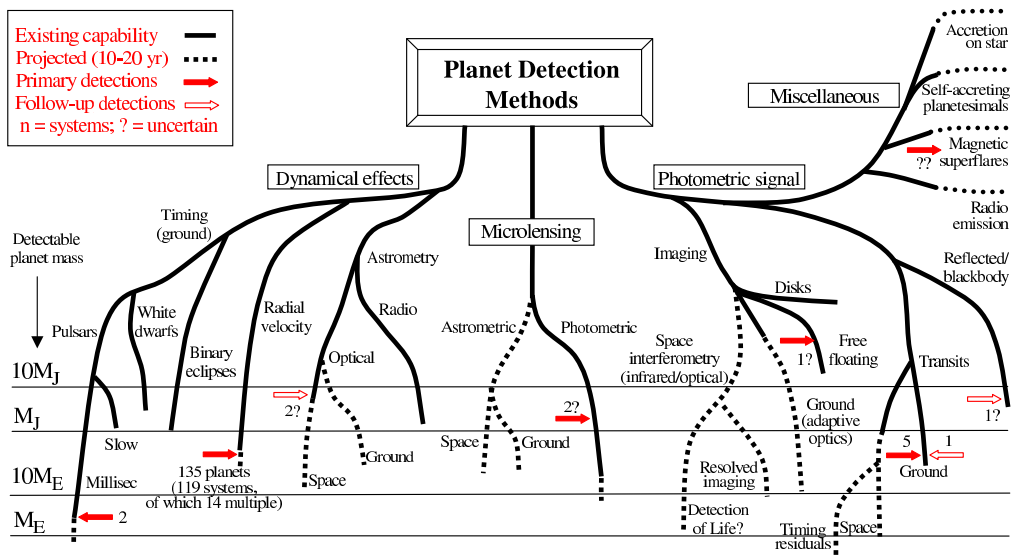,height=1.0\textwidth,angle=270}}
\end{center}
\vspace{-20pt}
\caption[]{\footnotesize
  Detection methods for extra-solar planets, updated from
  \cite{per00}.  The lower extent of the lines indicates, roughly, the
  detectable masses that are in principle within reach of present
  measurements (solid lines), and those that might be expected within
  the next 10--20~years (dashed).  The (logarithmic) mass scale is
  shown at left.  The miscellaneous signatures to the upper right are
  less well quantified in mass terms. Solid arrows indicate (original)
  detections according to approximate mass, while open arrows indicate
  further measurements of previously-detected systems.  `?'~indicates
  uncertain or unconfirmed detections. The figure takes no account of
  the numbers of planets that may be detectable by each method.  }
\label{fig:methods}
\end{figure}

Detection methods considered to date are summarised in
Figure~\ref{fig:methods}, which also gives an indication of the lower
mass limits which are likely to be reached in the foreseeable future
for each method.  More information and ongoing projects are given in
Jean Schneider's www page: http://www.obspm.fr/encycl/searches.html.

An earlier ESO Working Group on the `Detection of Extra-Solar Planets'
submitted a report with detailed recommendations in 1997
\citep{eso-wg97}.  A summary and status of these recommendations is
attached as Appendix~C.

\subsection{The Search for Earth-Mass Planets and Habitability}
\label{sec:habitability}

The search for planets around stars in general, and Earth-mass planets
in particular, is motivated by efforts to understand their frequency
of occurrence (as a function of mass, semi-major axis, eccentricity,
etc.) and their formation mechanism and, by analogy, to gain an
improved understanding of the formation of our own Solar System.
Search accuracies will progressively improve to the point that the
detection of telluric planets in the `habitable zone' will become
feasible, and there is presently no reason to assume that such planets
will not exist in large numbers. Improvements in spectroscopic
abundance determinations, whether from Earth or space, and
developments of atmospheric modelling, will lead to searches for
planets which are progressively habitable, inhabited by
micro-organisms, and ultimately by intelligent life (these may or may
not prove fruitful).  Search strategies will be assisted by improved
understanding of the conditions required for development of life on
Earth.

Very broadly, the search for potentially habitable planets is being
concentrated around Sun-like stars (spectral type and age), focussing
on Earth-mass planets, in low-eccentricity orbits at about 1\,AU
representing the `continuously habitable zone' (the habitable zone is
the distance range from the parent star over which liquid water is
likely to be present; the continuously habitable zone is the region
throughout which liquid water should have been present over a
significant fraction of the star's main-sequence lifetime). Further
details and potential spectral diagnostics of life are given in this
section.  Such considerations may imply that the fraction of
habitable planets is small, but they should add to the
knowledge of where to look.

Assessment of the suitability of a planet for supporting life, or
habitability, is based on our knowledge of life on Earth. With the
general consensus among biologists that carbon-based life requires
water for its self-sustaining chemical reactions, the search for
habitable planets has therefore focused on identifying environments in
which liquid water is stable over billions of years. Earth's
habitability over early geological time scales is complex, but its
atmosphere is thought to have experienced an evolution in the
greenhouse blanket of CO$_2$ and H$_2$O to accommodate the 30\%
increase in the Sun's luminosity over the last 4.6~billion years in
order to sustain the presence of liquid water evident from geological
records.  In the future, the Sun will increase to roughly three times
its present luminosity by the time it leaves the main sequence, in
about 5~Gyr.

The habitable zone is consequently presently defined by the range of
distances from a star where liquid water can exist on the planet's
surface.  This is primarily controlled by the star-planet separation,
but is affected by factors such as planet rotation combined with
atmospheric convection. For Earth-like planets orbiting main-sequence
stars, the inner edge is bounded by water loss and the runaway
greenhouse effect, as exemplified by the CO$_2$-rich atmosphere and
resulting temperature of Venus.  The outer boundary is determined by
CO$_2$ condensation and runaway glaciation, but it may be extended
outwards by factors such as internal heat sources including long-lived
radionuclides (U$^{235}$, U$^{238}$, K$^{40}$ etc., as on Earth),
tidal heating due to gravitational interactions (as in the case of
Jupiter's moon Io), and pressure-induced far-infrared opacity of
H$_2$, since even for effective temperatures as low as 30~K,
atmospheric basal temperatures can exceed the melting point of water.
These considerations result, for a $1\,M_\odot$ star, in an inner
habitability boundary at about 0.7\,AU and an outer boundary at around
1.5\,AU or beyond. The habitable zone evolves outwards with time
because of the increasing luminosity of the Sun with age, resulting in a
narrower width of the continuously habitable zone over $\sim4$~Gyr of
around 0.95--1.15\,AU.  Positive feedback due to the greenhouse effect
and planetary albedo variations, and negative feedback due to the link
between atmospheric CO$_2$ level and surface temperature may limit
these boundaries further. Migration of the habitable zone to much
larger distances, 5--50\,AU, during the short period of
post-main-sequence evolution corresponding to the sub-giant and red
giant phases, has been considered.

Within the $\sim1$\,AU habitability zone, Earth `class' planets can be
considered as those with masses between about 0.5--10\,$M_\oplus$ or,
equivalently assuming Earth density, radii between
0.8--2.2\,$R_\oplus$. Planets below this mass in the habitable zone
are likely to lose their life-supporting atmospheres because of their
low gravity and lack of plate tectonics, while more massive systems
are unlikely to be habitable because they can attract a H-He
atmosphere and become gas giants.  Habitability is also likely to be
governed by the range of stellar types for which life has enough time
to evolve, i.e.\ stars not more massive than spectral type~A.
However, even F~stars have narrower continuously habitable zones
because they evolve more strongly (and rapidly), while planets
orbiting in the habitable zones of late~K and M~stars become trapped
in synchronous rotation due to tidal damping, which may preclude life
apart from close to the light-shadow line.  Mid- to early-K and G
stars may therefore be optimal for the development of life.

\cite{owe80} argued that large-scale biological activity on a telluric
planet necessarily produces a large quantity of O$_2$.  Photosynthesis
builds organic molecules from CO$_2$ and H$_2$O, with the help of H$^+$ ions
which can be provided from different sources. In the case of oxygenic
bacteria on Earth, H$^+$ ions are provided by the photodissociation of
H$_2$O, in which case oxygen is produced as a by-product. However,
this is not the case for anoxygenic bacteria, and thus O$_2$ is to be
considered as a possible but not a necessary by-product of life (for
this signature of biological activity, as well as for any other, a key
issue is that of false positives, i.e.\ cases where the signature is
detected but there is no actual life on the planet, while the case of
false negatives, when there is some life on the planet but the
signature is absent, is significantly less `serious').  Indeed,
Earth's atmosphere was O$_2$-free until about 2~billion years ago,
suppressed for more than 1.5~billion years after life originated.
\cite{owe80} noted the possibility, quantified by \cite{sch94} based
on transit measurements, of using the 760-nm band of oxygen as a
spectroscopic tracer of life on another planet since, being highly
reactive with reducing rocks and volcanic gases, it would disappear in
a short time in the absence of a continuous production mechanism.
Plate tectonics and volcanic activity provide a sink for free O$_2$,
and are the result of internal planet heating by radioactive uranium
and of silicate fluidity, both of which are expected to be generic
whenever the mass of the planet is sufficient and when liquid water is
present.  For small enough planet masses, volcanic activity disappears
some time after planet formation, as do the associated oxygen sinks.

O$_3$ is itself a tracer of O$_2$ and, with a prominent spectral
signature at 9.6\,$\mu$m in the infrared where the planet/star
contrast is significantly stronger than in the optical
($1.4\times10^{-7}$ rather than $2\times10^{-10}$ for the Earth/Sun
case), should be easier to detect than the visible wavelength lines.
These considerations are motivating the development of infrared space
interferometers for the study of bands such as H$_2$O at 6--8\,$\mu$m,
CH$_4$ at 7.7\,$\mu$m, O$_3$ at 9.6\,$\mu$m, CO$_2$ at 15\,$\mu$m and
H$_2$O at 18\,$\mu$m.  Higher resolution studies might reveal the
presence of CH$_4$, its presence on Earth resulting from a balance
between anaerobic decomposition of organic matter and its interaction
with atmospheric oxygen; its highly disequilibrium co-existence with
O$_2$ could be strong evidence for the existence of life.

The possibility that O$_2$ and O$_3$ are not unambiguous
identifications of Earth-like biology, but rather a result of abiotic
processes, has been considered in detail by \cite{loa+99} and
\cite{sdp02}.  They considered various production processes such as
abiotic photodissociation of CO$_2$ and H$_2$O followed by the
preferential escape of hydrogen from the atmosphere.  In addition,
cometary bombardment could bring O$_2$ and O$_3$ sputtered from H$_2$O
by energetic particles, depending on the temperature, greenhouse
blanketing, and presence of volcanic activity.  They concluded that a
simultaneous detection of significant amounts of H$_2$O and O$_3$ in
the atmosphere of a planet in the habitable zone presently stands as a
criterion for large-scale photosynthetic activity on the planet.  Such
an activity on a planet illuminated by a star similar to the Sun, or
cooler, is likely to be a significant indication that there is local
biological activity, because this synthesis requires the storage of
the energy of at least 2~photons (8~in the case on Earth) prior to the
synthesis of organic molecules from H$_2$O and CO$_2$. This is likely
to require delicate systems that have developed during a biological
evolutionary process.  The biosignature based on O$_3$ seems to be
robust because no counter example has been demonstrated. It is not the
case for the biosignature based on O$_2$ \citep{sdp02}, where false
positives can be encountered. This puts a hierarchy between
observations that can detect O$_2$ and those that can detect O$_3$.

Habitability may be further confined within a narrow range of [Fe/H]
of the parent star \citep{gon99b}. If the occurrence of gas giants
decreases at lower metallicities, their shielding of inner planets in
the habitable zone from frequent cometary impacts, as occurs in our
Solar System, would also be diminished. At higher metallicity,
asteroid and cometary debris left over from planetary formation may be
more plentiful, enhancing impact probabilities.  \cite{gon99a} has
also investigated whether the anomalously small motion of the Sun with
respect to the local standard of rest, both in terms of its
pseudo-elliptical component within the Galactic plane, and its
vertical excursion with respect to the mid-plane, may be explicable in
anthropic terms.  Such an orbit could provide effective shielding from
high-energy ionising photons and cosmic rays from nearby supernovae,
from the X-ray background by neutral hydrogen in the Galactic plane,
and from temporary increases in the perturbed Oort comet impact rate.

\begin{figure}[t]
\begin{center}
\centerline{\epsfig{file=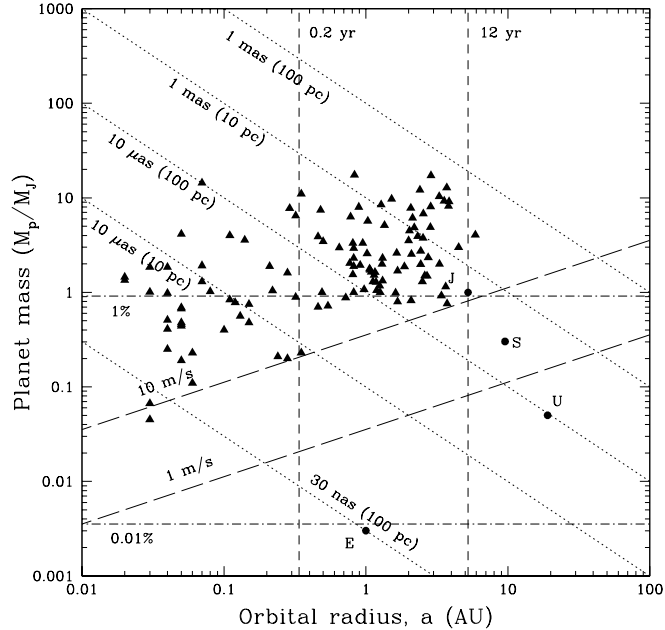,width=0.6\textwidth}}
\end{center}
\vspace{-20pt}
\caption[]{\footnotesize
  Detection domains for methods exploiting planet orbital motion, as a
  function of planet mass and orbital radius, assuming $M_*=M_\odot$.
  Lines from top left to bottom right show the locus of astrometric
  signatures of 1~milli-arcsec and 10\,micro-arcsec at distances of 10 and
  100\,pc; a measurement accuracy 3--4 times better would be needed to
  detect a given signature.  
  Vertical lines show limits corresponding to orbital periods of 
  0.2~and 12~years, relevant for Gaia (where very short and very long
  periods cannot be detected) although not for SIM.  
  Lines from top right to bottom left show radial
  velocities corresponding to $K=10$ and $K=1$~m\,s$^{-1}$; a
  measurement accuracy 3--4 times better would be needed to detect a
  given value of~$K$.  Horizontal lines indicate photometric
  detection thresholds for planetary transits, of 1\% and 0.01\%,
  corresponding roughly to Jupiter and Earth radius planets
  respectively (neglecting the effects of orbital inclination, which
  will diminish the probability of observing a transit as $a$
  increases).  The positions of Earth (E), Jupiter (J), Saturn (S) and
  Uranus (U) are shown, as are the lower limits on the masses of known
  planetary systems as of December 2004 (triangles).  }
\label{fig:domains}
\end{figure}

\subsection{Present Limits: Ground and Space}
\label{sec:limits}

Figure~\ref{fig:domains} illustrates the detection domains for the
radial velocity, astrometry, and transit methods as a function of
achievable accuracy. It also shows the location of the exo-planets
known to date, in a mass-orbital radius (period) diagram.

The fundamental accuracy limits of each method are not yet firmly
established, although such knowledge is necessary to predict the real
performances of dedicated surveys on ground and in space. Granular
flows and star spots on the surface of late-type stars place specific
limits on the photometric stability, the stability of the photocentric
position, and the stability of spectroscopically-derived radial
velocities, whether these observations are made from ground or space.
A series of hydrodynamical convection models covering stellar objects
from white dwarfs to red giants has been used to give estimates of the
photometric and photocentric stellar variability in
wavelength-integrated light across the HR~diagram \citep{sl05}.

(a) Radial velocity experiment accuracies are close to the values of
around 1--3~m\,s$^{-1}$ at which atmospheric circulation and
oscillations limit measurement precision, implying mass detection
limits only down to 0.01--0.1~$M_{\rm J}$ (depending on orbital
period); detection of an Earth in the habitable zone would require
accuracies of $\sim0.03-0.1$~m\,s$^{-1}$. Observations from space will
not improve these limits, and no high-precision radial velocity
measurements from space have been proposed.

The idea of `stacking up' many radial velocity observations to average
the effects of stellar oscillations is appealing, but faces several
complications: (i) even if p-mode oscillation effects can be
minimised, beating amongst these modes may induce large
radial-velocity variations (up to 10\,m\,s$^{-1}$ peak-to-peak) over
timescales of a few hours, specifically some 5--6~hours for $\mu$~Arae
(Bouchy, private communication).  The star will therefore need to be
observed over several hours for each epoch (radial velocity point);
(ii) simulations by Bouchy (private communication) show that a precision of
$\sim1$\,m\,s$^{-1}$ is reached in about 15--20~min, while the gain is
much less rapid with increasing observation time. A precision of
0.1\,m\,s$^{-1}$ (still insufficient for the detection of the Earth
around the Sun) will therefore be very expensive in terms of telescope
time; (iii) a wavelength calibration precision from night-to-night is
then needed at the level of the long-term precision targeted.
Reference calibration to 0.1\,m\,s$^{-1}$ will require further
improvements in calibration techniques.  With HARPS, a precision of
about 0.5\,m\,s$^{-1}$ is reached, as illustrated by asteroseismology
results on $\mu$~Arae with 250~observations each.  Investigations are
ongoing (Udry, private communication) into the possibility of having a
reference at the 0.02\,m\,s$^{-1}$ level for an instrument on OWL,
while the HARPS GTO programme anyway pushing in this direction will
soon help to better characterise the question.  In conclusion, a very
high radial velocity precision seems possible, but at a very high
cost.
  
There is a significant difference in the case of transiting
candidates: now the period and phase are known, and with $e\sim0$ for
short period planets, a series of accumulated measurements can be used
to constrain the radial velocity semi-amplitude.  With HARPS at a
precision of 1\,m\,s$^{-1}$, for short-period planets, it is expected
that limits of a few Earth-masses, for $P<10$~days, can be reached. If
the transiting object is larger, then the radial velocity effect will
be larger and easier to detect.  False positive detections will be the
main problem.

(b) Photometric (transit) limits below the Earth's atmosphere are
typically a little below the 1\% photometric precision, limited by
variations in extinction, scintillation and background noise
(depending on telescope aperture size), corresponding to masses of
about 1~$M_{\rm J}$ for solar-type stars.  One main challenge is to
reach differential photometric accuracies of around 1~mmag over a wide
field of view, in which airmass, transparency, differential refraction
and seeing all vary significantly. The situation improves above the
atmosphere, and a number of space experiments are planned to reach the
0.01\% limits required for the detection of Earth-mass planets.  HST
can place much better limits on transit photometry than is possible
from the ground, as exemplified by HD~209458 (see
Sections~\ref{sec:ground-transits} and~\ref{sec:space-transits}).
Simulations have been made by the COROT teams in order to estimate the
transit detection threshold due to stellar activity.  In the case of a
very active star, the detection of an Earth (80~ppm) is not possible.
In the case of a quiet star (like the Sun), it is possible if several
transits are summed. In the case of COROT, 1.6~$M_\oplus$ is detected
after 10--30 transits.  Another complication is again false-positives,
where statistical effects, stellar activity, and background binaries
can all mimic transit events, and which call for independent
confirmation of detections in general.

(c) Astrometric measurements do not yet extend below the
1~milli-arcsec of Hipparcos, implying current detectability limits
typically above 1--10~$M_{\rm J}$.  Even with the expected advent of
narrow-field ground-based astrometry at 10~micro-arcsec (e.g.\ PRIMA),
detections would be well short of Earth-mass planets, even within
10\,pc. Above the atmosphere, astrometric accuracy limits improve
significantly. The studies of \cite{sl05} indicate that for $\log
g\sim4.4$, resulting displacements are around $10^{-7}-10^{-8}$\,AU
suggesting, for example, that this effect will not degrade the Gaia
measurements, with the exception of nearby ($<100$\,pc) red giants.
Nevertheless, that work treats only the variability caused by the
evolution of stellar surface inhomogeneities driven by thermal
convection (stellar granulation).  At lower temporal frequencies, the
variability is much higher (but not yet treatable by hydrodynamic
models), caused by magnetic stellar activity, spottiness, and
rotation, all of which may make substantial additional contributions
to the astrometric (and photometric) variability.

(d) Microlensing searches are not limited by current measurement
accuracies for Earth-mass planets, which can produce relatively large
amplitude photometric signals (a few tenths of a magnitude or larger),
though small amplitude signals are more frequent. The limitations of
this method are rather of statistical nature: even if all stars acting
as microlenses have planets, only a small subset of them would show up
in the microlensed lightcurve, depending on the projected separation
and the exact geometry between relative path and planetary caustic.
Space measurements help significantly by reducing the photometric
confusion effects resulting from observations in very crowded regions
(such as the Galactic bulge) which are favoured fields to improve the
statistics of detectable events.

\clearpage
\section{The Period 2005--2015}

\subsection{Ground Observations: 2005--2015}

There are many ongoing ground-based surveys. At the time of writing,
the Planets Encyclopaedia www page
(http://www.obspm.fr/encycl/searches.html) lists as either ongoing or
planned: 18~radial velocity searches, 15~transit searches,
5~microlensing programmes, 10~imaging/direct detection programmes,
2~radio surveys, and 3~astrometric efforts. An overview of efforts and
expected results is given in this section, with a particular focus on
the ESA and ESO contributions.

\subsubsection{Radial Velocity Searches}
\label{sec:ground-radvel}

A summary of ongoing or planned radial velocity experiments is given
in Table~\ref{tab:radvel}.  It is certainly incomplete, and is
intended only to give a flavour for the activity in the field. The
vast majority of extra-solar planets discovered so far have been found
by radial velocity searches, which have a natural bias for the
discovery of massive planets orbiting close to their central star (hot
Jupiters).  As the surveys continue for longer periods of time they
become more and more sensitive to planets having longer periods, and to
additional planets in systems in which one hot Jupiter is already
known.  Dedicated designs have brought spectrographs very close to
the practical accuracy limit for ground-based radial velocity searches
of $\sim1$~m\,s$^{-1}$, allowing detection of lower mass planets.
HARPS has recently detected a second planet around $\mu$~Arae, with a
period of 9.5~days, a velocity semi-amplitude of less than
5~m\,s$^{-1}$ \citep{sbm+04}, and a derived $M\sin i$ of only 14~$M_\oplus$
(about one Uranus mass), making it the lowest mass planet found so far
(as of December 2004).  A planet of $\sim17\,M_\oplus$ has been
discovered orbiting the M~dwarf GJ~436 based on Keck data
\citep{bvm+04}, and of $\sim18\,M_\oplus$ reported for 55~Cnc using
HET observations \citep{mec+04}.

\begin{table}
\footnotesize
\caption[]{\footnotesize A summary of completed, operational or planned 
radial velocity searches (as of December 2004). The table is certainly
incomplete, and includes both large successful surveys as well as 
more uncertain plans; it is intended to give a flavour of the activity in
this field.}
\begin{center}
\begin{tabular}{llll} 
\hline
\noalign{\smallskip}
Name & PI & Telescope & Comments \\ 
\noalign{\smallskip}
\hline
\noalign{\smallskip}
AAA & Connes & 1.52 m  & EMILIE will provide m\,s$^{-1}$ accuracy\\
AFOE & Korzennik & 1.5 m & Relocating to Mount Wilson 2.5 m in 2004 \\
AAPS & Butler & 3.8 m & Precision 3 m\,s$^{-1}$, to operate to 2010 \\
California \& Carnegie & Marcy & 3/10 m & Precision 3 m\,s$^{-1}$, plan for dedicated 2.4 m \\
CES & K\"urster & 3.6 m & 1992--98 \\
Coralie & Mayor & 1.2 m & Precision 3 m\,s$^{-1}$, $>$1650 stars, 200 nights/year \\
ELODIE & Mayor  & 1.93 m & Since 1993 \\
Exoplanet Tracker & Ge & 11/5 m & Interferometry to $V\sim12$, test runs 2001--02\\
Fringing Spectrometer & Erskine  &  ? &  Precision  3~m\,s$^{-1}$ targeted\\
HARPS & Mayor & 3.6 m &  Precision 1~m\,s$^{-1}$, operational since 2003\\
HET & Cochran  & 11 m & Survey of 170 F--K stars $V<10$~mag \\
Magellan & --  & 6.5 m & Observes stars with OGLE transits, since 2003 \\
SARG  & Gratton & 3.6 m & Binary stars, study effect of metallicity \\
SOPHIE & Gillet & 1.93 m & Precision $<$3 m\,s$^{-1}$, northern HARPS counterpart\\
Spectrashift & Amateurs & 0.4 m & 1.1\,m, under construction\\
TOPS & Hatzes & 2 m & First planet recently announced \\ 
UCSD/Leiden/Berkeley & Quirrenbach & 0.6 m & K-giant survey at Lick CAT \\ 
UVES/FLAMES & ESO & 8 m & Available to community \\
\noalign{\smallskip}
\hline
\end{tabular}
\end{center}
\label{tab:radvel}
\end{table}

Another trend is that larger telescopes are being used for the radial
velocity searches. As a result the limiting magnitude of such searches
has increased from typically $V=7.5$ a few years ago to $V\sim12$,
with the number of stars thus available for radial velocity
study increasing by almost two orders of magnitude. For example,
the N2K Consortium is using the Keck, Magellan and Subaru telescopes
to track the next 2000 (N2K) closest ($<$110\,pc), brightest, and most
metal-rich FGK stars not on current Doppler surveys for new hot
Jupiters.  Started in early 2004, and with a precision of
4--7~m\,s$^{-1}$, the first Saturn-mass planet from this survey has
recently been reported \citep{flb+05}.

Table~\ref{tab:prospects-pre2015} summarises predictions of the
numbers of planets which might be detected by each of the methods
discussed in this section, including radial velocity measurements, out
to 2008--10. Although such a table is open to misinterpretation and
debate (being sensitive to planetary and instrumental hypotheses) it
provides some indication of the development of exo-planet statistics
over the next few years.

To derive these estimates for radial velocity observations, the
following arguments have been used: (a) during the past year (2003)
the number of target stars monitored has at least doubled (HARPS, new
Elodie programme, HET, others). New programmes will also probably
start in the coming years (e.g.\ Sophie at OHP).  The number of
presently known planets with masses in the range $0.5-10\,M_{\rm J}$
(`easy planets') can then probably be multiplied by $\sim3-4$; (b) the
typical precision is improving, and the number of planets with lower
masses will thus further increase.  Uncertainty remains on the
existence or frequency of low-mass planets in short-period orbits
(presently unknown), and in the detailed effects of stellar jitter;
(c) the number of long-period planets in the present surveys already
ongoing for several years will increase, as the distribution of planet
numbers increases with period. However, the maximum mass of detected
planets also increases with period, such that many high-mass planets
will be found, probably with masses $>10\,M_{\rm J}$.  Earth-masses
will be out of reach due to stellar jitter, except maybe for
short-$P$, low-mass transiting planets with $P$ and $e$ fixed, for
which a large number of measurements can be stacked at the appropriate
phase (see Section~\ref{sec:limits}).

More details are included of the ESO contribution, HARPS, since it
typifies the state-of-the-art technical and scientific objectives:

{\bf HARPS:} The Observatoire de Gen\`eve together with the
Physikalisches Institut der Universit\"at Bern, the Observatoire de
Haute-Provence, and the Service d'A\'eronomie du CNRS and in
collaboration with ESO, developed the HARPS spectrograph installed on
ESO's 3.6-m Telescope at La Silla.  The instrument is a
high-resolution, high-efficiency fibre-fed echelle spectrograph
designed to efficiently search for extra-solar planets reaching a
precision of 1~m\,s$^{-1}$ on radial-velocity measurements. Typically,
this precision is reached in 1~minute for a $V=7.5$ G-dwarf star. The
long-term precision is ensured by the spectrograph's stability: the
spectrograph resides in a pressure- and temperature-controlled vacuum
tank, with a drift usually well below 1~m\,s$^{-1}$ during one night,
which can be further corrected using the simultaneous thorium
technique.  HARPS has been available to the community since October
2003.

For the development of this instrument, the HARPS consortium has been
granted 500 guaranteed nights over 5~years (100~nights per year).  The
HARPS survey is designed to address several specific questions:

(a) only a few of the hundred detected planets have masses less than
the mass of Saturn, and due to the present precision of radial
velocity surveys the distribution of planetary masses is heavily
biased (or completely unknown) for masses less than half the mass of
Jupiter.  The high precision of HARPS will allow searches for low-mass
planets: for a sample of preselected non-active solar-type stars (from
the Coralie planet-search sample), the aim is to explore the domain of
the mass-function for short-period planets below the mass of
Saturn down to a few Earth masses;

(b) in a continuation of the planet-search programmes conducted over
$\sim$10 years, a quick screening of a large volume-limited sample of
$\sim$1000 still unobserved stars will be performed in order to
identify new `hot Jupiters' and other Jovian-type planets.  Increasing
the list of `hot Jupiters' will improve the prospects of finding
further stars with a planetary transit among relatively bright stars.
Better statistics are needed to identify new properties of the
distribution of exo-planet parameters.  This part of the programme has
already revealed two new short-period planets \citep{pmq+04};

(c) a systematic search for planets will be made for a volume-limited
sample of slowly-rotating non-binary M-dwarfs closer than 11\,pc.
Such a survey of very low mass stars should constrain the frequency of
planets as a function of stellar mass.  Up to now only one
planetary system orbiting an M-dwarf is known.  For the less massive
stars short-period planets of only a few times the mass of the Earth
could be detected.  Since most of these objects are faint, high
efficiency is required.  These objects are of prime importance for
future astrometric studies to be carried out with the VLTI or SIM;
 
(d) stars with detected giant planets exhibit an impressive excess of
metallicity in contrast to stellar samples without giant planets.  The
excess of metallicity does not seem related to the mass of the
convective zone and probably originates in the chemical composition of
the primordial molecular cloud.  To add new constraints to the link
between star chemical composition and frequency (or properties) of
exo-planets, two programmes are being carried out. The first is a
search for exo-planets orbiting solar-type stars with notable
metal deficiency (for most of them [Fe/H] between $-0.5$ and $-1.0$).  Among
the existing detections of exo-planets only two or three have been
found with metallicity in that range. The aim is to estimate the
frequency of exo-planets in that domain of metallicity and, if
possible, to compare their characteristics (masses, orbits) to planets
orbiting metal-rich stars;

(e) the second `abundance-related' programme aims at exploring the
link between stellar metallicity and properties of exo-planets.  Visual
binaries with solar-type stars of almost identical magnitudes have
been selected. For those including giant planets a detailed chemical
analysis will be done for both stellar components to search for
possible differences in their chemical compositions;

(f) follow-up radial velocity measurements for stars with planetary
transits detected by COROT will be made with HARPS (where 
the photometric transit provides an estimate of the radius of the
transiting planet as well as the orbital period and phase).
Complementary ground-based spectroscopic measurements with HARPS will
constrain the planetary mass and thus the planet mean density.  The
main scientific return for the planetary programme of the COROT
mission will come from this combination of photometric and radial
velocity data.

\subsubsection{Transit Searches}
\label{sec:ground-transits}

The transit method aims at detecting the dimming of the stellar light
by occultation due to an orbiting planet.  Transit experiments offer a
number of very important contributions: (i) searches can be conducted
over wide fields over long periods (5~years or more), and are
therefore potentially efficient at detecting previously unknown
systems; (ii) from the ground they are able to detect massive
transiting planets, especially the `hot Jupiters', while from space
planets down to Earth-mass or below can be detected; (iii)
spectroscopy during the planet transit can yield physical diagnostics
of the transiting planets.  A summary of ongoing or planned transit
experiments is given in Table~\ref{tab:transits}; see also the recent
review by \cite{hor03}.  Again, Table~\ref{tab:prospects-pre2015}
summarises predictions of the numbers of planets that might be
detected by this method out to 2008--12.

Transit measurements can only detect planets with a favourable
orientation of their orbital plane, implying that only a small
fraction of planets can ever be detected or monitored using this
technique. In particular a nearby census can only reveal a small
fraction of existing systems.  The probability of viewing a planetary
system edge-on depends on the distance of the planet to the central
star. For close-in orbits it is about 10\%, decreasing for more
distant planets. Transit searches therefore try to maximise the number
of stars they can observe simultaneously.  This can be reached by
either small telescopes with wide field of view observing relatively
bright stars, or large telescopes providing deep exposures to increase
the number of targets monitored. The current transit search surveys
therefore group into two classes, using small or larger ($>70$~cm)
telescopes. Support for the implicit hypothesis that the orbits of
planetary systems are randomly distributed in the Galaxy comes from
the fact that stellar rotation axes are themselves randomly
distributed (obtained by $v\sin i$ distributions, and independently
confirmed by magnetic-field orientation studies).

The wide-angle survey teams follow STARE and Vulcan in using small
(10~cm) wide-angle (10$^\circ$) CCD cameras with a pixel size of order
1~arcsec or larger, sacrificing angular resolution to expand the field
of view. The faint limit, at $V\sim12-13$ reaches to
$d\sim300-500$\,pc, comparable to the disk scale height, so that
target fields cover the entire sky, which may contain some 1000 hot
transiting Jupiters to this limit \citep{hor03}. The deep surveys use
(mosaic) cameras on 1--4~m telescopes, reaching $V\sim19-21$ and
$d\sim4-5$~kpc, so that Galactic plane and open cluster fields are
primary targets. \cite{hor03} predicts up to 200~hot Jupiters per
month being discovered by ongoing ground transit surveys in the future
(perhaps by the year 2010).  

The size of planets detectable from transits with ground-based
searches is limited by the Earth's atmosphere
(Section~\ref{sec:limits}). The photometric precision of typical
lightcurves is a little under 1\%, which corresponds to about
Jupiter-sized planets for solar-type stars. Ground-based surveys are
further limited in their time coverage of potential transiting planets
by daytime and bad weather periods. The small number of transit
surveys operating over more than a few months so far, and the lack of
continuous observations on the target fields, are the most likely
reason why only few planets have been discovered through transit
detections so far, somewhat in contrast with the large number of
surveys listed in Table~\ref{tab:transits}.  The situation could be
significantly improved by ground-based networks of telescopes spanning
a range of longitudes to ensure continuous observational coverage of
target fields.

\begin{table}[p]
\footnotesize
\caption[]{\footnotesize A summary of planned or operational transit 
searches, from space and ground.}
\begin{center}
\setlength\tabcolsep{5pt}
\begin{tabular}{lrrl}
\hline
\noalign{\smallskip}
Name &    D(cm)  &  FOV (deg) & Comments/Status \\
\noalign{\smallskip}
\hline
\noalign{\smallskip}
\quad Space: \\
MOST      &          15 & $2.5\times2.5$ & Primarily asteroseismology \\
COROT     &          30 & $2.8\times2.8$ & Asteroseismology and transits  \\ 
Kepler    &          95 &   $10\times10$ & Transits: Earth-size planets or larger \\
Eddington & $3\times70$ & $4.5\times4.5$ & Transits: not approved \\
MPF       &         120 & $1.2\times1.2$ & Microlensing and transits (previously GEST)\\
HST       &         240 & $0.05\times0.05$ & Specific transit observations possible \\
\noalign{\smallskip}
\hline
\noalign{\smallskip}
\quad Ground: \\
PASS &
3.6 &
15$\times$ (28$\times$28) &
Observations planned at Tenerife \\
KELT &
4.2 &
26$\times$26 &
Tests in New Mexico \\
WASP-0 &
6.4 &
8.8$\times$8.8 &
Tests on La Palma and in Greece \\
ASAS-3 &
7.1 &
2$\times$ (8.8$\times$8.8) &
Operational since Aug 2000 at Las Campanas \\
SLEUTH (TrES) &
10.0 &
5.66$\times$5.66 &
Operational since May 2003, Palomar Observatory \\
STARE (TrES)  &
10.0 &
6.03$\times$6.03 &
Operational since 1999, Tenerife \\
PSST (TrES)  &
10.7 &
5.29$\times$5.29 &
Operational, Arizona \\
HATnet &
11.0 &
5$\times$ (8.2$\times$8.2) &
       HAT-5 operational since Feb 2003 \\
 & & & HAT-6 \& HAT-7 operational since Sep 2003 \\
 & & & HAT-8 operational since Nov 2003 at Mauna Kea \\
Super-WASP &
11.1 &
4$\times$ (15.9$\times$15.9) &
       Operational since Apr 2004 on La Palma; \\
 & & & \qquad second system under construction \\
RAPTOR-P &
14.0 &
5$\times$ (4.2$\times$4.2) &
Under construction in Fenton Hill, New Mexico \\
Vulcan &
12.0 &
7.04$\times$7.04 &
Operational since 1999 at Mount Hamilton, California \\
BEST &
20.0 &
3.1$\times$3.1 &
       Operational since Jul 2001 in Tautenburg, \\
 & & & Relocation to OHP (France) in 2004 \\
Vulcan-South &
20.3 &
6.94$\times$6.94 &
       First light 2004, 1 field in Carina observed, Antarctic\\
APT &
50.0 &
2$\times$3 &
       Operational at Siding Spring, temporarily used\\
TeMPEST &
76.0 &
0.77$\times$0.77 &
Operational at McDonald Observatory, Texas \\
STELLA-2 &
80.0 &
0.50$\times$0.50 &
Commissioning planned for 2005, Tenerife \\
EXPLORE-OC &
101.6 &
0.25$\times$0.40 &
Operational in Las Campanas, temporarily used \\
PISCES &
120.0 &
0.38$\times$0.38 &
Operational \\
STELLA-1 &
120.0 &
0.37$\times$0.37 &
Comissioning starts at the end of 2004 at Tenerife \\
MONET &
120.0 &
various &
       Under construction at McDonald Observatory, \\
 & & & \qquad Texas, and in Sutherland, South Africa \\
ASP &
130.0 &
0.17$\times$0.17 &
Operational at Kitt Peak \\
OGLE-III &
130.0 &
0.58$\times$0.58 &
Operational at Las Campanas, temporarily used \\
STEPPS &
240.0 &
0.41$\times$0.41 &
Operational at Kitt Peak, temporarily used \\
INT &
250.0 &
0.57$\times$0.57 &
Operational at La Palma, temporarily used \\
OmegaTranS &
260.0 &
1.00$\times$1.00 &
Planned for guaranteed time with OmegaCam at VST \\
EXPLORE-N &
360.0 &
0.57$\times$0.57 &
Operational at Mauna Kea, temporarily used \\
EXPLORE-S &
400.0 &
0.61$\times$0.61 &
Operational at Kitt Peak/CTIO, temporarily used \\
\noalign{\smallskip}
\hline
\end{tabular}
\end{center}
\label{tab:transits}
\end{table}

The discovery of a temporary dimming of the stellar lightcurve alone
is not sufficient to secure the detection of a transiting planet.
Grazing eclipsing binary stars, background binaries, brown dwarfs and
stellar spots can cause lightcurves similar to transiting planets.
Follow-up measurements, in particular radial velocity measurements,
and determination of stellar parameters, therefore play an
important role in the detection of planetary transits to exclude other
causes of light dimming.

Up to now, five confirmed planets have been {\it discovered\/} by
ground-based transit searches: four using the 1.3~m OGLE telescope,
and one with a 10~cm ground-based system (TrES-1). The OGLE experiment
(Optical Gravitational Lensing Experiment) uses the 1.3~m Warsaw
Telescope at Las Campanas Observatory, Chile. It is equipped with a
mosaic of 8~CCDs of 2k$\times$4k each, giving a field of view of
35~arcmin square with 0.26~arcsec/pixel. The telescope is primarily
used to search for microlensing events by viewing near the Galactic
centre, but significant time (more than three months) was made
available for transit searches.  It has monitored some 52\,000 disk
stars for 32~nights, reporting some 100 transit candidates with
periods ranging from 1--9~days (e.g.\ Udalski et al.\ 2002a;
2002b) based solely on the dimming of stellar lightcurves. Most of
them were quickly identified as stellar binary systems. Radial
velocity follow-up for many of the candidate stars was difficult
because of the faintness of the stars (down to $I\sim16$~mag).
Nevertheless, four events were confirmed as transiting planets by
radial velocity follow-up measurements \citep{ktj+03, bps+04, pbq+04}.

The TrES telescopes belong to the class of small telescopes dedicated
to transit searches. All three telescopes of this transit search
programme are small aperture (10~cm) wide-field ($6^\circ$) systems.
They are located at Tenerife, Lowell Observatory, and Palomar
Mountain, and thus span a range of longitudes \citep{abt+04}.
Recently, the first planet found by this system has been announced.
The discovery is again based on the lightcurves and radial velocity
confirmation, showing that small-scale systems indeed have the
potential to find transiting planets, providing their observational
coverage is sufficiently high.

These examples show the potential of the transit method to find a
large number of planets, including small planets, in an unbiased
sample of stars. The full potential will be exploited in future space
missions, from which Earth-mass planets can be detected 
(Section~\ref{sec:space2005-15}).

In addition to the geometric information derived rather directly from
accurate photometric measurements of planetary transits, high-cadence,
high S/N spectroscopy of transit events can reveal properties of the
planetary atmosphere and exosphere.  Extensive work on the first
transiting planet, HD~209458b, has shown the level of current
photometric and spectroscopic capabilities, principally using HST
before the failure of STIS. Ground-based photometry was able to reach
a precision of ~0.2\% \citep{hmb+00, cbn+02, jcg+00, dgc01} while
HST/STIS has achieved $\sim0.01$\% \citep{bcg+01}. The high-cadence
capability of the HST Fine Guidance Sensor (FGS) is also being
exploited for transit timing \citep{skk+04}.

The use of time-resolved spectroscopy by \cite{cbn+02} showed that the
HD~209458b transit was $2.3\times10^{-4}$ deeper when observed at the
sodium~D lines. Again using STIS, \cite{vld+03} detect a very large
(15\%) transit depth at Ly-$\alpha$, showing that the planet is losing
mass.  Subsequently the same group \citep{vdl+04} reported a detection
of C~and~O in the exosphere. In addition to these lines, the
possibility exists, in the optical band, of looking for the effects of
water (longward of 500\,nm) and of Rayleigh scattering in the blue.
\cite{mcs+03} searched unsuccessfully for He~I 1083\,nm absorption
using the VLT with ISAAC. Their upper limit of 0.5\% at 3$\sigma$ for
a 0.3\,nm bandwidth was limited by the detector fringing properties at
this wavelength.  An alternative approach is to search for an infrared
signature during secondary eclipse. This method, applied by
\cite{rdw+03} which they call `occultation spectroscopy', searches for
the disappearance and reappearance of weak spectral features due to
the exo-planet as it passes behind the star.  They argue that at the
longest infrared wavelengths, this technique becomes preferable to
conventional transit spectroscopy. They observed the system in the
wing of the strong $\nu_3$~band of methane near 3.6\,$\mu$m during two
secondary eclipses, using the VLT/ISAAC spectrometer at a spectral
resolution of 3300 but were unable to detect a signal.

A recent study by \cite{hm05} has shown that for many planets discovered
by transit surveys, accurate timing measurements between successive
transits (of accuracies between 0.1--100~minutes) will allow for the 
detection of additional planets in the system (not necessarily
transiting) via their gravitational interaction with the transiting
planet. The transit time variations depend on the mass of the additional
planet, and in some cases Earth-mass planets will produce a measurable 
effect. This effect is particularly prominent for long-period
transiting systems, where the `perturber' (e.g.\ an Earth-mass
planet) is close to orbital resonance \citep{ass+05}.

The possibilities for future follow-up studies of this nature at
optical and UV wavelengths are seriously compromised by the failure of
STIS on HST. Longer wavelength absorption spectroscopy (1--28\,$\mu$m)
should be possible with the NIRSpec and MIRI instruments on JWST
provided that these are configured to allow efficient high-cadence and
high S/N observations. If HST is followed by another, similar aperture
optical UV telescope before 2015, it is likely that strong arguments
will be made to equip it to allow STIS-type transit spectroscopy.

\vspace{10pt}
{\bf CRIRES:} CRIRES is a cryogenic high-resolution infrared echelle
spectrometer to be installed at the UT1 of the VLT in 2005. It covers
the wavelength region 950--5200~nm at a maximum resolution of 100\,000
(0.2~arcsec slit). The instrument has been designed for stability, and
will be suited for radial velocity studies. Furthermore it may be the
most powerful ground-based instrument for transit spectrosocopy in the
infrared.  Interference from the atmosphere is a severe complication
at infrared wavelengths and the gain in resolution of factor 30 with
respect to ISAAC will help alleviate this problem. \cite{kau02}
discusses the use of OH~lines in the K~band for the detection of
extra-solar planet atmospheres. Hydrocarbons like C$_2$H$_2$ or CH$_4$
are prominent constituents in the atmospheres of Jupiter-like planets
in the solar system, and also provide lines in the operating range of
CRIRES. The isotope shift of $^{12}$CO and $^{13}$CO will be well
resolved \citep{bbt02, bbo04}. CRIRES will also allow analysis of the
atmospheres of planets, moons and comets in our solar system, some of
which have a rich organic chemistry.  Performed in close collaboration
with the solar system community, a survey at high-spectral resolution
will result in a reference library for study of extra-solar planets.
Comparsion with measurements from space will result in a much better
understanding of the relation between integrated spectrum and local
physical conditions in the atmospheres.

\subsubsection{Reflected Light} 

Additional phase-dependent effects, such as the modulation of light
reflected from a planet, should also be detectable by accurate
photometric satellites with a precision of order $10^{-4}$. The method
can be used, in principle, both for independent detection of planets,
and for studying known hot Jupiters. It might be more effective than
transit searches in some cases, because the effect may be observable
for a wide range of inclinations, and is not confined to a narrow
angle around $90^\circ$.  The technique is applicable in two cases:

(a) the stellar light reflected by the planet. The ratio of reflected
light to the stellar light is of the order of $A (R/2d)^2 F(\phi)$,
where $A$ is the albedo of the planet, $d$ is the planetary orbital
distance, and $F(\phi)$ is a function of the orbital phase, of the
order of unity. For a 2.5-day period, Jupiter-radius planet at an orbital
separation of $7R_\odot$ around a solar-type star, this ratio is
$5\times10^{-5}A$, which is almost detectable with COROT or Kepler.
For $R=1.4\,R_{\rm J}$, as in HD~209458, the ratio is $10^{-4}A$.
\cite{as04} point out that the shape of the modulation is sensitive to
whether the planet has rings.

(b) a stronger effect might be the black-body emission of a close
planet.  If the planet rotation is synchronised with the orbital
motion, one side of the planet faces the parent star all the time,
resulting in different temperatures of the two sides of the planet.
Estimates suggest differences of up to 1500~K or more, depending on
the distance from the star and on the circulation streams in the
planetary atmosphere. The hot side of the planet can contribute some
fraction of the light of the system, again with a sinusoidal
modulation. In the infrared tail of the black-body radiation, the
ratio of the planet to star emission could be of the order of 1/400
(Mazeh, private communication), assuming that the far side of the
planet is cold.  The effect can be seen only in the infrared, and only
if there is a large temperature difference between the two sides of
the planet.

\subsubsection{Microlensing Searches}
\label{sec:ground-microlensing}

Microlensing searches for extra-solar planets take advantage of the
very characteristic temporal magnification of a (bright) background
star, due to a (faint) star-plus-planet system passing in front of it,
as seen from Earth.  This method is very different from all the other
techniques used for planet searching.  There are a number of apparent
disadvantages: (i) very small probability for a planetary microlensing
event, even if all stars have planets (of order $10^{-8}$ for
background stars in the Galactic bulge); (ii) potential planets are
much more distant than those found with other techniques (of order a
few kpc), which means subsequent more detailed investigations of the
planet are close to impossible; (iii) the duration of the
planet-induced deviation in the microlensing lightcurve can be very
short (typically hours to days), and the measurement is not
repeatable: it is a once-and-only event; (iv) lightcurve shapes caused
by extra-solar planets can be very diverse and do not always yield a
unique planet mass/separation fit; (v) the derived property is not the
planet mass, but the mass ratio between host star and planet.  However
these apparent disadvantages of the microlensing method can largely be
`overcome', and are more than balanced by its many advantages:

(i) no bias for pre-selected nearby host stars: microlensing will
provide a fair `mass-selected' sample of the planet population in the
Milky Way;

(ii) no strong bias for planets with large masses: the duration of the
planetary signal in the lightcurve is roughly proportional to the
square root of the planet mass (with a wide spread); its amplitude,
however, is independent of the planet mass to first order (though
affected by the finite size of the source star);

(iii) Earth-bound method sensitive down to (almost) Earth-masses:
Microlensing is sensitive to lower-mass planets than most other
methods (except pulsar searches).  In principle, it is possible to
even detect Earth-mass planets with ground-based monitoring via
microlensing.  In practice, however, this would mean extremely high
monitoring frequency and photometric accuracy;

(iv) most sensitive for planets within `lensing zone', which overlaps
with habitable zone: microlensing is not sensitive to very close-in
planets: the signal would be undistinguishable from a star with the
combined mass.  In the current mode of operation (planet searching
with high monitoring frequency of `alerted' events), far out planets
are not detected either.  The most likely detection range is the
so-called lensing zone (between 0.6--1.6 Einstein radii), roughly
corresponding to projected separations of a couple of AU;

(v) multiple planet systems detectable: The detection of more than one
planet per system is certainly possible with microlensing, though its
probability is probably another order of magnitude smaller than for a
single planet;

(vi) `instantanous' detection of large semi-major axes possible: The
measured (projected) distance between planet and host-star of
typically a few~AU is, though, only a lower limit to the real
semi-major axis;

(vii) detection of free-floating planets (i.e.\ isolated bodies of
planetary mass) possible: space-based searches will have high enough
photometric accuracy and monitoring frequency to detect and
characterise any existing free-floating planets;

(viii) ultimately best statistics of galactic population of planets:
Gravitational microlensing will ultimately provide the best
statistics for planets in the Milky Way; it is not bias-free, but
the biases in the search technique are of very different character
from those of all other methods, can easily be quantified and are 
more favorable for global (i.e.\ Galactic) statistics.

These characteristics of the microlensing technique mean that
it is complementary to the other search methods.

The first planet detection with the microlensing technique was
published in May~2004 \citep{buj+04}: The OGLE- and MOA-teams detected
a clear caustic-crossing microlensing signal in event
OGLE-2003-BLG-235 or MOA-2003-BLG-53 which could only be reproduced with a
binary-lens model involving a mass ratio of $q = 0.0039^{+11}_{-07}$.
The planetary deviation lasted about one week, with a measured maximum
magnification of more than a factor of 12.  Assuming a low-mass main
sequence primary, this would correspond to a planet of about $m_{\rm
P} \sim 1.5\,M_{\rm J}$ at a projected separation of about
$d\sim3$~AU.

A summary of ongoing or planned microlensing experiments is given in
Table~\ref{tab:microlensing}.  (Stellar) microlensing events continue
to be observed at large rates, particularly towards the Galactic bulge
(which is the prime search direction due to the high density of
background stars).  Both MOA and OGLE regularly post their
microlensing alerts on their web sites.  In the context of exo-planets,
specialized networks have been established (PLANET, MicroFUN) which
perform follow-up observations of `alerted' events at high time
resolution and look for possible planetary perturbations in the
stellar microlensing light curves.  In the 2004 observing season, OGLE
alone had detected and alerted on more than 600 stellar microlensing
events. With the MOA inauguration of a new 1.8-m dedicated telescope in
December 2004, more than 1000 stellar microlensing events will be
found per season from 2005 onwards.

A number of studies looked into the statistics of planets from
microlensing searches.  They come in two kinds, either providing
detection/exclusion probabilities for planets in individual
lightcurves or for ensembles of events:

In the case of MACHO 98--BLG--3, the estimated probability for
explaining the data without a planet is $<1$\%.  The best planetary
model has a planet of $0.4-1.5\,M_\oplus$ at a projected radius of
either 1.5 or 2.3\,AU \citep{brs+02}.

Very high magnification events are well-suited for showing signatures
of planets, because the relative track is very close to the central
caustic which should be slightly perturbed by the existence of any
planet (at the same time, such planets are difficult to characterise
uniquely): In the case of MOA~2003--BLG--32 = OGLE~2003--BLG--219
\citep{abb+04}, with a peak magnification of more than 500, continuous
observations around the maximum did not show any planetary
signature. This enabled the authors to put very stringent limits on
the probability of a companion: planets of $m_{\rm P} = 1.3\, M_\oplus$
are excluded from more than 50\% of the projected annular region from
$\sim2.3-3.6$\,AU surrounding the lens star, Uranus-mass planets from
0.9--8.7\,AU, and planets 1.3 more massive than Saturn are excluded
from 0.2--60\,AU.

The best published statistical limits on the frequency of Jupiter-mass
planets from (lack of) microlensing signatures can be found in
\cite{gaa+02}, based on the first five years (1995--99) of PLANET team
data.  They concluded that less than 33\% of the M-dwarfs in the
Galactic bulge have Jupiter-mass companions with a projected
separation between 1.5--4\,AU. Many more stellar events have been
monitored subsequently, and improved limits should become available
soon.

The availability of the VST may soon provide a means for ESO to
support a massive microlensing search for planets (cf. Sackett 1997,
Appendix~C of the Final report of the ESO Working Group on the
Detection of Planets).  

A space-based microlensing mission (MPF/GEST) is discussed in
Section~\ref{sec:space-microlensing}.

\begin{table}[t]
\footnotesize
\caption[]{\footnotesize A summary of completed, operational or planned
microlensing searches.}
\begin{center}
\begin{tabular}{llll} 
\hline
\noalign{\smallskip}
Name & PI & Telescope & Comments \\ 
\noalign{\smallskip}
\hline
\noalign{\smallskip}
MACHO & Alcock & 1.3 m & 1992--99: $\sim10^7$ stars in LMC, $\sim10^7$ in Milky Way \\
MOA & Network & 0.6 m & 1.8 m with $2^\circ$ field planned for 2005\\
MPS & Bennett & 1.9 m & \\
PLANET &  --& 0.6/0.9/1.0/2.2 m & Collaboration of 16 institutes, 10 countries \\
MicroFUN & --& multiple & Microlensing follow-up\\
MPF/GEST & Bennett & 1--1.5 m & Proposal for NASA Discovery Mission\\ 
\noalign{\smallskip}
\hline
\end{tabular}
\end{center}
\label{tab:microlensing}
\end{table}

\subsubsection{Astrometry}
\label{sec:ground-astrometry}

The principle of planet detection with astrometry is similar to that
underlying the Doppler technique: the presence of a planet is inferred
from the motion of its parent star around the common centre of
gravity. In the case of astrometry the two components of this motion
are observed in the plane of the sky; this gives sufficient
information to solve for the orbital elements without $\sin i$
ambiguity. Astrometry also has advantages for a number of specific
questions, because this method is applicable to all types of stars,
and more sensitive to planets with larger orbital semi-major
axes. Astrometric surveys of young and old planetary systems will
therefore give unparalleled insight into the mechanisms of planet
formation, orbital migration and evolution, orbital resonances, and
interaction between planets. Interferometric techniques should improve
astrometric precision well beyond current capabilities. 

Specific applications are:

(a) mass determination for planets detected in radial velocity surveys
(without the $\sin i$ factor). The radial velocity method gives only a
lower limit to the mass, because the inclination of the orbit with
respect to the line-of-sight remains unknown.  Astrometry can resolve
this ambiguity, because it measures two components of the orbital
motion, from which the inclination can be derived;

(b) confirmation of hints for long-period planets in radial velocity
surveys.  Many of the stars with detected short-period planets also
show long-term trends in the velocity residuals \citep{fmb+01}.  These
are indicative of additional long-period planets, whose presence can
be confirmed astrometrically;

(c) inventory of planets around stars of all masses. The radial
velocity technique works well only for stars with a sufficient number
of narrow spectral lines, i.e., fairly old stars with
$M<1.2\,M_\odot$. Astrometry can detect planets around more massive
stars and complete a census of gas and ice giants around stars of all
masses;

(d) detection of gas giants around pre-main-sequence stars, signatures
of planet formation. Astrometry can detect giant planets around young
stars, and thus probe the time of planet formation and
migration. Observations of pre-main-sequence stars of different ages
can provide a test of the formation mechanism of gas giants. Whereas
gas accretion on $\sim 10\,M_\oplus$ cores requires $\sim 10$\,Myr,
formation by disk instabilities would proceed rapidly and thus produce
an astrometric signature even at very young stellar ages;
    
(e) detection of multiple systems with masses decreasing from the
inside out.  Whereas the astrometric signal increases linearly with
the semi-major axis $a$ of the planetary orbit, the radial velocity
signal scales with $1 / \sqrt{a}$. This leads to opposite detection
biases for the two methods. Systems in which the masses increase with
$a$ (e.g., $\upsilon$\,And) are easily detected by the radial velocity
technique because the planets' signatures are of similar
amplitudes. Conversely, systems with masses decreasing with $a$ are
more easily detected astrometrically;

(f) determine whether multiple systems are coplanar or not. Many of
the known extra-solar planets have highly eccentric orbits. A plausible
origin of these eccentricities is strong gravitational interaction
between two or several massive planets. This could also lead to orbits
that are not aligned with the equatorial plane of the star, and to
non-coplanar orbits in multiple systems.

\vspace{10pt} Astrometric observations by interferometry are based on
measurements of the delay $D = D_{\rm int} + (\lambda / 2 \pi) \phi$,
where $D_{\rm int} = D_2 - D_1$ is the internal delay measured by a
metrology system, and $\phi$ the observed fringe phase. Here $\phi$
has to be unwrapped, i.e., not restricted to the interval $[0, 2
\pi)$. In other words, one has to determine which of the sinusoidal
fringes was observed. This can, for example, be done with
dispersed-fringe techniques \citep{quir01}. $D$~is related to the
baseline $\vec{B}$ by $D = \vec{B} \cdot \hat{s} = B \cos \theta$,
where $\hat{s}$ is a unit vector in the direction towards the star,
and $\theta$ the angle between $\vec{B}$ and $\hat{s}$. Each data
point is thus a one-dimensional measurement of the position of the
star $\theta$, provided that the length and direction of the baseline
are accurately known. The second coordinate can be measured with a
separate baseline at a roughly orthogonal orientation.  The photon
noise limit for the precision $\sigma$ of an astrometric measurement
is given by $\sigma = (1/{\rm SNR}) \cdot (\lambda/(2\pi B))$.  Since
high signal-to-noise ratios can be obtained for bright stars, $\sigma$
can be orders of magnitude smaller than the resolution $\lambda / B$
of the interferometer. For example, the resolution of SIM ($B =
10$\,m) is about 10\,milli-arcsec, but the astrometric precision should
approach 1\,micro-arcsec; for PRIMA, $B=200$\,m, the resolution is
2\,milli-arcsec, and the astrometric precision should approach
10\,micro-arcsec.

Because of the short coherence time of the atmosphere, precise
astrometry from the ground requires simultaneous observations of the
target and an astrometric reference. In a dual-star interferometer, each
telescope accepts two small fields and sends two separate beams
through the delay lines. The delay difference between the two fields
is taken out with an additional short-stroke differential delay line;
an internal laser metrology system is used to monitor the delay
difference (which is equal to the phase difference multiplied with
$\lambda / 2 \pi$). For astrometric observations, this delay
difference $\Delta D$ is the observable of interest, because it is
directly related to the coordinate difference between the target and
reference stars; it follows that $\Delta D \equiv D_t - D_r = \vec{B}
\cdot \left( \hat{s}_t - \hat{s}_r \right) = B (\cos \theta_t - \cos
\theta_r)$, where the subscript $t$ is used for the target, and $r$
for the reference. To get robust two-dimensional position
measurements, observations of the target with respect to several
references and with a number of baseline orientations are required.

Measurements of the delay difference between two stars give relative
astrometric information; this means that the position information is
not obtained in a global reference frame, but only with respect to
nearby comparison stars, which define a local reference frame on a
small patch of sky.  This approach greatly reduces the atmospheric
errors, and some instrumental requirements are also relaxed.  The
downside is that the information that can be obtained in this way is
more restricted, because the local frame may have a motion and
rotation of its own. This makes it impossible to measure proper
motions. Moreover, all parallax ellipses have the same orientation and
axial ratio, which allows only relative parallaxes to be measured.

\vspace{10pt}
Specific instrument approaches are discussed in
Section~\ref{sec:ground-direct} (NAOS-CONICA, Planet Finder, PRIMA)
and Section~\ref{sec:space-astrometry} (Gaia, SIM, etc.).  

No planets have been discovered using this technique to date.

\subsubsection{Direct Detection}
\label{sec:ground-direct}

The light coming from an extra-solar planet is much fainter (of order
$10^9$ in the optical, and a factor 10--100 less in the infrared) than
the signal from the star. Therefore the challenge is to build
instruments that are able to provide extremely high contrast and
spatial resolution. The different approaches are summarised below and
in Table~\ref{tab:direct}.  The first direct detection of a young
planet may already have been achieved by a team using NACO on the VLT.
An object detected close to 2MASS~WJ1207334--393254 is either a planet
or possibly a brown dwarf \citep{cld+04}.  Regardless of the exact
nature of this particular object, it is likely that imaging of more
massive, young extra-solar planets will become more feasible in the
near future.

A number of programmes are using, or planning to use, interferometry
to achieve high spatial resolution.  Destructive interference can be
used to remove most of the light from the central star (nulling).  ESA
and ESO are collaborating on a nulling demonstrator for Darwin called
GENIE which will be used on the VLT.  Other searches use a
coronographic approach to block out the star's light.  The Lyot
project's coronograph has now been declared fully operational and will
conduct a survey of 300~nearby stars.

\begin{table}[t]
\footnotesize
\caption[]{\footnotesize A summary of planned or operational direct
searches. Space astrometry and transit searches are covered elsewhere,
and are excluded from this table. The first category are interferometric
or multi-telescope projects; the second are single telescope direct methods;
the third are in the radio or sub-mm.}
\begin{center}
\begin{tabular}{llll} 
\hline
\noalign{\smallskip}
Name & PI & Telecope & Comments \\ 
\noalign{\smallskip}
\hline
\noalign{\smallskip}
Darwin & ESA & 3 $\times$ 3.5 m  & IR interferometer in L2, launch goal 2015 \\
TPF-I & NASA & several 3--4 m & IR interferometer launch before 2020\\
AMBER & Petrov & VLT & Operations recently started \\ 
ANI & Hinz & 6.5 m  & Tests on nulling in 1997--98\\
AIC & Gay & 1.5 m & First test runs in 1997 \\
BLINC & Hinz & 6.5 m & Instrument ready for use at MMT\\
Carlina & Le Coroller & hypertelescope & First fringes with $2\times24$~cm mirrors, 2004\\
Keck i/f & & $2\times10$ m & First fringes 2001, shared risk science 2002 \\
KEOPS & & 1 km array & Time frame 2015+ \\
LBT & & $2\times8.4$ m & First light planned for Dec 2004\\
PTI & Colavita & 3 $\times$ 0.4 m &  K-band stellar interferometer, $\sim$100~$\mu$as  \\
VLTI & Glindemann & 8/1.8 m & MIDI operational since 2003 \\
\noalign{\smallskip}
\hline
\noalign{\smallskip}
TPF-C & NASA & 4--6 m  &  Visible light coronograph launch 2014 \\
CHEOPS & Feldt & 8 m & Proposal for VLT instrument \\
EXPORT & Eiroa & several & Observations in 1998/99\\
JWST & NASA & 6.5 m & IR-optimised telescope in L2, launch 2011\\
Lyot Project &  Oppenheimer & 3.6 m & First light April 2003 \\
MIRLIN & Ressler & 3, 5, 10 m & Operational since 2001\\
Trojan & Caton & 0.45/0.7 m & 2-colour photometry of 20 eclipsing binary \\
\noalign{\smallskip}
\hline
\noalign{\smallskip}
ALMA & ESO & $64\times12$ m & First light planned for 2007\\
LOFAR & & 100 antennae & Decametric emission: initial 2006/full 2008 \\
Nancay & & radio arrays & \\
Pulsar Planets & Wolszczan &305 m  & Planets around PSR~B1257+12\\
\noalign{\smallskip}
\hline
\end{tabular}
\end{center}
\label{tab:direct} 
\end{table}

{\bf NAOS-CONICA:} NAOS-CONICA (NACO) is installed on UT3 at the VLT.
It is an adaptive optics system working in the 1--5\,$\mu$m range,
with a Shack-Hartmann wavefront analyser operating in the visible or
near-infrared. The instrument is equipped with a large collection of
broad- and narrow-band filters for imaging, and a set of grisms for
low-dispersion spectroscopy. It also features a polarimetric system.
In good conditions, with a bright reference star, a Strehl ratio of
$\sim0.5$ is achievable.

Two modes are of special interest for planetary observations: (1)~a
coronographic mode: a classical Lyot-type instrument with a circular
focal spot (of 0.7 or 1.4~arcsec) is used together with an undersized
pupil mask. A four quadrant phase mask (which introduces a shift of
$\pi$ to the wavefront) is being commissioned. It reduces the light of
the central star by a factor of $\sim70$, and permits observations
within 0.35~arcsec of the centre, i.e.\ much closer than with a
classic mask; (2)~simultaneous differential imaging over
$5\times5$~arcsec$^2$: four images are obtained simultaneously through
3~narrow-band filters. Two are taken in the 1.625\,$\mu$m methane
feature, and the two others at 1.575 and 1.600\,$\mu$m, outside the
spectral line. The data are registered simultaneously in the four
channels, and the point-spread function (including all its residual
aberrations, and the speckles, including super-speckles) are identical
in all four images. This mode was designed to search for methane-rich
objects near very bright stars, with a contrast of 50\,000 accessible.
With these characteristics, a Jupiter-like planet in a Jupiter-like
orbit around a very nearby star (within 5\,pc) should be just
detectable at its largest elongation. Detection performances are
limited by uncorrected phase residuals (mainly low-order aberrations).
NACO should be considered as a prototype permitting the study of novel
techniques that will be used in dedicated instruments such as ESO's
Planet Finder.

\vspace{10pt}
{\bf Planet Finder:} 
\label{sec:planet-finder}
The next step beyond the VLT AO facility NAOS-CONICA (NACO) would be a
dedicated VLT instrument optimised for the detection of extra-solar
planets. Two independent design studies are currently underway for
such a Planet Finder instrument at the VLT. The prime goal is to gain
at least an order of magnitude with respect to NACO in the detection
of faint objects very close to a bright star, ideally reaching giant
planets.  Much higher Strehl ratios than NAOS, around 0.9 in the
K-band, are targetted. Planet Finder will combine high-order adaptive
optics with differential detection techniques; multi-waveband imaging,
integral-field spectroscopy, and imaging polarimetry are foreseen for
the focal plane instruments.  Planet Finder could become operational
around 2009. A review board for the assessment of the Phase~A studies
met on 16--17~December 2004.

Planet Finder may discover giant planets in different phases of their
evolution. During the ongoing contraction and accretion phases, the
internal luminosity of these planets exceeds the reflected light
contribution by several orders of magnitude, e.g.\ a Jupiter mass
planet will be 10$^3$ times brighter at 1\,Myr than at 1\,Gyr.  This
raises the possibility of detecting young planets around the closest
young stars, in spite of their relatively large distances.  Planet
Finder will also search for old planets in the Solar
neighbourhood. The S/N for the detection of exo-planets drops rapidly
with distance, due to the combined effects of inverse-square
brightness losses and the reduction in stellar-planet angular
separation. The most promising targets for old systems are therefore
within 5--10\,pc, exploring the range in separation down to
$\simeq3-5$\,AU. For this reason, possible targets for Planet Finder
may be found among stars known to have planetary systems from
high-precision radial velocity surveys. Even more promising is the
synergy with planet searches using astrometric perturbations: giant
planets detected by Planet Finder should give a signal in the tens of
milli-arcsecond range, clearly measurable with PRIMA and/or future
space missions (SIM, Gaia). This will provide an independent estimate
of planetary masses. 

The detailed science cases will probably differ between both groups
because of different AO system performances and focal instruments. In
the French-led proposal, typical targets are: (a) stars in young
associations (more than 100~candidates), which will be looked at much
closer than with NACO (with an interference coronagraph, at
1--2\,$\lambda/D$, achievable contrast $>10^{-5}$); (b) close-by
solar-type stars at moderate ages (1--2~Gyr): contrast: a few
$10^{-5}$; about~200 objects; for some nearby objects performances
should be much better; (c) late-type stars of all ages: with better
performances on the youngest and closest ones; (d) other science:
disks and stellar environments. In the German-led proposal (CHEOPS)
the primary goals are to find mature (old, Jupiter-like) planets
in nearby systems (within 15~pc), with polarimetry possible; and
young, still-warm planets in the nearest star-forming regions
(within 100~pc) with an integral-field spectrograph operating in the 
J and H~bands. Secondary goals are to observe brown dwarfs, young stellar
object disks, debris disks, etc.

\vspace{10pt}
{\bf PRIMA:}  
\label{sec:prima}
The ESO VLT Interferometer consists of four 8\,m Unit Telescopes and
four moveable 1.8\,m Auxiliary Telescopes, which can form baselines up
to 200\,m in length. The PRIMA (Phase-Reference Imaging and
Micro-Arcsecond Astrometry, \cite{qcd+98}) facility will implement
dual-star astrometry at the VLTI; it is expected to become operational
in 2007. Its goal is to measure the masses and orbital inclinations
of planets already known from radial velocity surveys. In addition, a survey
with PRIMA will be conducted to establish the frequency of planets along the 
main sequence and through time.

\begin{figure}[t]
\begin{center}
\centerline{\epsfig{file=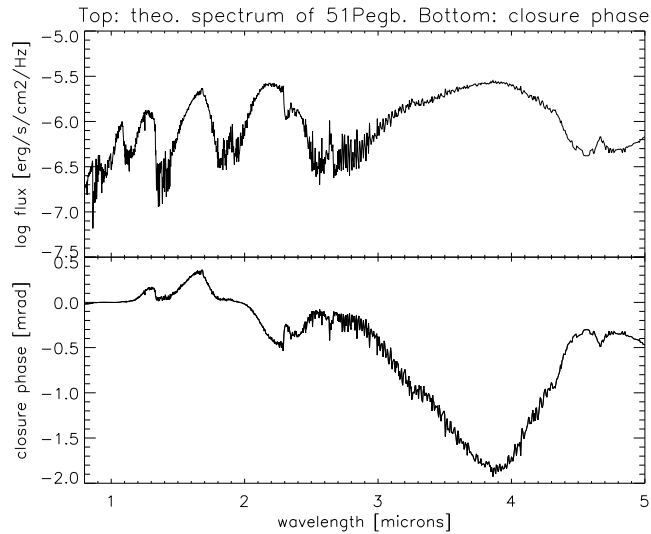,width=0.6\textwidth}}
\end{center}
\vspace{-40pt}
\caption[]{\footnotesize
  Spectral signal of planet in the closure phase.  Top panel:
  theoretical spectrum of the giant irradiated planet orbiting the
  solar-like star 51~Peg \protect\citep{sbh03}.  Clearly visible are CO
  and H$_2$O absorption bands in the near-IR. Bottom panel: closure
  phases in milli-radian based on the theoretical spectrum of 51~Peg~b
  as well as of the host star and a simulated observation with the
  near-IR instrument AMBER at the VLTI using the three telescopes UT1,
  UT3 and UT4 (baseline lengths 102\,m, 62\,m, and 130\,m). It is
  evident that the spectrally-resolved closure phases contain a wealth
  of spectral information of the planet.}
\label{fig:clphase}
\end{figure}

The principles of interferometric astrrometry are summarised in
Section~\ref{sec:ground-astrometry}.  Differential phase observations
with a near-IR interferometer offer a way to obtain spectra of
extra-solar planets. The method makes use of the wavelength dependence
of the interferometer phase of the planet/star system, which depends
both on the interferometer geometry and on the brightness ratio
between the planet and the star. The differential phase is strongly
affected by instrumental and atmospheric dispersion effects.
Difficulties in calibrating these effects might prevent the
application of the differential phase method to systems with a very
high contrast, such as extra-solar planets.  A promising alternative is
the use of spectrally resolved closure phases, which are immune to
many of the systematic and random errors affecting the single-baseline
phases.

Figure~\ref{fig:clphase} shows the predicted response of the AMBER
instrument at the VLTI to a realistic model of the 51~Peg system,
taking into account a theoretical spectrum of the planet as well as
the geometry of the VLTI.  \cite{jq04} have presented a strategy to
determine the geometry of the planetary system and the spectrum of the
extra-solar planet from such closure phase observations in two steps.
First, there is a close relation between the nulls in the closure
phase and the nulls in the corresponding single-baseline phases: every
second null of a single-baseline phase is also a null in the closure
phase. This means that the nulls in the closure phase do not depend on
the spectrum but only on the geometry, so that the geometry of the
system can be determined by measuring the nulls in the closure phase
at three or more different hour angles. In the second step, the known
geometry can then be used to extract the planet spectrum directly from
the closure phases.

\vspace{10pt}
{\bf ALMA:} 
\label{sec:alma}
ALMA is an interferometer in the mm-wavelength range, which will
consist of sixty-four 12~m diameter antennae located in northern
Chile, at 5050\,m altitude. The antennae can be spaced from a compact
configuration with a maximum separation of 150~m to a very extended
configuration where the maximum spacing is 16~km, providing a
resolution of 10~milli-arcsec at shortest wavelengths. The receivers
will cover the atmospheric windows in the 35--1000~GHz range
(350\,$\mu$m -- 7\,mm) with a bandwidth of 8~GHz in two polarizations,
and a resolution of 32000 channels.  ALMA will be powerful for
studying the disks around young stars, able to image such disks out to
several hundred parsecs, providing density and temperature profiles
(through measurements of thermal dust emission), and providing
constraints on disk dynamics and chemistry (through measurements of
spectral lines). In the case of protoplanetary disks, ALMA will be
able to image gaps and holes caused by protoplanets.

In terms of direct detection of the planet themselves, however, ALMA
is rather limited. At its best resolution (widest configuration), the
system will be able to resolve a Jupiter-like planet from its star out
to 100--150\,pc. The main limitation comes from the flux of the
planet. The best frequency for planet observation, which optimises the
combination of expected detector noise characteristics, the spectrum
of the objects, and the site characteristics, is at 350~GHz. The flux
density of the planet is directly related to their temperature, size
and distance as $F_{350} = 6.10^{-8}\, T R_{\rm J}^2/D^2$, where the
distance $D$ is in pc, $R_{\rm J}$ the radius expressed in Jupiter
units, and $T$ the temperature in~K. Together with the expected
sensitivity performances of ALMA, this indicates that a Jupiter would
be detectable only out to about 1\,pc.  In the case of a `hot Jupiter'
($R=1.5\,R_{\rm J}$, $T=1000$\,K), this limit is pushed to a few
parsecs, but still not far enough to actually encompass any useful
star. On the other hand, a proto-Jupiter (with $R=30\,R_{\rm J}$ and
$T=2500$\,K) would be detectable in a matter of minutes to hours out
to several tens of parsecs.

The contrast between the star and the surrounding bodies, a critical
factor at shorter wavelengths (visible--infrared), becomes an
advantage in the case of ALMA. The contrast factor is of the order of
1000, which is well within the dynamic range of the detectors, and the
bright central source helps maintaining the optimal coherence of the
interferometer.  So, while ALMA is not expected to contribute
significantly to the study of mature planets, its contribution will be
very significant for studying early stages of planet formation, from
nebula to protoplanet.

\subsubsection{Other Searches}

The upper right-hand section of Figure~\ref{fig:methods} indicates
some miscellaneous search methods.  Jupiter's magnetosphere is known
to produce strong emission of radio waves.  These decametric bursts
are targeted by a number of collaborative efforts in the radio
community, as summarised in Table~\ref{tab:direct}. As noted in
Section~\ref{sec:space-astrometry}, Gaia might also detect a few 
protoplanetary collisions photometrically \citep{zs03}, although
whether they could ever be recognised as such, buried within such 
a vast volume of other variables, has not been assessed.

\clearpage
\subsection{Space Observations: 2005--2015}
\label{sec:space2005-15}

In the near-term future of space missions, there are two principal
detection approaches: transits (exemplified by the COROT, Kepler and
Eddington missions), and astrometry (exemplified by Gaia and SIM). A
non-approved concept, MPF (originally GEST), uses microlensing to
expand on the parameter space for which statistical information on
planet frequency would be provided.

\subsubsection{Space Transit Measurements: COROT, Kepler and Eddington}
\label{sec:space-transits}

{\bf COROT:} COROT is a French-European-ESA collaboration, led by
CNES, comprising a 27~cm telescope with a CCD camera, with a launch
planned for June 2006. After MOST, it will be the second satellite
dedicated to long-term high-accuracy photometric monitoring from
space. COROT will combine the study of asteroseismology with the
search for exo-planetary transits.  The observation of 60\,000 stars
(12\,000 stars simultaneously for 150~days each) is expected to result
in the detection of a few hot telluric planets.

As a general remark, which applies to the other predictions in this
section, as well as to the radial velocity and ground-based transit
searches discussed previously, it should be noted that the expected
rate of exo-planet detection is difficult to quantify, given both
their unknown frequency of occurrence, and detection uncertainties due
to stellar activity.

The numbers in Table~\ref{tab:prospects-pre2015} are based on the
number of dwarf stars in the COROT fields, the mass and period
distribution of known exo-planets, the probability of transits, the
fact that no more than 1~short-period exo-planet per star is expected,
and the expected accuracy of COROT as a function of magnitude.
Table~3 in \cite{brl03} assumes one planet per dwarf star, and a
uniform orbital distribution law. This leads to a significantly larger
number of predicted detections, and is presumably an overestimate of
the actual number of planets expected (as they also note). In
practice, the critical factor in all these predictions is the unknown
number of low-mass planets per star.
  
\ignore{As a consistency check, it can be argued that Kepler will
  observe 3~times more stars than COROT during 4~years (8~times longer
  than COROT), implying of order $3\sqrt{8}=8.5$ times more
  candidates, broadly in agreement with their estimation for the giant
  planets.}

{\bf Kepler:} improved prospects for photometric transit detections
will come with NASA's Kepler mission, due for launch around 2007.
Kepler is a 0.95~m aperture, differential photometer with a 105~square
degree field of view. It focuses on the detection of Earth-size
planets or larger in or near the habitable zone of a wide variety of
stellar spectral types, monitoring some $10^5$ main-sequence stars
brighter than 14~mag.  Detection of some 50--640 terrestrial
inner-orbit planet transits are predicted, depending on whether their
typical radii lie in the range $R\sim1.0-2.2 R_{\rm E}$, determining
the distribution of sizes and orbital characteristics.  Kepler will
assist the preparation of future programmes like SIM and Darwin/TPF by
identifying the common stellar characteristics of host stars, and
defining the volume of space needed to search.

The numbers in Table~\ref{tab:prospects-pre2015} are taken from the
Kepler www site, and are based on the following assumptions: 100\,000
main-sequence stars observed with a precision of better than
$\sim5\times10^{-5}$; typical variability of 75\% of the stars is
similar to that of our Sun; most main-sequence stars, including
binaries, have terrestrial planets in or near the habitable zone; on
average, two Earth-size or larger planets exist between 0.5--1.5\,AU;
transit probability for planets in the habitable zone is 0.5\% per
planet; the transit is near grazing in a 1-year orbit; each star has
one giant planet in an outer orbit; on average 1\% of the
main-sequence stars have giant planets in orbits shorter than 1~week
and comparable numbers in periods of 1--4~weeks and 1--12~months;
mission life time of 4~years. Results for giant planets are expected
around 2007, with those on terrestrial planets (which will require
more careful verification) around 2010.

{\bf Eddington:} ESA's Eddington mission was originally proposed for
launch around 2008.  It entered ESA's science programme as a `reserve'
mission, was approved in 2002, but cancelled in November 2003 due to
overall financial constraints.  The Eddington payload was composed of
three identical, co-aligned telescopes with a $\simeq0.7$~m aperture
and identical $3\times2$ mosaic CCD cameras, a total collecting area
of $\simeq 0.75$~m$^2$ and a field of view of $\simeq 20$~deg$^2$.
Each telescope had a slightly different bandpass, allowing colour
information to be derived for high S/N transits.

The baseline lifetime was 5~years (extended operations possible), of
which three would be dedicated to a single long observation (currently
baselined in Lacerta), and two would be used for short (one to a few
months) observations of other fields.  Planet searches were to be
conducted during the three years observation, in $\sim10^5$ stars, of
which $\sim10^4$ would be observed with sufficient accuracy to detect
Earth-like planets.  During the shorter observations, a further
$\sim4\times 10^5$ stars would be searched for planets, allowing many
shorter-period planets to be discovered.

The numbers in Table~\ref{tab:prospects-pre2015} are the results of
detailed Monte Carlo simulations, which assume a `standard' planet
function of the form $f(a, m| M) \propto a^\alpha m^\beta M^\gamma$
where $a$ is the orbital radius of the planet, $m$ its mass, and $M$
the mass of the parent star, and using the mass-radius relationship
obtained from solar system objects.  The simulations used $\gamma =
0$, i.e.\ the planet function is independent of the stellar mass. As
both the Doppler planet population and the solar system objects are
consistent with $\alpha = -1$, $\beta = -1$, the simulations have also
used $f(a, m) \propto a^{-1} m^{-1}$ normalised to 0.01~hot Jupiters
per star (as derived from the Doppler surveys). In this assumption,
each star has 0.25 normal Jupiters ($r>R_{\rm J}$), 0.01~hot Jupiters
($r>R_{\rm J}$, $P=3-5$~day), 0.85~Earths ($R>R_\oplus$) at any
orbital distance, and 0.07 `habitable planets' (with liquid water
temperatures). This is considered as a conservative mass function, as
the typical star probably has fewer rocky planets than the solar
system. Also, the simulations are stopped at $R=1R_{\rm J}$ (because
of the emphasis on the smaller planets), so that the number of gas
giants is under-predicted.  The result of the Monte Carlo simulation
predicts, for the single 3-year observation: 14\,000 planets in total
(in 12\,500 planetary systems), of which 8000~hot planets, of which
5600~with $R>5 R_\oplus$, 660~Earths ($0.5<R<2.0 R_\oplus$);
160~habitable zone planets, of which 20~`Earths'.  A larger number of
hot Jupiters would also be found in the short (asteroseismology)
observations. Predictions for multiple systems depend sensitively on
the assumed distribution of relative orbital inclinations.

\vspace{10pt}
The following missions, not dedicated to transits, can also be
noted:

{\bf HST:} although suitably placed above the atmosphere such that
low-mass planets could be detected by HST using this technique in
principle, HST is not a dedicated transit discovery instrument, and
its discovery efficiency is constrained by its limited field of view,
and available observing time. Any searches using HST will therefore
likely be restricted to observations of especially high-surface
density regions.  Observations of 47~Tuc (34\,000 main-sequence stars
monitored for 8.3~days) failed to detect planets \citep{gbg+00},
whereas 17~would have been expected based on radial velocity surveys.
This non-result is currently attributed to effects of metallicity
\citep{gon98}, ultraviolet evaporation \citep{arm00}, or collisional
disruption \citep{bsd+01} of the protoplanetary disks in this crowded
stellar environment.  An advance in transit statistics should come
from observations of the Galaxy bulge with HST by Sahu et al.,
monitoring 100\,000 stars to $V=23$ over 7~days in February 2004, with
results expected in spring 2005.  

Nevertheless HST, and its successor JWST, have considerable potential
for follow-up observations of transiting systems discovered by other
methods, for example by Kepler. This issue is developed further in
Section~\ref{sec:eso-esa}.  

{\bf MOST:} the Canadian satellite MOST is a 15~cm telescope launched on
30~June 2003. Dedicated to the long-term photometric monitoring of a
small number of stars primarily for asteroseismology studies, it has a
photometric performance just a factor of~2 better than from ground
\citep{mkg+04}. Although with limited transit discovery potential, it
will nevertheless aim to detect reflected light of a few known hot
Jupiters.

\subsubsection{Space Astrometry Missions: Gaia and SIM}
\label{sec:space-astrometry}

As noted previously, astrometric planet detection involves detecting
the system's photocentric motion on the plane of the sky, in the same
way that radial velocity detection involves detecting the system's
photocentric motion along the line of sight.  The amplitude of the
displacement, and therefore the system's detectability, can be
characterised by the system's `astrometric signature', $\alpha=(M_{\rm
  Planet}/M_{\rm star})\cdot (a/d)$. This signature is measured in
arcsec when the orbital radius $a$ is measured in AU and the distance
$d$ is measured in\,pc.  Astrometric measurements can provide the
planetary mass directly rather than $M\sin i$ (as provided by radial
velocity techniques) if $d$ is determined and if $M_{\rm star}$ is
estimated from stellar evolutionary theory.
 
\begin{figure}[t]
\begin{center}
\centerline{\epsfig{file=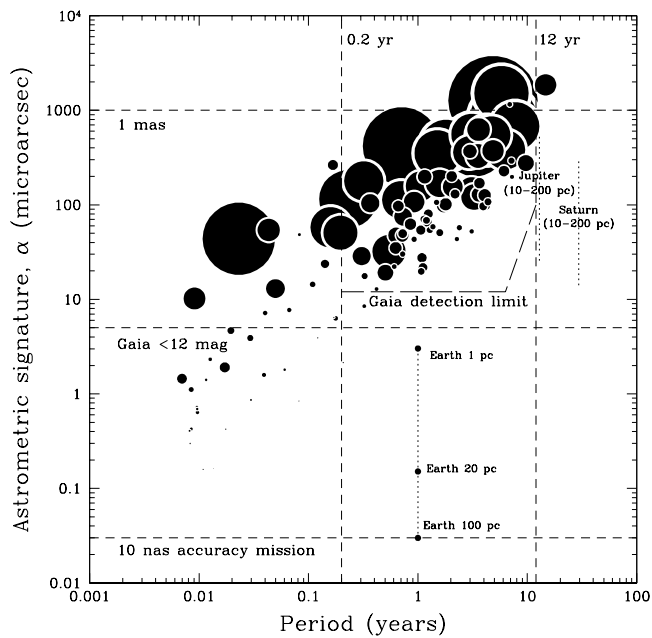,width=0.6\textwidth}}
\end{center}
\vspace{-20pt}
\caption[]{\footnotesize
  Astrometric signature, $\alpha$,
  induced on the parent star for the known planetary systems, as a
  function of orbital period.  Circles are shown with a radius
  proportional to $M_p\sin i$.  Astrometry at the milli-arcsec level
  has negligible power in detecting these systems, while the situation
  changes dramatically for micro-arcsec measurements.  Short-period
  systems to which radial velocity measurements are sensitive are
  difficult to detect astrometrically, while the longest period
  systems will be straightforward for micro-arcsec positional
  measurements.  Effects of Earth, Jupiter, and Saturn are shown at
  the distances indicated.}
\label{fig:astrometry}
\end{figure}

Figure~\ref{fig:astrometry} shows the astrometric signature versus
orbital period for the known exo-planets, where the size of the
circles indicates the planetary mass.  The horizontal line at the top
of the figure indicates the Hipparcos astrometric accuracy, and shows
immediately why Hipparcos was unable to detect new planetary systems.
Neverthess, Hipparcos data was useful for placing some constraints 
on the masses of planet candidates \citep{zm01}. \ignore{Mazeh commented
that this null result superseded their earlier claim in mzt+99}

Gaia (ESA) and SIM (NASA) are two very different approaches to space
astrometry, both approved and under development:

{\bf Gaia:} Gaia is a scanning, survey-type instrument, with a launch
around 2011 \citep{pbg+01}. Its detectability domains are shown in
Figure~\ref{fig:astrometry}: periods below about 0.2~yr will not be
detectable because of the relatively long times between successive
observations dictated by the scanning law, while periods longer than
about 12~yr will result in photocentric motions indistinguishable from
rectilinear motion over the mission's measurement duration (about
5~years). As seen in the figure, Gaia will therefore contribute
substantially to the large-scale systematic detection of Jupiter-mass
planets (or above) in Jupiter-period orbits (or smaller); some
10--20\,000 detections out to 150--200\,pc are expected \citep{lss+00,
  scl+01}, including confirmation of most of the (longer-period)
radial velocity detections known to date.  Planetary masses, $M$,
rather than $M\sin i$, will be obtained. Full orbital parameters will
be obtained for some 5000 (higher S/N) systems.  Relative inclinations
can in principle be obtained for multiple systems with favourable
orbits \citep{scl+01}, important for studies of formation scenarios
and orbital stability of multiple systems.  Some 4--5000 transit
systems, of the hot-Jupiter type, might also be detected
\citep{rob02}.  Gaia might also detect a handful of protoplanetary
collisions photometrically \citep{zs03}. Gaia cannot observe systems
at epochs other than those determined by its fully deterministic
scanning law, and will not detect planets with masses much below
10--20~$M_\oplus$ unless such systems exist within 10--20\,pc.

{\bf SIM:} SIM is a pointed interferometer with a launch around 2010
\citep{du99}: accuracies of a few micro-arcsec down to 20~mag are
projected. Such faint observations will be expensive in terms of
observing time, and brighter target stars are likely to be the rule.
Of 15 key projects and mission scientist programmes currently studied,
three focus on planetary systems: (1) A Search for Young Planetary
Systems (Beichman): this will survey 200~stars with ages from
1--100~Myr (mostly in star-forming regions at 125--140~pc, but
including TW~Hya at 50~pc) which expects to find anywhere between
10--200 planets, depending on whether the occurrence rate is the
canonical 5--7\,\% from current radial velocity surveys, or 100\,\% of
all young stars. The survey will be sensitive to $M_{\rm J}$ planets
at orbital distances of 1--5~AU.  (2) Discovery of Planetary Systems
(Marcy): this will focus on searches for 1--3\,$M_\oplus$ planets
within 8~pc, and for 3--20\,$M_\oplus$ planets within 8--30~pc. The
survey is also considered as a reconnaissance for TPF. Target stars
will be selected from ongoing surveys of the nearest 900~GKM
main-sequence stars in the northern hemisphere with the Lick 3-m and
the Keck 10-m telescopes, the nearest 200~GK stars in the south with
the AAO 3.9-m, and the planned 6.5-m Magellan survey extending to a
further 600~GKM stars in the south.  (3) Extra-Solar Planet
Interferometric Surveys (Shao): this major SIM survey programme
comprises a deep survey of about 75 nearby main-sequence stars within
about 10~pc of the Sun, of which one third are G~dwarfs and the
remainder are inactive main-sequence stars of other spectral types
(mostly K~and M but including a few A~and~F). Over the mission
lifetime, each target will be observed some 70~times, each of twenty
1-minute observations resulting in a final accuracy of about
1\,micro-arcsec.  Each pointing will be accompanied by the observation
of typically 28 additional bright nearby stars (within 25~pc), with
single 1-minute observations leading to some 2000 stars observed with
accuracies of about 4\,micro-arcsec. These will be from diverse types:
all main-sequence spectral types, binaries, a broad range of age and
metallicity, dust disks, white dwarfs, planets from radial velocity
surveys, etc.  Preparatory programmes for this survey include radial
velocity monitoring and adaptive optic imaging.  The total expected
number of new detections from SIM, for any given planetary mass and
orbital radius is again not straightforward to predict, and depends on
the (unknown) mass distribution of exo-planets versus orbital radius
at $a\sim1$\,AU. Estimates are given in
Table~\ref{tab:prospects-pre2015}.

The NASA SMEX proposal AMEX (which followed on from the FAME study)
aimed at 150\,micro-arcsec accuracy at 9~mag and 3~milli-arcsec at
15~mag and, with a proposed launch in 2007--08, would have provided
limited prospects for planet detection through the comparison of
proper motions with Hipparcos, including some 600 detections to 30\,pc
down to K5V stars, and transits to V~=11~mag. AMEX was not selected by
NASA in 2003.  In mid-2004 NASA announced the selection of nine
studies for future mission concepts within its `Astronomical Search
for Origins Program', including the `Origins Billion Star Survey'
(OBSS) focussing on a census of giant extra-solar planets using the
principles of Gaia. If OBSS is selected, its contribution to
astrometric exo-planet research would not surpass those of Gaia.

The Japanese mission JASMINE, and a potential prototype nano-JASMINE,
have been under discussion at a low level in Japan for several years
\citep{Vilnius-2001-14}.  Originally conceived as a mini-Gaia but able
to concentrate on the Galactic centre by operating in the infrared,
the mission's technical feasibility has been improved in the past few
months (although its scientific niche has been weakened) with the move
to CCD detector technology.

\subsubsection{Space-Based Microlensing: MPF}
\label{sec:space-microlensing}

Proposals for exo-planet detection through their microlensing
signatures have been made. GEST (Galactic Exo-Planet Survey Telescope
\citep{br02}) was proposed for a NASA mission in 2001--02 (a Survey
for Terrestrial Exo-Planets (STEP) was also submitted to NASA's
Extra-Solar Planets Advanced Concepts Program at the same time).  It
was not selected in 2002, but was re-submitted during 2004 under the
name of Microlensing Planet Finder (MPF), using HgCdTe and Si-PIN
detectors in place of the earlier CCDs \citep{bbb+04}.

A 1.2-m aperture telescope with a 2~deg$^2$ field of view continuously
monitors $10^8$ Galactic bulge main-sequence stars. In about one case
out of a million, sources in the bulge are lensed by foreground (bulge
or disk) stars which are accompanied by the planets being sought.
Observing high surface-density sky regions improves lensing
probabilities to sufficient levels that successful detections can be
expected over reasonable observing times. Space observations are
considered mandatory to permit the high photometric accuracy required
for detection even in very crowded regions where seeing limits the
achievable photometric accuracy and hence detectability achievable
from the ground.

Microlensing probes particular exo-planet domains: for example,
low-mass planets can be detected, albeit usually at very large
distances of typically 5--8~kpc.  The sensitivity of such measurements
is highest at (projected) orbital separations of 0.7--10\,AU, but it
will also detect systems with larger separations, masses as low as
that of Mars, large moons of terrestrial planets, and some 50\,000
giant planets via transits with orbital separations of up to 20\,AU
(the prime sensitivity of a transit survey extends inward from 1\,AU,
while the sensitivity of microlensing extends outwards).  There are
theoretical reasons to believe that free-floating planets may be
abundant as a by-product of planetary formation, and MPF/GEST will
also detect these.

The planetary lensing events have a typical duration of 2--20~hr
(compared to the typical 2--20~weeks duration for lensing events due to
stars), and must be sampled by photometry of $\sim1$\% accuracy
several times per hour over a period of several days, and with high
angular resolution because of the high density of bright main-sequence
stars in the central bulge. The proposed polar orbit is oriented to
keep the Galactic bulge in the continuous viewing zone.
Most of the multiple-planet detections in the simulations of
\cite{br02} are systems in which both `Jupiter' and `Saturn' planets
are detected. Since multiple orbits are generally stable only if they
are close to circular, a microlensing survey will be able to provide
information on the abundance of giant planets with nearly circular
orbits by measuring the frequency of double-planet detections and the
ratios of their separations.

Just over 100 Earths would be detected if each lens star has one in a
1\,AU orbit. The peak sensitivity is at an orbital distance of 2.5\,AU,
with 230 expected detections if each lens star had a planet in such an
orbit. Although the prime quantity obtained from a microlensing
detection is the mass ratio between planet and star, additional
information or hypotheses can be combined to estimate the mass of the
host star, the planetary mass, the distance to the host star, and the
planet-star separation in the plane of the sky.

One of the disadvantages of lensing experiments is that a planet
event, once observed, can never (in practice) be seen again --
follow-up observations for further characterisations are not feasible
(unlike the case for any of the other principal detection methods).
Nevertheless, a mission like MPF/GEST will provide important
observational and statistical data on the occurrence of low-mass
planets (Earth to Jupiter masses), low-mass planets at larger orbital
radii, multiple systems and, significantly, free-floating planets
formed as a by-product of the system formation.

Detection by microlensing could in principle also be included in the
Eddington mission, but the Eddington team has made no detailed
evaluation of feasibility, and its inclusion would be likely to drive
instrumental requirements in a non-trivial manner.

\begin{figure}[t]
\begin{center}
\centerline{
\epsfig{file=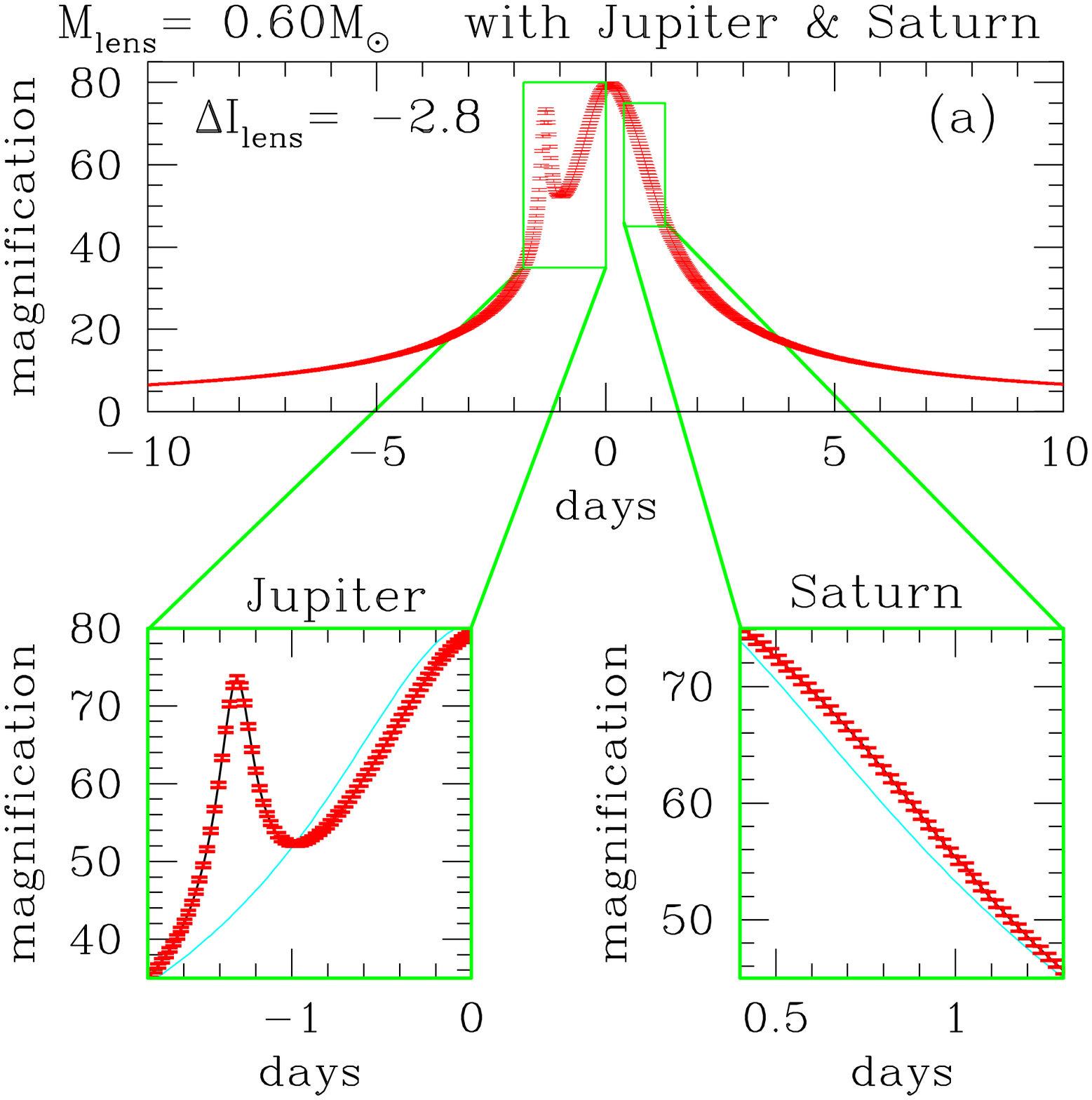,width=0.45\textwidth} \hfill
\epsfig{file=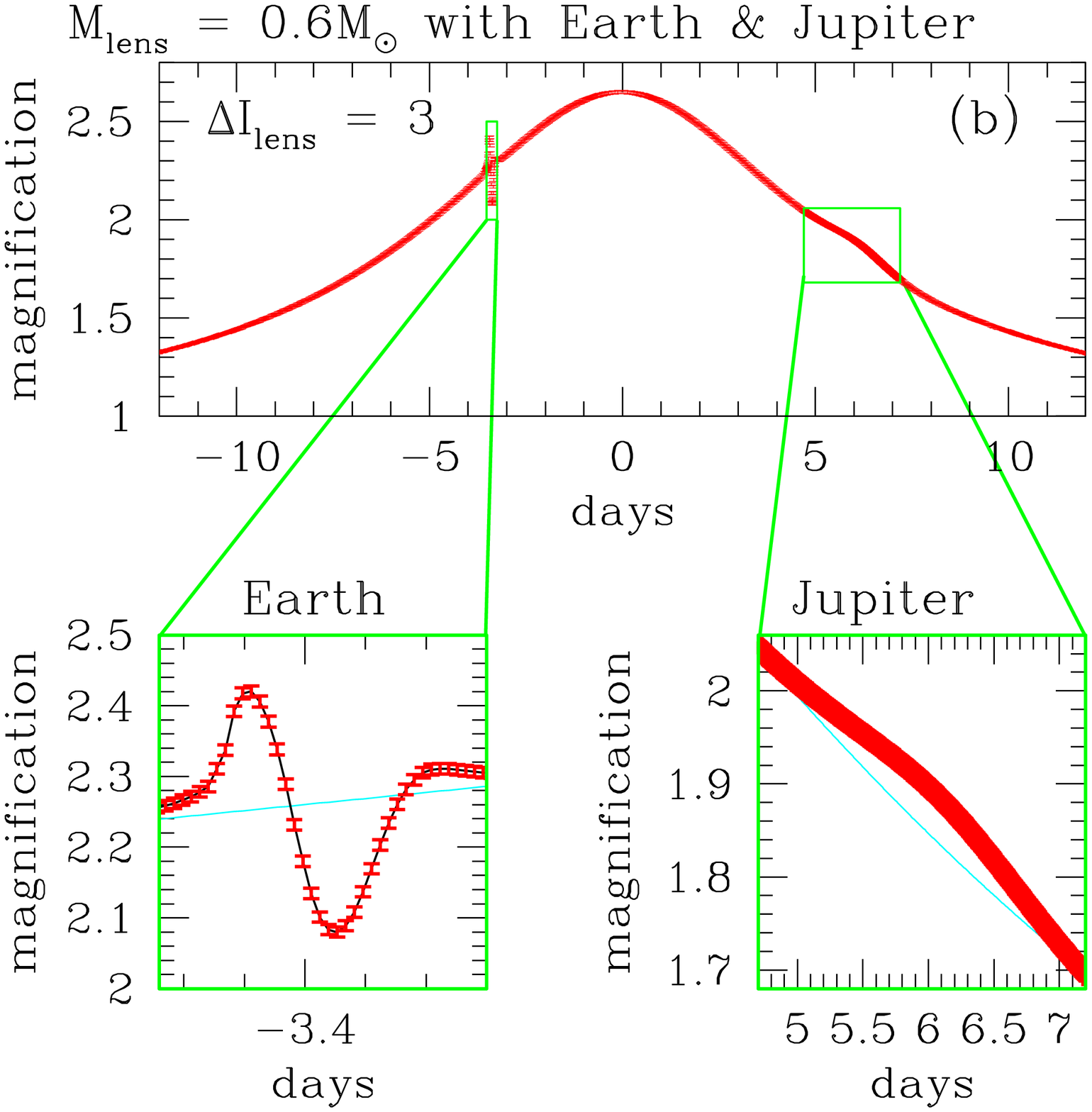,width=0.45\textwidth}
}
\end{center}
\vspace{-30pt}
\caption[]{\footnotesize
  Example of multiple-planet microlensing light curves from the
  simulation of planetary systems with the same planetary mass ratios
  and separations as in our solar system, from the MPF/GEST studies
  of \protect\cite{br02}.  Left: an example of a Jupiter/Saturn
  detection. Right: an example of the detection of an Earth and a
  Jupiter.}
\label{fig:lensing}
\end{figure}

\subsubsection{Other Space Missions: JWST, Spitzer, SOFIA}
\label{sec:other-space}

{\bf JWST: }
JWST (http://www.jwst.nasa.gov/) is a collaboration between NASA, ESA
and the CSA. It will be a passively-cooled (40--50~K) observatory
spacecraft with an 18-segment primary mirror having an effective
aperture of about 6.5~m and diffraction-limited performance at
2\,$\mu$m, equipped with four principal science instruments and a fine
guidance sensor. It is scheduled for launch in 2011 into an L2
Lissajous orbit.  The observatory is optimised for the 1--5\,$\mu$m
band, but will be equipped to cover 0.6--28\,$\mu$m with a combination
of imaging (through fixed and tunable filters) and low- to
moderate-resolution ($100<R<3000$) spectroscopy. The `Origin and
Evolution of Planetary Systems' is one of JWST's five science themes,
and the science requirements are drafted accordingly.

The Design Reference Mission
(http://www.jwst.nasa.gov/ScienceGoals.htm) describes the exo-planet
programmes currently foreseen for JWST. The survey programmes are
aimed at finding giant planets and isolated objects using direct
imaging, and bound planets using coronography.  Follow-up studies are
planned using tunable filter imaging ($R\sim100$) and slit
spectroscopy. For isolated sources, objects at AB~=~30~mag can be
reached using the near-infrared camera (NIRCam).  The tunable filter
can reach AB~=~27~mag, while mid-infrared spectroscopy (with MIRI) can
reach AB~=~23~mag at $R\sim3000$.  For widely-separated giant planets,
$R\sim100$ coronography will provide preliminary temperature estimates
and for these and for isolated systems, $R\sim1000$ near-IR
spectroscopy will access metallicity indicators.  Synoptic
observations of bodies in our own Solar System, such as Titan, over
the 10-year lifetime of JWST will begin the study of secular surface
and atmospheric changes.
  
A report on `Astrobiology and JWST' \citep{slun04} listed three areas
where the technical capabilities of JWST should be optimised for the
follow-up of transit events: (1) in principle, JWST can measure the
transmission spectra of giant planet atmospheres during planet
transits of bright stars (7--14~mag) but this requires capabilities of
rapid detector readout and high instrument duty-cycle in order to
achieve very high S/N over a typical transit time (12~hr). If
Earth-sized planets are common and detected in transit around stars
brighter than 6~mag, the JWST near-IR spectrograph (NIRSpec) could
detect atmospheric biomarker signatures; (2) the collection of
$\sim10^8$~photons per image for NIRSpec, by spreading photons over
$10^5$ spatial+spectral pixels, would enable JWST to characterise
atmospheres during the transit of a terrestrial planet in the
habitable-zone of a solar-type star; (3) NIRSpec is important for
characterising transiting extra-solar planets.  Many transiting
extra-solar planets are expected to be found in the next several years
with both ground-based and space-based telescopes (including Kepler).
The short wavelength end is especially important for detecting
scattered light and characterising planetary albedos. In particular,
NIRSpec's long-slit configuration is essential for these observations.
Thus JWST, even more so than HST, has considerable potential for
follow-up observations of transiting systems discovered by other
methods, for example by Kepler. This issue is developed further in
Section~\ref{sec:eso-esa}.

\vspace{10pt}
{\bf Spitzer:} Spitzer (ex-SIRTF, http://www.spitzer.caltech.edu/)
is an 85~cm aperture, liquid helium cooled telescope in an
Earth-trailing heliocentric orbit. Launched in August 2003, it has a
projected lifetime (minimum) of 2.5~years with a goal of 5~years or
more. The instrument complement provides the capabilities for
imaging/photometry from 3--180\,$\mu$m, spectroscopy from
5--40\,$\mu$m and spectrophotometry from 50--100\,$\mu$m.  Spitzer's
expected contribution to the field of exo-planet research lies in
its ability to measure excess radiation from dust disks over the
critical mid-infrared wavelength range. The imaging capability is
determined by the diffraction limit of the relatively small telescope
(1.5~arcsec at 6.5\,$\mu$m). The sensitivity, however, allows the
detection of dust masses to below the mass in small grains inferred in
our Kuiper Belt ($6\times10^{22}$~gm) surrounding a Solar-type star at
30\,pc.

One of the six Legacy Programs is concerned explicitly with the
formation and evolution of planetary systems (see: `The Formation and
Evolution of Planetary Systems: Placing Our Solar System in Context'
http://feps.as.arizona.edu/). This uses 350~hr of photometric and
spectroscopic Spitzer time to detect and characterise the dust disk
emission from two samples of solar-like stars, the first consisting of
objects within 50\,pc spanning an age range from 100--3000~Myr and the
second containing objects between 15--180\,pc spanning ages from
3--100~Myr.

The first call for Guest Observer programmes resulted in 8~accepted
exo-planet proposals out of a total of 202 programmes in all subject
areas.  This includes one to characterise the atmosphere and evolution
of the transiting extra-solar planet HD~209458b. The Cycle~2 call for
proposals had a deadline of 12~Feb 2005. The accepted Guest Observer
proposals related to exo-planets were: 
\parskip 4pt

* Ultracool Brown Dwarfs and Massive Planets Around Nearby White Dwarfs

* The SIM/TPF Sample: Comparative Planetology of Neighbouring Solar Systems

* Evolution of Gaseous Disks and Formation of Giant and Terrestrial Planets

* Searching the Stellar Graveyard for Planets and Brown Dwarfs with SST

* A Search for Planetary Systems around White Dwarf Merger Remnants

* Characterising the Atmosphere and Evolution of HD 209458b

* Survey for Planets and Exozodiacal Dust Around White Dwarfs

* Mineralogy, Grain Growth and Dust Settling in Brown Dwarf Disks

\parskip 10pt

\vspace{10pt}
{\bf SOFIA:} SOFIA (Stratospheric Observatory for Infrared Astronomy)
is a joint endeavour of NASA and the German DLR
(http://www.sofia.arc.nasa.gov/). A modified Boeing 747SP carrying a
2.7-m telescope will operate at an altitude of about 12~km.  The first
call for proposals will be issued in August 2005 for the first
observing cycle starting in January 2006. A total of nine instrument
have been selected, providing imaging and spectroscopy in the range
1--600~$\mu$m.  Operating out of NASA Ames Research Center, the
facility is to observe three or four nights a week for at least twenty
years.  Its location above the bulk of Earth's atmosphere will provide
access to the mid-infrared region without the limitations of
observatories on the ground. Its long projected lifetime will make it
possible to conduct muli-epoch observations for variable or evolving
objects and the ability to easily exchange instruments will ensure
that the latest technology can be incorporated as it becomes
available.

The science with SOFIA will revolve around cold matter in our solar
system, the interstellar medium, stars and galaxies. Similar to
Spitzer, SOFIA is not expected to contribute directly to the discovery
of extra-solar planets, but it will enhance understanding of planet
formation by studying circumstellar disks.  Furthermore it will
provide interesting information on the infrared spectra of solar
system bodies including the chemistry of atmospheres.

\clearpage
\subsection{Summary of Prospects 2005--2015}

The prospects for the main search experiments described in this
section are summarised in Table~\ref{tab:prospects-pre2015}.  This
table presents only a simplified picture of planet detection
capabilities, ignoring the comparative importance of finding large
numbers of exo-planets with only an estimation of $M\sin i$ or $r/R$,
or more comprehensively characterising a smaller number of planets
(with mass, radius, albedo, and age). It also ignores the fact that
different objects will be detected by different methods, and that
different methods supply complementary astrophysical information.

\vfill
\begin{table}[h]
\footnotesize
\caption[]{\footnotesize
Predictions for the numbers of planet detections out to 2015 according
to the major experiments currently planned, and the planet mass range
given in the first column ($M_\oplus\sim0.003M_{\rm J}$).  
The predictions depend sensitively on the (unknown) frequency of 
occurrence of (especially the lower-mass) planets, uncertain sample 
sizes, stellar jitter, etc. These are generally assessed somewhat 
differently for each project, and details are given in the accompanying
text. The numbers must be understood as indications of possible developments
only.
}
\vspace{-10pt}
\begin{center}
\setlength\tabcolsep{5pt}
\begin{tabular}{cccccccccc}
\hline
\noalign{\smallskip}
$M_{\rm Planet}$&     2004 &  2008--10&    2008--10 &    2008 &   2010&     2010--12&   2015& 2016 & 2016 \\
($M_{\rm J}$)  & \multicolumn{2}{c}{Radial Velocity} & Transits&  COROT &  Kepler&  Eddington&   SIM&  \multicolumn{2}{c}{Gaia} \\
               & \multicolumn{2}{c}{(ground)} &    (ground) &   &        &         &          &   astrom & photom \\
               &  (a) &   (b) &         (c) &    (d) &     (e) &      (f)&     (g)&   (h) &  (i) \\
\noalign{\smallskip}
\hline
\noalign{\smallskip}
1--10 &      90 & 200--250 & 100--1000&   5--15 &  \multicolumn{2}{c}{see below} &  200 &           15\,000 & 3000 \\ 
0.1--1 &     30 & 150--200 &         0& 50--150 &  \multicolumn{2}{c}{see below} &   50 & \phantom{1}5\,000 &    0 \\
0.01--0.1 &   0 & 10--20   &         0&  10--30 &  \multicolumn{2}{c}{see below} &   20 &                 0 &    0 \\
0.003--0.01 & 0 & 0        &         0&    0--3 &  0--640 & 0--660 &  0--5 &               0 &    0 \\
\noalign{\smallskip}
\hline
\noalign{\smallskip}
\end{tabular}
\end{center}
\label{tab:prospects-pre2015}
\vspace{0pt}
The columns of the table give results expected for:
(a) the present situation, dominated by radial velocity detections;
(b) the situation projected in the year 2010 from ground-based radial
velocity observations (estimates by Udry, see
Section~\ref{sec:ground-radvel});
(c) ground-based transit detections: hot Jupiters extrapolated from
\cite{hor03};
(d) COROT (estimates by Bouchy \& Rauer, see Section~\ref{sec:space-transits});
(e) Kepler (taken from the Kepler www site, see Section~\ref{sec:space-transits});
(f) Eddington (estimates by Favata, see Section~\ref{sec:space-transits});
(g) SIM (estimates by Perryman from the 
surveys noted in Section~\ref{sec:space-astrometry});
(h--i) Gaia astrometric and photometric detections (see
Section~\ref{sec:space-astrometry}).

\vspace{10pt}
The estimated detections for Eddington are not available 
as a simple function of mass, but have been 
classified (as detailed under the relevant section) as follows,
with the same assessment then made for the Kepler mission using the 
same assumptions:
\vspace{5pt}\\
-- Eddington: 14\,000 planets in total (in 12\,500 systems)\\
-- Kepler: 51\,000 planets in total \\
\vspace{-5pt}\\
-- Eddington: 8000~hot planets, 5600~with $R>5R_\oplus$\\
-- Kepler: 30\,000~hot planets, 22\,000~with $R>5R_\oplus$ \\
\vspace{-5pt}\\
-- Eddington: 660~Earths ($0.5<R<2.0 R_\oplus$)\\
-- Kepler: 19\,000~Earths ($0.5<R<2.0 R_\oplus$) \\
\vspace{-5pt}\\
-- Eddington: 160~habitable zone planets, of which 20~`Earths'\\
-- Kepler: 530~habitable zone planets, of which 35~`Earths' 
\end{table}

\clearpage

While confidence is developing in statistical distributions of planets
above about 0.05\,$M_{\rm J}$, the occurrence of lower-mass, and
specifically Earth-mass planets remains a matter for speculation. In
this spirit, the lower range of detection limits for low-mass planets
for Kepler, Eddington and SIM is indicated in the table as~0.

While these transit and astrometric discovery predictions must
therefore be taken with certain caveats, they promise a major advance
in the detection and knowledge of the statistical properties of a wide
range of exo-planets: ranging from massive (Jupiter-mass) planets in
long-period (Jupiter-type) orbits, and the occurrence and properties
of multiple systems (via astrometry), through to Earth-mass planets in
the habitable zone (via transits or microlensing).

\clearpage
\section{The Period 2015--2025}

\subsection{Ground Observations: 2015--2025}

\subsubsection{OWL/ELT}
\label{sec:owl} 

The 100-m diameter Overwhelmingly Large Telescope (OWL) is being
studied by ESO.  S/N~=~10 will be reached at 35~mag ($t=1$~hr) for
imaging, and at 30~mag for $R=1000$ spectroscopy ($t=3$~hr). The
resolution, $\lambda/D\sim1$~milli-arcsec in the V-band, is a
factor~40 improvement over HST.  Detection of Earth-mass exo-planets
is part of the scientific case for the 100~m OWL/ELT project (see the
on-line case for the European Large Telescope, ELT, at
www-astro.physics.ox.ac.uk/$\sim$imh/ELT, from which some of the
following has been taken).

This section focuses on the science case and technical issues for
Earth-mass planet detection.  For this, OWL must be equipped with
advanced adaptive optics systems to compensate for atmospheric seeing
and the production of a near-diffraction limited image or image core,
expected to deliver Strehl ratios $S$ (the peak image brightness
relative to that in a near-perfect image) from several to many tens of
percent. In the near-infrared (1--5\,$\mu$m), values near 90\% may be
attainable, and are needed for planet detection.  Ultimate detection
capabilities depend sensitively on whether such high Strehl ratios can
be delivered in practice (such advanced AO systems do not yet exist).
In addition, the telescope design (e.g.\ number and shape of elements)
has an impact on the final image quality, even with co-phasing
techniques, so that the exo-planet objectives must be considered early
in the project.

Applied to a nearby solar-type star, adaptive optics produces an image
with several components. A central `spike' resembling a
diffraction-limited image contains some $S$~per cent of the total
light, less a modest fraction diverted into the other, wider-spread
components of the telescope point-spread function. The spike is
surrounded by a residual halo containing the remaining light
distributed like an unmodified seeing disc, i.e.\ with a generally
Lorentzian distribution.  This halo overlies, and within the seeing
disk generally dominates, the fainter wings of the telescope
point-spread function. These wings combine the light diffused by the
small-scale imperfections of the optics and the accumulated dust, as
well as the light diffracted by the central obstruction and geometric
edges (mirrors, and supporting structures in the beam).  The
diffraction pattern is strongly dominated by the secondary mirror
supporting structure, and by the edges of the mirror segments, which
also produce strong secondary peaks.

The diffracted light can locally strongly dominate the uncorrected
seeing light.  The detailed structure of the halos in adaptive
optics-corrected images is still being explored. At large radii they
are composed of the rapidly varying diffraction-spot-sized `classical'
speckles. Less well-understood `super-speckles' occur closer to the
first 2--3 diffraction rings; larger, brighter, less rapidly variable
and therefore less likely to average out on a long-exposure image. At
the wavelengths of interest they should not, however, occur more than
10~milli-arcsec from the image centre, corresponding to 0.1\,AU at a
distance of 10\,pc.  Additionally, techniques such as simultaneous
differential imaging may completely cancel these
super-speckles. Nevertheless, their noise contribution remains even
after subtraction, and is typically the strongest noise source in the
5--15\,$\lambda/D$ region. This can only be reduced by improving the
Strehl ratio of the adaptive optics system.

Detection of a terrestrial-like planet in the presence of the stellar
glare is made possible in principle by the relatively large angular
separation between the image of a habitable planet and the central
diffraction peak of its parent star.  Thus at separations of
100~milli-arcsec (1\,AU at 10\,pc) only the sky background, the wider
scattering components of the intrinsic point-spread function, and the
adaptive optics halo contribute to the background. The orbital radius
at which an exo-planet is in practice detectable is then bounded by
two effects.  At the inner extreme, the bright inner
structures of the stellar image will drastically reduce sensitivity at
angular separations below 10--20~milli-arcsec, corresponding to an
orbital radius of 1\,AU at 50\,pc or 5\,AU at 250\,pc, so that beyond
these distances only self-luminous exo-planets (`young Jupiters')
could be detected.  At the outer extremes, the brightness of the
starlight reflected by the planet falls off with increasing orbital
distance from the parent star, even though it is well separated from
it. Some relevant scales are shown in Table~\ref{tab:owl}.

Various techniques to increase the contrast of the images are at
different stages of development, from theoretical concepts to working
prototypes. The most promising are: (i)~classic coronography, which
involves masking the star in the focal and pupil planes; advanced Lyot
stop studies taking into account the segmented mirror suggest that
contrasts of $10^{-9}$ are achievable (neglecting diffusion by dust,
mirror micro-roughness and the atmosphere); (ii)~nulling
interferometry (using the coherence of the star light to eliminate it
interferometrically), (iii)~extreme adaptive optics and
multi-conjugated adaptive optics (which result in a higher Strehl
ratio and cleaner point-spread function), (iv)~simultaneous
differential imaging (using the contrast of the target between nearby
spectral bands and/or the polarisation of the target). Some of these
methods can be combined to reach yet higher contrast \citep{ca04}.

The wavelength range where a broad range of planets may best be
detected and studied with OWL is probably the near infrared J-band at
1.25\,$\mu$m (where achievable Strehl ratios should approach 90\% and
where strong spectral features of water are available as diagnostics),
and the far red Z and I bands extending down to 700\,nm (less
favourable for adaptive optics, but containing the critical O$_2$
B-band absorption complex at 760\,nm). In the centres of the stronger
absorption bands saturated lines will obscure the signal from an
exo-Earth. However, in the wings numerous narrow unsaturated but
detectable lines will move in and out of coincidence as the two
planets move around their respective suns with a modulated Doppler
shift of up to 50~km\,s$^{-1}$. As discussed in
Section~\ref{sec:habitability}, free oxygen in the exo-planet
atmosphere would be a strong indicator of life (i.e.\ of
photosynthetic biochemistry), while its absence would not be
conclusive evidence for an abiotic world.

\begin{table}[t]
\footnotesize
\caption[]{\footnotesize
Magnitudes and separations for exo-planets predicted for OWL
(estimated by Hainaut).  
Separations are given in arcsec. Contrast ratios between the planets
and the parent star are of order $10^{-8}$ to $10^{-9}$. 
The magnitudes of the planets were taken as follows:
Jupiter: real observations;
Earth: real observations of the Earth, and for some wavelength where 
no observations were available, constructed by scaling observations of 
Venus and/or Mars;
hot Jupiter: models by \cite{sbh03} and \cite{bsh04}, for a planet with 
$M=1M_{\rm J}$ and $a=0.2$\,AU.}
\begin{center}
\setlength\tabcolsep{15pt}
\begin{tabular}{ccccc}
\hline
\noalign{\smallskip}
Distance &    Star &     Hot Jupiter &    Earth &   Jupiter \\
(pc)&              (mag)&      0.2\,AU &        1\,AU &      5\,AU \\
\noalign{\smallskip}
\hline
\noalign{\smallskip}
\phantom{1}10 &      4.8&       V = 24.1 &    27.9 &      25.8     \\
              &         &       sep = 0.020 &  0.100 &      0.500    \\
\noalign{\smallskip}
\phantom{1}25 &      6.8&       V = 26.1 &    29.9 &      27.8 \\
              &         &       sep = 0.008 &   0.040 &     0.200 \\
\noalign{\smallskip}
\phantom{1}50 &      8.3&       V = 27.6 &    31.4 &      29.3 \\
              &         &       sep = 0.004 &   0.020 &    0.100 \\
\noalign{\smallskip}
          100 &      9.8&       V = 29.1 &    32.9 &      30.8 \\
              &         &       sep = 0.002 &   0.010 &     0.050 \\
\noalign{\smallskip}
\hline
\end{tabular}
\end{center}
\label{tab:owl}
\end{table}

Table~\ref{tab:angel} indicates the S/N that can be expected from a
24~hr OWL observation of an Earth at 10\,pc, in the optical and
near-infrared. Clearly, even if Earths exist at distances as close as
only 10\,pc, detection and spectroscopy will be challenging even with
OWL.  Concerning the number of target stars accessible for
investigation by OWL, the Darwin study identifies some 500 F5--K9
stars out to 25\,pc, of which some 285 are single. There are only some
2--5 G0V--G2V~stars within 10\,pc, and some 21~single, non-variable
G0V--G4V within 15~pc.  

The distance at which an Earth will be observable will depend on
various parameters (Strehl ratio achievable, the performance of
nulling/subtraction techniques, e.g.\ phase-induced apodization
coronography could yield attenuation of $10^{-9}$, etc.) but in any
case is a strong function of mirror diameter.  Assuming, as above,
that a telescope can see a planet beyond an angular distance of
$5\lambda/D$ from the parent star, the volume of space explored (i.e.\
the number of stars) is proportional to $D^3$, and the numbers for
G~stars go from about 25 for a 30\,m telescope to about 900 for a
100\,m. The time to achieve the same S/N for different size telescopes
scales as (S/N)$^{-2}$; the object lies within the uncorrected light
from the star, so measurements are in the background-limited regime,
i.e.\ $S\propto D^2$, background $\propto D^2$, pixel size $\propto
D^{-2}$ (being diffraction limited) so S/N~$\propto
D^2/\sqrt{D^2\times D^{-2}}=D^2$ and $t\propto D^{-4}$.  Thus a 30\,m
aperture would take 123~times longer than a 100\,m to observe an
object that both would separate from the parent star. This is also the
reason why a lower limit of about 70\,m diameter is considered useful
for spectroscopy of the very nearest exo-solar systems, and why 30\,m
telescopes do not include Earth-like exo-planets as part of their
scientific goals.

The limiting distances at which imaging and spectroscopic observations
can be performed must take into account the photon noise from the
star, sky and planet, the speckle noise for a realistically high
Strehl ratio, and slowly varying aberrations that contaminate the
image subtraction. From these distances, the number of host candidates
are obtained from star catalogues (e.g., keeping only single G and
early K~stars). Results are shown in Table~\ref{tab:owl-numbers}. The
actual number of planets which will be discovered is a function of the
(unknown) fraction of planets per star. The conclusions of the OWL
studies are that the number of stars accessible to a 100~m telescope
is large enough to secure spectroscopic measurements of a large number
of planets. In the case of spectroscopy of Earths at 1~AU, however,
the number of accessible stars is just large enough to guarantee that
a few planets should be observable. These observations will be
difficult (but hopefully feasible) for a 100\,m telescope, but
are out of reach of smaller telescopes. These preliminary conclusions
clearly all require further evaluation.

\begin{table}[t]
\footnotesize
\caption[]{\footnotesize
Distance limits and numbers of stars searchable by ELT/OWL as a function 
of planetary mass and primary mirror diameter (estimated by Hainaut).
The detailed assumptions (noise sources, exposure times, etc.) have not
been documented, and these results should be taken as indicative and 
preliminary. They also rest on the feasibility of achieving the high
Strehl ratios referred to in the text.}
\begin{center}
\setlength\tabcolsep{15pt}
\begin{tabular}{rlcccc}
\hline
\noalign{\smallskip}
D(m) && \multicolumn{2}{c}{Earth-mass}& \multicolumn{2}{c}{Jupiter-mass} \\
     &&   Imaging &   Spectroscopy &   Imaging &  Spectroscopy \\
\noalign{\smallskip}
\hline
\noalign{\smallskip}
 30& d(pc)&     10   &    --      &           70     & 5 \\
   & n(star)&   22      &    0       &           6800      & 3 \\
\noalign{\smallskip}
 60& d(pc)&   22   &    --      &           120    & 18 \\
   & n(star)&   210     &    0       &           35\,000   & 170 \\
\noalign{\smallskip}
100& d(pc)&  40   &    15   &           500    & 35 \\
   & n(star)&  1200    &    67      &         2\,500\,000 &   860 \\
\noalign{\smallskip}
\hline
\end{tabular}
\end{center}
\label{tab:owl-numbers}
\end{table}

After the mechanical assembly of the OWL telescope is completed, the
mirror cell will start to be populated with segments. It is expected
that that phase will take a few years ($\sim 3$), during which the
telescope will already be available for scientific observations,
although with a reduced collecting area. The configuration of the
segments in the cell during this filling phase is still under
discussion, but an attractive option is to place them in such way that
OWL could be used as a `hyper-telescope', i.e.\ an interferometer with
densified pupil, assuming that proper phasing can be achieved in such
a scheme. In such configuration, the resolution and imaging
characteristics are very similar to that of a filled aperture
telescope with the full diameter (cf. \cite{rbg+02} and Appendix~B),
but with a very small accessible field (of the order of $\lambda/d$ of
a single segment), i.e.\ $\sim0.1$~arcsec, which is very suitable for
a planet detection.  Science time during the mirror assembly phase
could then be used to perform a broad survey for planets around nearby
stars. Assuming 100~nights of observations per year for 2--3 years,
1~hr per single observation, and 8--10 epochs per star in order to
sample the (potential) planet orbit, of the order of 300 stars could
be searched for planets.  In that way, a catalogue of targets would be
made available for further studies.  While the detection limit would
not be as deep as that of the full, completed OWL, objects discovered
with the early, hypertelescope version would have characteristics
(separation and intrinsic magnitude) that would permit further
physical studies with the completed telescope.

In conclusion, it is stressed that there remain technical
uncertainties in inferring that OWL can indeed observe Earth-type
planets: the major assumptions which must be clarified over the coming
years are that adaptive optics can work as needed, producing the 
required Strehl ratios, and providing adequate S/N ratios
in both continuum and spectral diagnostic lines.  Tracking is not
seen as a comparable issue: if the telescope can track to sub-arcsec
accuracy (it does according to analysis, at some 0.3~arcsec rms with
10\,m\,s$^{-1}$ wind and taking into account friction in drives, and
without field stabilisation with~M6), then the problem of pointing
stability is a control loop issue comparable to adaptive optics but
with a much lower frequency.

Even though smaller ELTs, such as the 30-m TMT (Thirty-Metre
Telescope), will not reach Earth-like planets, there is still a
scientific niche for them between the objectives of VLT-Planet Finder
and OWL: notably, detection and characterisation of giant planets, of
all ages, and close to their parents stars. Evidently, their technical
challenges will be easier to reach than those for a 100-m aperture.

The astrometric capabilities of OWL remain to be assessed, but
narrow-angle astrometric prospects should be very good, and provision
of an astrometric facility in the instrument plan may be appropriate.

\subsubsection{Observations at an Antarctic Site}
\label{sec:antarctic}

Excellent ground-based astronomical sites in terms of telescope
sensitivity at infrared and submillimeter wavelengths are located on
the Antarctic Plateau, where high atmospheric transparency and low sky
emission result from the extremely cold and dry air; these benefits
are well characterised at the South Pole station. The relative
advantages offered by three potentially superior sites, Dome~C,
Dome~F, and Dome~A, located higher on the Antarctic Plateau, are
quantified by \cite{law04a}. In the near- to mid-infrared, sensitivity
gains relative to the South Pole of up to a factor of 10 are predicted
at Dome~A, and a factor of 2 for Dome~C. In the mid- to far-infrared,
sensitivity gains relative to the South Pole up to a factor of 100 are
predicted for Dome~A and 10 for Dome~C.  These values correspond to
even larger gains (up to 3~orders of magnitude) compared to the best
mid-latitude sites, such as Mauna Kea and the Chajnantor Plateau.

Thus these Antarctic sites appear to offer extremely good conditions
for future interferometric projects and for (robotic) photometric
surveys, yielding one long observing `night' per year (completely dark
in June--July, decreasing to 7~hr darkness during March and October,
although with an extended `twilight' period due to the fact that the
Sun does not set very far below the horizon).  Table~\ref{tab:angel}
shows that an Antarctic OWL might achieve comparable performances to
Darwin/TPF in the near-infrared; currently this is viewed as somewhat
hypothetical, given the complex logistics and formidable
meteorological conditions for construction and operation.

Dome~C (elevation 3250~m, latitude $-75^\circ$) is the site of the
French-Italian Concordia station, whose main characteristics have been
studied over the last few years. It provides: (i) an ambient
temperature ranging from 195~K (winter) to 235~K (summer), resulting
in low, stable thermal emission; (ii) extremely dry air (250~$\mu$m
precipitable water vapour typical), resulting in enlarged and improved
infrared transmission windows; (iii) very low surface winds (median
wind speed of 2.7\,m\,s$^{-1}$, and below 5\,m\,s$^{-1}$ for more than
90\% of the time), resulting in very low free-air turbulence, with a
quasi-absence of jet streams since Dome~C is located inside the polar
vortex; (iv) some 80\% of clear skies.  The combination of coldness
and dryness for the atmosphere results in infrared photometric gains
that peak at about a factor of 25 in the K and L~bands, i.e.\ an
Antarctic 1.8-m telescope is more efficient than an 8-m telescope at a
temperate site. In the H and N bands the gain (of order~3) is also
notable although less dramatic.

Exceptional winter conditions were confirmed by teams from the
University of Nice \citep{aav+03} and Australia UNSW \citep{law04b}:
over a 3-month period, the median seeing was 0.27~arcsec, with
0.15~arcsec achieved for more than 25\% of the time.  As most of the
turbulence is located near the ground, the isoplanatic angle is
enlarged (5.9~arcsec under normalised conditions, compared with
2.9~arcsec for Paranal) which improves the field of view (and the
likelihood of finding a reference star) for adaptive optics. This has
consequences in the feasibility of adaptive optics for ELTs, where
multiconjugate systems and laser guide stars may no longer be needed.
Because the turbulence is generated by low-velocity winds, it is slow,
meaning better sensitivity and improved phase correction for adaptive
optics systems.  There is still some debate about the normalised
coherence time (whose median value is 2.6~ms on Paranal): indirect
measurements with the MASS scintillometer indicate a median value of
5.9~ms, while direct DIMM measurements show a correlation of the image
motions beyond 250~ms (as expected from $r_0=50$~cm and a typical
2.5\,m\,s$^{-1}$ wind speed).

Even taking the most pessimistic value (5.9~ms), the coherence volume
that drives the sensitivity of adaptive optics and interferometric
systems is improved by a typical factor of~20, if the AO wavefront
sensor operates in the K-band. When combined with the factor of~25
photometric gain, this results in an expected sensitivity for
interferometers and AO systems 500 times better than a similarly
equipped temperate site. For bright sources, higher dynamic range
observations (either by coronography or nulling) can be achieved due
to the reduced phase and background noise.

Current adaptive optics mainly concerns the correction of the phase of
the wavefront to achieve diffraction-limited imaging.  Even after a
perfectly plane wavefront has been produced, diffraction effects
caused by scintillation may dominate in the far wings and halos of the
image.  Scintillation, originating in inhomogeneities of the higher
atmosphere, causes patterns of `flying shadows' on the ground (with
intensity contrasts of perhaps only a few percent), carried across the
telescope pupil at the windspeed of the originating atmospheric layer.
In principle, this could be corrected by second-order adaptive optics
(modulating both the phase and amplitude), but the task is not
trivial. An additional advantage for Dome~C (assuming that `ordinary'
adaptive optics will be fully operational) is that such sites also
appear to have very much less scintillation: not only do the low
windspeeds (no jet streams such as prevail over Chile and Hawaii)
imply much less energy deposited in atmospheric turbulence, but the
winter-time atmospheric structure is such that the tropopause
effectively reaches the ground, with no significant temperature
discontinuities in the higher atmosphere.

In the most favourable cases (K band and an L band extended to
2.8--4.2~$\mu$m) Dome~C thus provides a near space-quality
environment, presumably at a small fraction of the cost, and with
almost no limitations in weight or volume. The logistics for the
station were developed by the French (IPEV) and Italian (PNRA) polar
institutes: humans reach the site by plane with a total travel time of
about 48~hr from Europe, while the heavy equipment is shipped in
standard containers by boat to the coast and then by pack trains of
Caterpillar trucks onto the glacier slopes up to Concordia.

It is still the accepted wisdom that a census of exo-Earths and the
characterisation of their atmosphere (in search for biomarkers) will
require a space mission such as Darwin.  The survey part of the Darwin
programme could perhaps be undertaken from Dome~C, given the extreme
phase and background stability of the site. A preliminary study should
be undertaken to determine whether this could be achieved, either with
a coronographic ELT in the visible or near-IR, or a large
interferometer in the thermal infrared. This would enable Darwin to
concentrate on the spectroscopy of identified targets, i.e.\ at
wavelengths 6--18~$\mu$m where clear biomarkers exist and which for
the most part are not accessible from Dome~C.  A simplified
interferometer (two 1~m telescopes) equipped with a GENIE-type
instrument could provide good-quality spectra of hot (Jupiter- or
Neptune-class) bodies and the characterisation of exo-zodiacal light
around main-sequence nearby stars -- precursor science for Darwin that
otherwise necessitates extensive use of VLT UT time.  Other
applications of Dome~C for exo-planets include astrometry and transit
photometry. The ultimate astrometric accuracy of an interferometer was
estimated by \cite{sws+03} to be a factor of some~30 better at Dome~C
than at Paranal.  However, before this potential gain is realised, the
technique must have matured to the point where astrometry on temperate
sites is limited by the atmosphere only, and not by systematic
effects.  Transit photometry could benefit from lower scintillation,
and longer continuous time coverage (polar night), than on temperate
sites.

The nature of the site, halfway between ground and space, makes it an
especially interesting area for collaboration between ESO and ESA.
Although Europe is naturally in a leadership position due to the
presence of the Concordia station, Dome~C operation is made in a fully
international (Antarctica Treaty), politically interesting (`continent
of science') context, which may help to gather resources and foster
international cooperation (Australia, USA, China) for a large project.
A major facility could be proposed at the EU level in 2007, as part of
the FP7 programme, and in coincidence with the International Polar
Year.  

While an interferometer along the lines of GENIE seems very
appealing, building OWL at Dome~C is a different matter. Exo-planet
studies are going to be one of several science drivers for any
ELT. Limited sky coverage, and logistics, may be the limiting factors.

Meanwhile, specific concepts being studied include:

(a) a 0.8-m Italian project, featuring a mid-infrared imager
(5--25~$\mu$m) and possibly a 1--2.5~$\mu$m spectrograph. Construction
at Dome~C has started, and first light is scheduled for December 2005;

(b) a large Australian Dome~C development proposal including various
groups involved in a 2-m optical/IR telescope (PILOT). Associated
structures at Dome~C are already advancing, with the first winter-over
starting in 2005;

(c) a robotic reflective Schmidt telescope for Dome~C by \cite{sas04};

(d) a successful bid for NSF study money for Phase~1 of an Antarctic
Planetary Interferometer (API), to be placed at Dome~C (PI: Mark
Swain).  Phase~1 is intended to be capable of atmospheric spectroscopy
of gas giants, whereas it will take a full Phase~2 system to target
Earth-like planets;

(e) a French concept for a Dome~C interferometer called KEOPS
(Kiloparsec Explorer for Optical Planet Search, PI: Farrokh Vakili).
Proposed timescales cover the current study phase (telescopes, beam
stabilisation, co-phasing, delay lines, recombiner, coronograph); a
single 1.5-m telescope operation in 2007; a simple interferometer with
2~telescopes in 2008--09; first 6-telescope ring system, possibly in
association with API, around 2015; and a full 36-telescope system in
the more distant future.

\clearpage
\subsection{Space Observations: 2015--2025}
\label{sec:space-2015}

The literature makes no reference to transit missions beyond Kepler
and Eddington, nor astrometric missions beyond SIM and Gaia, all due
for completion around 2015.  Rather, space missions projected for
2015--20 and beyond fall into the category of `imaging' or `direct
detection' concepts, notably Darwin and/or TPF. `Imaging' here
generally refers to imaging of the exo-planetary system, i.e.\ direct
detection of the exo-planet as a point source of light distinct from
that of the parent star, and {\it not\/} to resolved imaging of the
exo-planet surface.

The next major break-through in exo-planetary science will be the
detection and detailed characterisation of Earth-like planets in
habitable zones. The prime goals would be to detect light from
Earth-like planets and to perform low-resolution spectroscopy of their
atmospheres in order to characterise their physical and chemical
properties. The target samples would include about 200 stars in the
Solar neighbourhood. Follow-up spectroscopy covering the molecular
bands of CO$_2$, H$_2$O, O$_3$, and CH$_4$ will deepen understanding
of Earth-like planets in general, and may lead to the identification
of unique biomarkers. The search for life on other planets will enable
us to place life as it exists today on Earth in the context of
planetary and biological evolution and survival.

In the more distant future, perhaps well beyond 2025, a successful
Darwin/TPF would logically be followed by `life finders' and true
`planet imagers'.  At present they appear only as more distant goals,
and a brief discussion of them is given in
Appendix~\ref{sec:beyond-2025}.  They are unlikely to affect ESA/NASA
policy over the next decade or more, at least until the prospects for
the success of Darwin/TPF can be quantified, except in the areas of
advanced technology studies.  Similarly, ideas beyond Darwin/TPF 
are unlikely to influence the choices or prospects for the very large
(50--100~m) telescopes on ground.

\subsubsection{Darwin}

\begin{figure}[t]
\begin{center}
\centerline{\epsfig{file=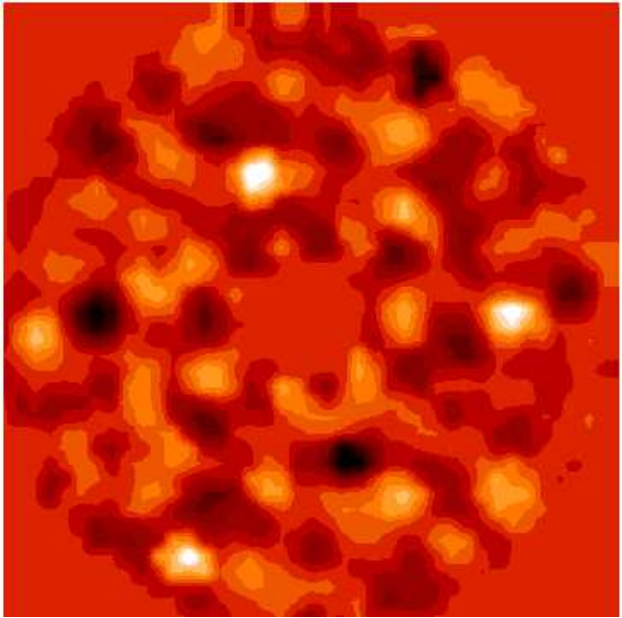,width=0.55\textwidth}}
\end{center}
\vspace{-20pt}
\caption[]{\footnotesize
  A $2\times2$ arcsec$^2$ simulated image of our Solar System viewed
  from 10\,pc using the (former) 6-telescope Darwin configuration.
  Observations are over 6--18\,$\mu$m in a 10-hour exposure, with the
  system viewed at 30$^\circ$ inclination, and assuming zodiacal
  emission as for our Solar System.  The Sun is located at the centre.
  Venus, Earth and Mars are visible.}
\label{fig:darwin}
\end{figure}

Darwin is the ESA mission concept aiming at the direct detection of
exo-planets, and is focused on an interferometer configuration. It was
originally conceived as a set of eight spacecraft at L2 (6~telescopes,
one beam combination unit, and one communication unit) that would
survey 100 of the closest stars in the infrared, searching for
Earth-like planets and analysing their atmospheres for the chemical
signature of life \citep{fri00}, scientific objectives in common with
those of TPF.  More recent studies have identified a simplified
option, employing 4~telescopes separated by up to 50--100~m operated
in a `dual-Bracewell' configuration (or possibly $3\times3.5$~m
telescopes), requiring a dual Soyuz-Fregat launch, and a target launch
date of 2015. The mission foresees a detection phase of 2~years
(allowing the follow-up of 150--200~stars), and a spectroscopy phase
of 3~years.  Specific precursor efforts include a possible space
mission (Smart-3) to demonstrate the concept of formation flying. The
emphasis on the infrared rather than the optical is guided by: (i) the
less-stringent requirements on technology (contrast and resolution);
(ii) infrared spectroscopy allows characterisation in a direct and
unambiguous manner; (iii) infrared interferometry allows a larger
sample of objects to be surveyed for Earth-like planets (150--200
versus 32~for visual coronography); (iv) it provides the heritage for
the next generation of missions which will demand higher resolution
spatial imaging. Figure~\ref{fig:darwin} is a simulated image of our
Solar System viewed from 10\,pc.

Table~\ref{tab:angel}, from \cite{ang03}, is a summary of detection
capabilities for an Earth-Sun system at 10\,pc for various experiments
being studied, including TPF and Darwin. It shows that even for the
most ambitious projects being planned at present, direct detection of
even a nearby Earth represents a huge challenge. 

\begin{table}[t]
\footnotesize
\caption{\footnotesize
Detection capabilities: Earth at 10\,pc, from \protect\cite{ang03}.
$\Delta\theta=0.1$~arcsec, $t_{\rm int}=24$~hr, QE~=~0.2, 
$\Delta\lambda/\lambda=0.2$. Mode~=~N~corresponds to a nulling
system, C~to a coronograph. The ground-based results assume that
long-term averaging is realistic, with fast atmospheric correction.
For further details of assumptions, see \protect\cite{ang03}}
\begin{center}
\setlength\tabcolsep{5pt}
\begin{tabular}{lrccrl}
\hline\noalign{\smallskip}
Telescope& Size (m)& $\lambda (\mu$m)& Mode& S/N& Comment\\
\noalign{\smallskip}
\hline
\noalign{\smallskip}
Darwin/TPF-I &  $4\times2$&    11&  N&   8& See also Table~\ref{tab:darwin} \\
TPF-C &         3.5&          0.5&  C&  11& Typical launcher diameter \\
\quad  ''  &           7&          0.8&  C&  5--34& \\
Antarctic &      21&           11&  N&  0.5& \\
\quad  ''  &            &          0.8&  C&    6& \\
CELT, GMT &      30&           11&  N&  0.3& 30~m [C] too small at 11\,$\mu$m \\
\quad  ''  &            &          0.8&  C&    4& \\
OWL &           100&           11&  C&    4& Large $\Phi$ [C] for IR suppression \\
\quad  ''  &            &          0.8&  C&   46& Optical spectroscopy possible \\
Antarctic OWL & 100&           11&  C&   17& Comparable to Darwin/TPF, but \\
 & & & & &                  O$_3$ (9.6\,$\mu$m) and CO$_2$ (15\,$\mu$m) \\
 & & & & &                  not accessible (atmosphere opaque) \\
\quad  ''  &            &          0.8&  C&   90& Water bands at 1.1--1.4\,$\mu$m \\
\hline
\end{tabular}
\end{center}
\label{tab:angel}
\end{table}

Table~\ref{tab:darwin} summarises the estimated integration times for
a variety of stellar types (target stellar types are F--G with some
K--M), and a range of distances. Broadly, these are consistent with
the simplified summary given in Table~\ref{tab:angel}.  Typical
distance limits are out to 25\,pc. A terrestrial planet is considered
as $R_{\rm max}\sim2R_\oplus$.  The spectral range is 6--18 $\mu$m,
covering the key absorption lines, with a spectral resolution of 20
(50~is required for CH$_4$ in low abundance).  Integration time
estimates are given for a S/N~=~5 in imaging (i.e.\ detected planetary
signal/total noise) and S/N~=~7 in the faintest part of spectrum for
spectroscopy, for the following assumptions: total collecting area for
all telescopes of 40~m$^2$; telescopes at 40\,K; effective planetary
temperature 265~K (signal scales as $T_{\rm eff}^4$); $R=R_\oplus$;
integrated over 8--16 $\mu$m; 10~times the level of zodiacal dust than
in the inner solar system; all of the habitable zone is searched;
Spitzer Si-As detectors.  Times given are based on a 90\% confidence
level for a non-detection based on 3~observations (a positive
detection at S/N~=~5 then takes one third of the times given).  To
detect the Earth at 265~K (instead of 290~K) around the Sun at 20\,pc
would take 9~hr at S/N~=~5, and 36~hr at S/N~=~10. For a K5V star at
10\,pc these times are about 1--4~hr respectively.  For spectroscopy,
the S/N varies between 7 and significantly higher, depending on the
atmosphere. For an Earth atmosphere (but at 265~K) at 20\,pc, an
integration time of 54~hr provides a S/N$\sim7-15$.
  
One of the strongest noise sources is the leakage of photons from the
resolved stellar surface.  This is a function of both stellar
temperature and diameter (distance, spectral type).  It is most
clearly seen for nearby stars where the detection time drops
strongly. For 20\,pc detection times do not change much since the star
does not leak very much for any diameter.  Detection times rise
rapidly for K5V stars at 30\,pc, because of the 90\% confidence
requirement, which itself demands more changes in array size because
the habitable zone is very close to the star.

The number of candidates accessible and visible to Darwin (two
$\pm45^\circ$ caps near the ecliptic poles are inaccessible) is
estimated as follows.  Considering only single stars, there are 211~K
and 82~G candidates out to 25\,pc, with another 30~F-type stars plus
many M-dwarfs.  Combined with the integration time estimates, Darwin
should survey more than 150 stars in 2~years. Within 5~years, all
293~single and accessible solar-type stars stars out to 25\,pc can be
surveyed, with a spectrum obtained for each system in the case of a
planet prevalence of $\sim$10\%. Up to 1000 systems in the solar
neighbourhood could be surveyed if the planetary fraction is even
smaller.

\begin{table}[b]
\footnotesize
\caption{\footnotesize Darwin: integration times for detection of 
Earth-like planets at S/N~=~5,
and spectroscopy at S/N~=~7 in the faintest part of the spectrum (in hours).
See text for details.}
\begin{center}
\setlength\tabcolsep{5pt}
\begin{tabular}{cccc}
\hline\noalign{\smallskip}
Stellar type&   10\,pc&  20\,pc&          30\,pc \\
\noalign{\smallskip}
\hline
\noalign{\smallskip}
G2V&    18--33 &        28--54 &        109--173 \\
G5V&    12--22 &        27--46 &        105--166 \\
K2V&     4--9  &        26--37 &        104--157 \\
K5V&     4--6  &        26--35 &        249--155 \\
\hline
\end{tabular}
\end{center}
\label{tab:darwin}
\end{table}

\subsubsection{The Darwin Ground-Based Precursor: GENIE}
\label{sec:genie}

GENIE is a nulling interferometer under development by ESA and ESO for
the VLTI. GENIE will allow the development and demonstration of the
technology required for nulling interferometry, allowing testing of
the Darwin technology in an integrated and operational system
(amplitude control loops, high-accuracy optical path difference
control loop, dispersion control, polarisation compensation,
background subtraction, and internal modulation).  GENIE is considered
by the Darwin project as necessary for demonstrating these technical
concepts in advance of launch.

Two competitive instrument definition studies were due for completion
by the end of 2004, with the scientific case being prepared in
collaboration between ESO and ESA.  GENIE will be considered by the
ESO Council in April 2005, and could be operational by mid-2008.  If
not accepted by ESO, a separate laboratory technology demonstrator
would be needed to validate the Darwin concepts.

GENIE is also considered mandatory as a specific instrumental
configuration to survey southern-hemisphere target candidates, in
order to decide which targets are most suitable for observation by
Darwin, and specifically to characterise the level of exo-zodiacal
light, which must be below certain limits necessary for exo-planetary
detections. A corresponding programme is planned to be undertaken at
the Keck telescope in the northern hemisphere for preparations for
TPF. Studies by \cite{hag+04} show that the capabilities of GENIE are
sufficient to detect the planet around $\tau$~Boo within 1~hour, and
by inference some 5~other candidate `hot Jupiter' planets.  GENIE
could use either the VLT UTs or ATs, at a wavelength of 3.6\,$\mu$m,
and will require a significant number of observing nights (of
order~50).

The main limiting factors for GENIE are the phase and thermal
background stability on Paranal, as well as system complexity issues
involved with its integration into the existing VLTI environment. If
GENIE were located on the high Antarctic plateau (Dome~C) where the
thermal background is lower, and the seeing both better and slower
(see Section~\ref{sec:antarctic}), the performance required to perform
its science programme could perhaps be achieved using smaller (of
order 1~m) telescopes and a simpler, dedicated system architecture.
This has to be balanced against the reduced sky coverage at
$-75^\circ$ latitude, and probably more complex logistics.

\subsubsection{Terrestrial Planet Finder (TPF)}

NASA's TPF roughly parallels the ESA Darwin study, with close
discussions taking place between the two teams. TPF was conceived to
take the form of either a coronograph operating at visible wavelengths
or a large-baseline interferometer operating in the infrared
\citep{bei03}. There are two aspects of this choice which should be
distinguished: (a) the scientific aspect: whether reflected (visible
and near IR) light or thermal emission (mid-IR) is the best regime to
characterise planets (albedo, temperature, colour, key species that
can be identified e.g.\ CO$_2$ etc.; see \cite{sch03} for a recent
discussion); (b) the instrumental aspects: whether an interferometer
or a coronograph is the best.  Here, the NASA Technology Plan for TPF
stated that `Technology readiness, rather than a scientific preference
for any wavelength region, will probably be the determining factor in
the selection of a final architecture'.  In May 2002, two
architectural concepts were selected for further evaluation: an
infrared interferometer (multiple small telescopes on a fixed
structure or on separated spacecraft flying in precision formation and
utilising nulling), and a visible light coronograph (utilising a large
optical telescope, with a mirror three to four times bigger and at
least 10~times more precise in wave-front error than the Hubble Space
Telescope).

In April 2004, NASA announced that it would embark on a
$6\times3.5$~m$^2$ (more recently changed to $8\times3.5$~m$^2$)
visual coronograph in 2014 (TPF-C), with a wavelength range
0.6--1.06\,$\mu$m, and targeting a full search of 32~nearby stars and
an incomplete search for 130~stars (more recently quoted as~80). A
free-flying interferometer, in collaboration with ESA, would be
considered before 2020 (TPF-I).

A visible light system can be smaller (some 10~m aperture) than a
comparable infrared interferometer, however advances in mirror
technology are required: mirrors must be ultra-smooth
($\sim\lambda/15\,000$; a number stated in the TPF documentation,
although values an order of magnitude inferior appear more plausible)
to minimise scattered light, and in addition active optics would be
needed to maintain low and mid-spatial frequency mirror structure at
acceptable levels. Infrared interferometry would require either large
boom technology or formation flying, typically with separation
accuracies at the cm-level with short internal delay lines.  For the
detection of ozone at distances of 15\,pc and S/N$\sim$25, apertures
of about 40~m$^2$, and observing times of 2--8~weeks per object, are
indicated.

Many ideas for scientific and technological precursors for TPF have
been examined. The many possible solutions involve combinations of
adaptive wavefront correction, coronographs, apodization,
interferometers, and large free-flying occulters.  The main contenders
are summarised in Appendix~\ref{sec:space-precursors} for
completeness, although with the recent (April 2004) NASA announcement
on TPF strategy, it is not clear whether any of these concepts will be
developed further.

\clearpage
\subsection{ESA Themes: 2015--2025}
\label{sec:esa-themes}

In mid-2004, ESA solicited a call for ideas for its scientific 
programme beyond 2015, resulting in several outline proposals 
in the area of exo-planets:

{\bf A Large UV Telescope (Lecavelier):} the proposal considers a
large UV telescope as a promising way to conduct a deep search for
bio-markers in Earth-mass exo-planets, through the detection of
relevant atmospheric signatures (such as ozone, water, CO, and CO$_2$)
in a very large number of exo-planets, and planetary satellites, up to
several hundred parsec distant. It notes that the detection of
atmospheric signatures in HD~209458b has been made through space UV
observations, and stresses that large numbers of prime targets for
these observations will be detected by Gaia. It underlines the
possibility of time-resolved spectroscopy (over the 10-min typical
transit time of an Earth-mass planet) providing spatial information of
atmospheric constituents along the planet surface (poles versus
equator, presence of continents). A JWST-size telescope is proposed,
and some quantitative expectations are given.
  
{\bf Search for Planets and Life in the Universe (L\'eger):} the
proposal makes the general case for the continued development of an
ESA strategy for the improved detection and characterisation of
exo-planets, primarily through the techniques of transits and direct
detection. This proposal was submitted on behalf of more than 200
individuals and institutions.
  
{\bf Astrometric detection of Earth-Mass Planets (Perryman):} the
proposal points out that, beyond the micro-arcsec astrometric
accuracies of ESA's Gaia mission, 10~nano-arcsec accuracies would
permit the detection of astrometric perturbations due to Earth-mass
planets around Sun-like stars at 100\,pc. If a survey-type mission
were feasible, the concept could lead to the systematic survey of many
hundreds of thousands of stars for Earth-mass planets -- important for
the generation of target objects if the fraction of Earth-mass planets
turns out to be very small.  Earth-mass perturbations around a
solar-mass star are 300~nano-arcsec at 10\,pc, or 30~nano-arcsec at
100\,pc, the latter requiring an instantaneous measurement accuracy a
factor~3 better, i.e.\ 10~nano-arcsec at, say, 12~mag.  This is a
factor of some 1000 improvement with respect to Gaia.  Keeping all
other mission parameters (efficiency, transverse field of view,
mission duration, total observing time per star, and image pixel
sampling, etc.) unchanged, we can consider reaching this accuracy
simply through a scaling up of the primary mirror size. The Gaia
primary mirror has an along-scan dimension $D=1.4$~m and a transverse
dimension $H=0.5$~m; the final accuracy scales as $\sigma\propto
D^{-3/2} H^{-1/2}$. These desired accuracies would therefore require a
primary mirror size of order $50\times12$~m$^2$, and a focal length
(scaling with~$D$) of about 1600~m, similar to the scale of the optics
derived for the mini-version of Life Finder.  Accuracy levels of
$\sim10$~nano-arcsec are still above the noise floors due to
interplanetary and interstellar scintillation in the optical, or
stochastic gravitational wave noise.  Astrometric accuracy limits due
to surface granular structure and star spots were discussed in
Section~\ref{sec:limits}.  To reach the astrometric precision of
300~nano-arcsec at 10\,pc means that the photocentre of a star must be
determined to $\sim5\times10^5$~m, or $1.5\times10^{-4}\,R_\odot$.
This is still significantly above the photometric stability limits of
$10^{-7}-10^{-8}$\,AU, or $10^3-10^4$~m, derived by \cite{sl05} for
stars with $\log g=4.4$. Note, however, that this only concerns
thermally-driven granulation, thus determining only the lowest
possible level of stellar astrometric stability.
  
{\bf Understanding the Planetary Population of our Galaxy (Piotto):}
the proposal underlines the importance of the field to ESA and to
Europe.  It argues that most disciplines start with the discovery and
study of individual objects, before moving on to the more mature stage
in which large, unbiased samples of objects are studied. In the decade
2015--25, an open issue will be the discovery and characterisation of
large unbiased samples of planets down to habitable systems. This
should provide information necessary to understand how the environment
and nature of the parent star affects the resulting planetary
population; how stellar metallicity, multiplicity, crowding, and
population influence the nature, habitability and survival of
planetary systems. The proposal suggests a transit-type follow-up to
Kepler/Eddington, using a significantly larger aperture and improved
detector technologies to obtain much improved samples/statistics;
combined with the availability of large ($>25$~m) ground-based
telescopes for follow-up spectroscopic observations.
  
{\bf Exo-planet Detection and Characterisation (Surdej):} the proposal
underlines the general importance of the field to ESA and to Europe,
and proposes to intensify efforts towards a Darwin-like mission,
including further emphasis on coronography, perhaps as an ESA
instrument on TPF-C. The proposal also underlines the importance to
astrobiology of sample-return missions in the Solar System, for
example to Mars, Europa, and Titan.  Otherwise, no specific instrument
approach or design is considered.
  
{\bf Evolution of Atmospheres and Ionospheres of Planets and
  Exo-Planets (Andr\'e):} the proposal considers a combination of {\it
  in situ\/} observations of Solar System planets and remote sensing
of exo-planets in order to advance understanding of the long-term
evolution of the atmospheres and ionospheres of planets (and large
moons), and to identify the important sources and sinks of atmospheres
and ionospheres of planets during different parts of their evolution.
No specific instrument approach or design is considered.
  
{\bf Planetary Habitability in the Solar System and Beyond (Bertaux):}
the proposal underlines the general importance of the field to ESA and
to Europe. It includes suggestions for intensifying the search for
(past) life on Mars through robotic exploration in order to refine
probability estimates of life beyond the Solar System.  It also notes
the potential problem that ESA's Darwin mission is presently designed
as its own candidate surveyor, and argues that a preliminary
programme, possibly ground-based, should undertake the advanced search
so that Darwin can focus its observing time on spectral acquisition of
Earth-mass planets. The suggestion is to conduct this survey, under
ESA responsibility, using Doppler measurements, reaching the required
accuracies of 0.1--0.8~m\,s$^{-1}$ through the accumulation of large
number of individual 1--2~m\,s$^{-1}$ measurements. The feasibility of
this proposal is addressed in Section~\ref{sec:limits}.

{\bf The Hypertelescope Path (Labeyrie):}
Appendix~\ref{sec:beyond-2025} provides an introduction to the
concepts and goals of the `densified pupil multi-aperture imaging
interferometer' or `hypertelescope'. The proposal refers to the fact
that space testing of versions as small as 20~cm mirror diameter,
with mass below 0.5~kg, could be considered. Progressive versions
could be used to study stellar surface resolution (20~cm apertures
spanning 100~m in geostationary orbit); detection of exo-Earths at
visible and infrared wavelengths (spanning a few hundred metres
at~L2); much larger versions will be needed to resolve surface 
features of exo-Earths.

{\bf Lamarck -- an International Space Interferometer for Exo-Life
  Studies (Schneider):} the proposal stresses the importance of the
optical (rather than infrared) domain for space interferometry, and
outlines a concept for a temporaily `evolving' interferometric
station, with the participation of other countries like China, Japan,
India and Russia, employing an initial collecting area of around
40\,m$^2$, baselines above 3~km, a spectral range of 0.3--3~$\mu$m and
$R=100$. The mission strategy would be to detect Earth-like planets
with baselines up to 1~km, imaging of the most promising candidates
with very long baselines, then interferometer upgrades with
subsequently-launched free flyers.

\vspace{10pt}

The resulting recommendations of the ESA Astronomy Working Group are
contained in ASTRO(2004)18 of 19~Oct 2004. The 47 responses were
assigned to three themes: (1) Other worlds and life in the Universe;
(2) the early Universe; (3) the evolving violent Universe. The
relevant part of the document is reproduced here verbatim:
 
{\it \small
1.1 From exo-planets to biomarkers

After the first discovery of an extra-solar planet in 1995, there has
been steady progress towards detecting planets with ever smaller
masses, and towards the development of a broader suite of techniques
to characterise their properties. There is no doubt that this trend
will continue into the next two decades, as substantial technological
challenges are progressively overcome. After Corot will have opened
the way to telluric planet finding, the Eddington mission would get a
first census of the frequency of Earth-like planets. Gaia will deliver
important insights into the frequency of giant planets; the existence
and location of such planets is crucial for the possible existence of
Earth-like planets in the habitable zone. Gaia will also further
improve our understanding of the stellar and Galactic constraints on
planet formation and existence.

The next major break-through in exo-planetary science will be the
detection and detailed characterisation of Earth-like planets in
habitable zones. The prime goals would be to detect light from
Earth-like planets and to perform low-resolution spectroscopy of their
atmospheres in order to characterise their physical and chemical
properties. The target sample would include about 200 stars in the
Solar neighbourhood. Follow-up spectroscopy covering the molecular
bands of CO$_2$, H$_2$O, O$_3$, and CH$_4$ will deepen our understanding of
Earth-like planets in general, and may lead to the identification of
unique biomarkers. The search for life on other planets will enable us
to place life as it exists today on Earth in the context of planetary
and biological evolution and survival. Being aware of the technical
challenge to overcome the high brightness ratio between star and
planet and the advancements in technology reached during the past few
years by ESA studies, the AWG strongly and unanimously recommends the
implementation of a mission addressing these objectives through a
nulling interferometer for the wavelength range between 6--20~$\mu$m.
Such a mission should be implemented around 2015, making
Europe highly competitive in the field.

The next step in this well-defined roadmap would be a complete census
of all terrestrial extra-solar planets within 100~pc, for example
through the use of high-precision astrometry. Longer-term goals will
include the direct detection and high-resolution spectroscopy of these
planets with large telescopes in IR, optical and UV wavelengths and
finally the spatially resolved imaging of exo-Earths, leading to the
new field of comparative exo-planetology.

}

The conclusions of the document state:
{\it \small
  For the very first part of the 2015-2025 decade, two
  observatory-class missions are recommended:
  (1) a mid-infrared nulling interferometer for the detection and
  characterisation of Earth-like planets in the Solar neighbourhood.
  In order to ensure a timely implementation of such a mission, the
  AWG gives high priority to the continuation of the present fruitful
  technology program for this project, and strongly and unanimously
  recommends that all steps be taken to ensure that such a mission can
  be flown as early as possible in the 2015 time frame;
  (2) a far-infrared observatory, etc...

}

The conclusions also comment: {\it \small The AWG reiterates its
  unanimous support to a rapid implementation of the Eddington
  mission, which would constitute an important intermediate step
  towards the scientific goals formulated in theme~1.}

\subsection{Other Concepts and Future Plans}

Although it is of course impossible to look forward a number of years
and to predict the status then, more or less plausible novel (i.e., as
yet unproven) methods might conceivably be applied for exo-planet
studies.

The use of `new' physical principles for the imaging of exo-planets may
develop.  Not long ago, it was realised that light (photons) may carry
orbital angular momentum (in addition to the ordinary photon spin
associated with circular polarization).  Interference of such light
may produce a dark spot on the optical axis, but high intensities
outside (but still inside the `classical' diffraction spot).  By
suitable manipulation of starlight, different amounts of angular
momentum might be induced to light from the central star, relative to
its nearby planet, thus enhancing the observable contrast by several
orders of magnitude.  For an introduction to light's orbital
angular momentum, see \cite{pca04}; for a discussion of high-contrast
imaging using such methods, see \cite{swa01}; for reference in the 
context of exo-planet studies, see \cite{har03}.

\clearpage
\subsection{Summary of Prospects: 2015--2025}

The major space and ground-based experiments so far
considered for the 2015--25 time-frame do not include
survey missions beyond Kepler/Eddington and Gaia,
but focus on the detection and characterisation of 
a few low-mass planets in the Solar neighbourhood.

The detection prospects are summarised in Table~\ref{tab:angel},
compiled by \cite{ang03}, and include Darwin/TPF and OWL.  An
assessment of the S/N of planet detections with OWL has been made
independently by Hainaut \& Gilmozzi (ESO), and is in broad
agreement with those in the table. The Darwin S/N assessments
for both imaging and spectroscopy, detailed in Table~\ref{tab:darwin},
are also broadly in agreement with this summary table. 

Accepting the (unproven) hypothesis that there are Earth-mass
planets around solar-type stars within 10--20\,pc, their detection and
characterisation even with Darwin/TPF and/or OWL will be challenging.
At this time it would seem appropriate to continue the search with
both ground and space techniques, at least until their respective
technical limitations and costs are better understood.

\cleardoublepage

\section{The Role of ESO and ESA Facilities}
\label{sec:eso-esa}

\subsection{The Expected Direction of Research}
\label{sec:expected-direction}

As detailed in the previous sections, plans for the forthcoming decade
are focused on surveys that attempt to detect large numbers
of objects by various methods (transits, astrometry, microlensing).
The aim is to obtain information on thousands of systems, and thus to
characterise the population frequency of extra-solar planets according
to mass, orbital radius, etc., essential for refining theories of
planet formation and evolution.

Beyond 2015, programmes targeting further characterisation of the
large-scale statistical distribution of planets do not (yet) exist,
probably since the requirements for large-scale surveys of other
aspects of planetary parameter space have not yet been carefully
considered. Instead, post-2015 projects currently address the
existence and characterisation of Earth-mass planets around a 
number of carefully selected candidate stars, both from ground and
space, and questions related to habitability and the existence of
life. Both ground and space approaches will be highly challenging and
expensive, and their technical feasibility is still being evaluated.

A synthesis of prospects is presented in Figure~\ref{fig:strategy},
which tabulates the space (second column) and ground (fourth column)
techniques, the resulting candidate detections (third column) and
statistical knowledge (first column), and possibilities for follow-up
observations.

In this section the following questions are addressed:

(a) what follow-up observations and facilities are required to
characterise these systems more completely (Section~\ref{sec:role-a});

(b) what does the resulting (statistical) knowledge of exo-planet
distributions imply for the targeted observations of Darwin and OWL
(Section~\ref{sec:role-b});

(c) what information will be available, or should be anticipated,
for a deeper astrophysical characterisation of the host stars of planetary
systems (Section~\ref{sec:role-c});

(d) what is the potential overlap amongst the major facilities
currently planned or studied by ESO and ESA (Section~\ref{sec:role-d});

(e) are there specific long-lead time space or ground facilities which
should be considered to fill observational gaps anticipated over the next
10--20~years (Section~\ref{sec:role-e});

(f) are there other considerations that ESO/ESA should investigate for
proper interpretation of the data which will be generated by these two
European organisations, or others, and which might limit the
development of the field unless suitably coordinated
(Section~\ref{sec:role-f}).

\subsection{Follow-Up Observations}
\label{sec:role-a}

\begin{figure}[p]
\caption[]{\footnotesize Links between detection methods in space
  (second column) and ground (fourth column), resulting statistical
  information (first column) and candidate detections (third column).
  Experiments which are not approved are shown as dashed boxes. The
  large dash-dotted boxes show the cumulative statistical knowledge, and
  the cumulative candidate lists respectively. Solid arrows show the
  candidates resulting from a particular observation. Dotted arrows
  indicate the follow-up observations needed. Note that the Kepler
  detections will all be in the northern hemisphere, and thus
  unobservable from ESO.}  
\vspace{20pt}
\begin{center}
\epsfig{file=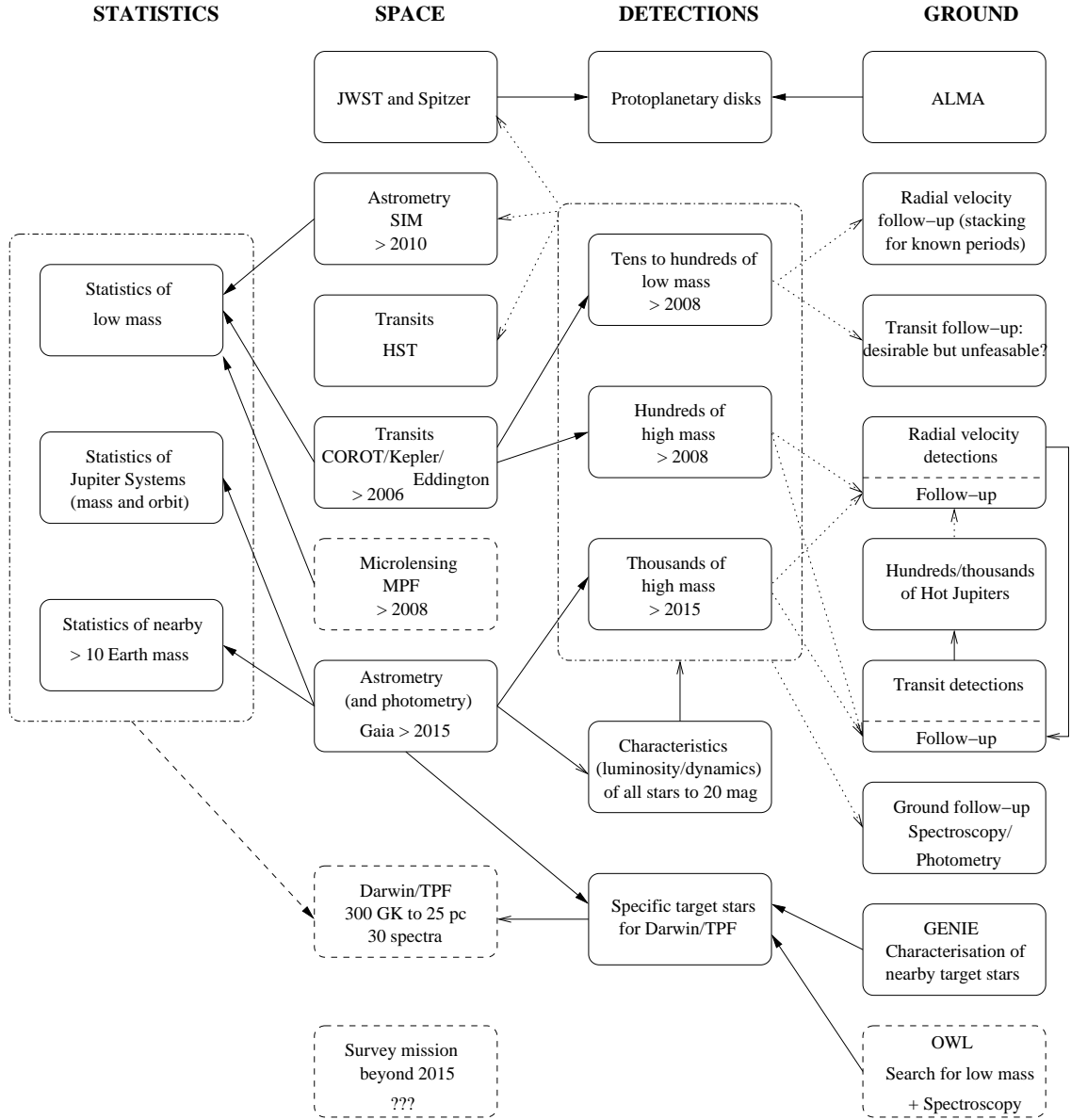,width=1.00\textwidth}
\end{center}
\label{fig:strategy}
\end{figure}

With reference to Figure~\ref{fig:strategy}, over the next 5--10~years
ongoing or approved survey experiments are expected to generate:

(a) high-mass ($\sim M_{\rm J}$) candidates: some hundreds from COROT,
Kepler and Eddington; thousands with Gaia astrometry and thousands of
transiting systems with Gaia photometry; some hundreds from ongoing
and future ground-based radial velocity surveys; and hundreds
(possibly thousands) of hot Jupiters from ground-based transit
surveys;

(b) low-mass ($\sim 1-3\, M_\oplus$) candidates: a small number of hot
terrestrial planets from COROT around 2008; with tens to hundreds from
Kepler expected to be available to the community after about 2010.

In principle, the information required for further characterisation of
a detected planetary system is independent of the star or planet mass:
as discussed elsewhere in this report: (a) radial velocity
measurements of transit detections to eliminate false alarms due to
grazing eclipsing binaries, triple stars, star spots, and false
positives; (b) transit spectroscopy from ground or space for the
determination of atmospheric properties; (c) the combination of
transit or astrometry data with radial velocity information allowing
the determination of the true mass of the planet; (d) photometric or
spectroscopic information needed to characterise the parent star; (e)
follow-up imaging of transiting candidates with high spatial
resolution adaptive optics to minimise the possibility that the object
is actually multiple, or that a foreground or background binary system
is causing the dimming of the light; (f) all the above will provide
candidates for ground-based imaging by VLT, VLTI, OWL, etc.

For microlensing candidates, no follow-up observations are generally
possible. Nevertheless, due to the relative proper motion, the lens
star will increase its angular separation to the source star and
become visible after some time. One case is known: MACHO LMC--5, in
which the lens star was imaged after about 8~years or so \citep{aaa+01}.
So in principle, there exists a possibility to study the host star of
a microlensing planet, depending on the mass/apparent brightness of
the lensing (i.e., host) star, and on the relative transverse
velocity.

In practice, the problem is distinct for high-mass or low-mass planets.

\subsubsection{High-Mass Planets}

For high-mass objects (of order $1\,M_{\rm J}$), ground-based
follow-up transit measurements will generally be technically feasible
with the current generation of instruments.  COROT, Kepler, Eddington,
Gaia, OGLE, possibly HST, and the ground-based transit networks will
supply thousands of targets which can be followed from the ground.
Small telescopes, preferably robotic, and substantial amounts of
observing time will be required.  Existing transit measurement
facilities and amateur networks could contribute, although some new
dedicated provision for follow-up is probably needed, in particular
enlarging networks in longitude to significantly improve the
efficiency of ground-based transit observations.

Radial velocity follow-up will be needed for all high-mass candidates,
of which Gaia will generate a very large number.  Thousands of the
more massive planets could be observed by ground-based radial velocity
instruments, assisting the determination of multiple planets, relative
orbital inclinations, etc. ESO could consider a coordinated
large-scale follow-up of such radial velocity observations, and
whether additional facilities will be needed.  An assessment of
telescope time/aperture needed for these projects has not been made
--- probably a combination of existing facilities and larger dedicated
instruments would be needed.

\subsubsection{Low-Mass Planets}

Low-mass (of order $1\,M_\oplus$) planets will hopefully be detected 
by Kepler and Eddington, perhaps in moderately large numbers 
(some hundreds). 

They will typically be too low-amplitude for ground-based photometric
follow-up, which in any case are unlikely to improve on the S/N of
weak transit events detected by Kepler or Eddington -- these missions
will have years of lightcurves on a star, and if the signal is still
weak after phasing and adding all data, it will be difficult to
improve from the ground in a reasonable time.  This is a potential
problem, since such follow-up observations will be needed: (i) to
confirm the reality of the lower S/N detections; (ii) to search for
planetary periods for candidates for which only one transit is
detected; (iii) to confirm candidate detections for which two or
possibly more transits were detected (since periods and transit times
are known, ground-based follow-up may be more feasible); (iv) to
search for period changes due to planetary moons etc.; (v) to
characterise the transit systems in terms of chemical abundances.
Probably the only prospect is follow up with HST and/or JWST (see
Section~\ref{sec:space-transits}) for transits, and possibly SIM for
astrometry, although many of the transit candidates will be too
distant even for these instruments.

Radial velocity measurements are again needed in principle to
supplement the orbital information. Improvements in radial velocity
precision for transits may be achieved by the stacking of repeated
observations at the known planet period, as described in
Section~\ref{sec:limits}.  If Eddington is approved (or for the study
of Kepler candidates), the development of new telescope facilities
should be considered, such as a high-precision spectrograph (like
HARPS) based on an 8-m (or larger) telescope. Given the low expected
surface density of accessible candidates, it is unlikely that a
multi-object instrument would be effective and so a highly-optimised
single object instrument offering a precision of $\sim1$\,m\,s$^{-1}$\ 
would be preferred. The HARPS detection of a 14\,$M_\oplus$ planet
demonstrates that it may be possible to characterise all the
exo-planets detected by COROT (massive Earths with short periods)
in this way.

\subsubsection{Summary of Follow-Up Facilities Required}
\label{sec:follow-up}

To summarise the requirements for these follow-up programmes, the
following facilities will be needed:

(a) high-precision radial velocity instrumentation for the follow-up
of astrometric and transit detections, to ensure the detection of a
planet by a second independent method, and to determine its true mass.
For Jupiter-mass planets, existing instrumentation may be technically
adequate but observing time may be inadequate; for Earth-mass
candidates, special-purpose instrumentation (like HARPS) on a large
telescope, would be required.  Requirements for automated telescopes
for precise radial-velocity measurements have been discussed by
\cite{pm04};

(b) photometric monitoring of large numbers of high-mass planet
 candidates coming from space experiments: a combination of existing
 and new facilities will be needed;

(c) very high-resolution (adaptive optics) imaging of transit
candidates to reject the `false-positives' generated by confusing
sources;

(d) photometric and spectroscopic characterisation of all candidate
planetary systems, providing fundamental stellar parameters such as
temperature, gravity and metallicity. A combination of planned (Gaia)
and new (spectroscopic survey) instrumentation will probably be
required;

(e) follow-up of candidates from microlensing studies will be
necessary to identify host-star properties that maximise the
probability of finding a planetary system.
 
A first (and incomplete) assessment of necessary telescope time can be
made as follows:

(a) for the photometric monitoring: assuming a telescope optimized for
fast photometric measurements (i.e.\ with low overhead), up to
6~measurements per hour can be expected, i.e.\ 15\,000 measurements
per year. A dedicated instrument, following the concept of the
`Gamma-Ray burst Optical/Near infrared Detector' (GROND,
http://www.mpe.mpg.de/$\sim$jcg/GROND/), permitting simultaneous
measurements in up to 7~photometric bands, could be duplicated for a
4-m and for a 1--2-m telescope in order to optimize the aperture of
the telescope with the brightness of the target. The photometric
monitoring would then be performed in about 1~year (full time)
assuming two telescopes.

(b) for the radial velocity measurements: assuming an instrument like
HARPS after optimization for fast observations, 4--5 stars could be
measured each hour (i.e.\ about 5~minutes preset and overhead, and
5--10~minutes exposure), thus 40--50 measurements per night, or up to
15\,000 measurements per year, assuming the instrument is operated
full time. This follow up will then require this telescope for a few
years in order to measure each star at least 3--4 times at different
orbital phases.

\subsection{Statistics of Exo-Planets: Implications for Darwin/OWL}
\label{sec:role-b}

As noted under Appendix~\ref{sec:space-precursors}, the McKee-Taylor
Decadal Survey Committee \citep{mt00} qualified its endorsement of the
TPF mission with the condition that the abundance of Earth-size
planets be determined prior to the start of the TPF mission. A similar
strategy will be desirable for Darwin.

The various missions and experiments discussed throughout this report
will contribute to detailed statistical information of planetary
distributions over different mass ranges, but cannot provide complete
information on the occurrence of Earth-mass planets in the Solar
neighbourhood (say, to 15--20~pc) in advance of the planned Darwin
launch: transit measurements are highly incomplete, the Gaia lower
mass limit is above 10\,$M_\oplus$, and the SIM survey will presumably
also be incomplete down to the levels of 1--2\,$M_\oplus$.  Gaia and
SIM may assist in identifying nearby stars accompanied by Jupiter-type
planets (i.e.,\ Jupiter mass in a Jupiter orbit), which may be a
representative pre-requisite for the development of life.  Nano-arcsec
astrometry could provide the necessary information in principle, but
practical implementation lies some years in the future.

Nevertheless the SIM and COROT/Kepler/Eddington results should clarify
whether Earth-mass planets are common or not.  If they are not
detected, and therefore not common, Darwin's task in identifying
Earth-mass planets within 20--30\,pc will be even more
challenging. Then, either additional identification strategies are
required (for example, the development of an Antarctic facility
dedicated to the purpose, such as a coronographic ELT in the
visible/near infrared, or a large interferometer in the thermal
infrared), or it is accepted that a significant fraction of Darwin
observations is devoted to a search for its own spectroscopic
candidates.  Thus Darwin could be launched relying on a statistical
estimate of the number, and size, of the terrestrial planets that are
expected around the stellar target list, combined with the GENIE
results on levels of exo-zodiacal emission. If statistics indicate
that the number of Earth-mass planets accessible to Darwin is of the
order of several tens, this strategy would be acceptable. Conversely,
it could also be argued that a null result from a 200-star survey
would be a valuable scientific result in its own right.

Current understanding suggests that metallicity is a crucial factor
influencing the formation of planets. Higher metallicity therefore
seems to be the most promising stellar characteristic that would
indicate whether a star is likely to harbour planets. The recently
published Geneva-Copenhagen survey \citep{nma+04}, mostly aimed at
stellar kinematics, has also provided valuable information on
metallicities for over 14\,000 F and G type dwarf stars using
Stromgren photometry. A more detailed measurement of stellar
parameters, including metallicities, requires medium- to
high-resolution spectra over a large wavelength range. For extra-solar
planet research an accurate knowledge of the properties of the host
star is essential, even more so when looking for the weak signatures
of the planetary atmosphere.  A full spectral characterisation of
stars in the solar neighbourhood would be a valuable step towards a
target list for major projects such as TPF, Darwin or OWL.

\subsection{Astrophysical Characterisation of Host Stars}
\label{sec:role-c}

Several thousand planetary systems should be discovered by a combination
of COROT, Kepler, Eddington, SIM and Gaia over the years 2006--2015.
There will be a need for detailed astrophysical characterisation of
large numbers of stars (some tens of thousands), in both the northern and
southern hemispheres. Specifically:

(a) physical characteristics of the host star via photometry: it will
be important to characterise, homogeneously, luminosity ($T_{\rm
  eff}$, $\log g$), metallicity, and micro-variability. Large-scale
multi-colour, multi-epoch photometry for this purpose may be available
from Pan-STARRS (Panoramic Survey Telescope and Rapid Response System)
or LSST (Large Synoptic Survey Telescope) and from Gaia's 11-band
multi-colour photometry, although not before 2012;

(b) physical characteristics of the host star via spectroscopy: the
goal is again to determine the spectroscopic parameters of the central
star: $T_{\rm eff}$, $\log g$, [Fe/H]. Medium-high spectral resolving
power is needed (20\,000--60\,000) in the visible range;

(c) spatial distribution and kinematics: Gaia will provide
highly-accurate distances essential for luminosity calibration (e.g.,\ 
1\% at 1~kpc at 15~mag), characterised as a function of the local
environment (e.g.,\ for stars within the core or halo of open
clusters), and kinematic motions (e.g.,\ with respect to the Local
Standard of Rest, velocity with respect to the Galactic mid-plane,
etc.). This characterisation is important because of possible links
with habitability (as discussed in Section~\ref{sec:habitability}).

(d) kinematic radial velocities for fainter stars. Hipparcos would
have greatly benefited from a coordinated acquisition of radial
velocities from the ground. Key programmes went part way towards this,
but only for half the sky, for a subset of spectral types, and on a
schedule out of phase with the Hipparcos Catalogue publication in
1997---the Geneva-Copenhagen survey of 14\,000 F and G~dwarfs has just
been published \citep{nma+04}. Gaia partly addresses the problem by
acquiring radial velocities on-board, but higher precision and a
fainter limiting magnitude would demand a concerted and coordinated
campaign on the ground.

With the caveats on the time-scales of data availability, planned
surveys (Gaia, if not others) should generally provide detailed
astrophysical characterisation of the host stars of all detected
systems, with the possible exception of kinematic radial velocities 
for faint stars ($>20$~mag), and a dedicated spectral survey.

\subsubsection{A Dedicated Spectral Survey}
\label{sec:dedicated-survey}

ESO has capabilities to perform a full spectral characterisation of
all F--K type stars in the southern hemisphere within a radius of say
50\,pc.  Facilities at ESO could be used to obtain a complete spectrum
of the stars from the atmospheric cut-off to 5\,$\mu$m.  By combining
the Echelle spectrographs FEROS (2.2-m), UVES, and CRIRES the complete
wavelength range can be observed at $R\sim50\,000$.  A cursory study
shows that such a project could be undertaken at the level of a (very)
large programme, of the order of 100~nights in total. Selecting stars
within 50\,pc (Hipparcos distances) and at declination south of 30
degrees North, a total of about 2500 stars would have to be observed.
75\% of these have V magnitudes between 5 and 9 making them ideal
targets for FEROS, while the remaining fainter ones would be observed
with UVES. Using the ESO exposure time estimators for the respective
instruments we found that a S/N of about 100 will be achieved within
minutes.  Using about 110\,hr of time on FEROS and about 40\,hr on
UVES per period to obtain the spectra from 350--980~nm the optical
survey could be completed within two years with existing facilities,
assuming an ESO `Large Programme' using only late twilight and periods
of bright moon and/or poor seeing.  CRIRES is being assembled at this
point, therefore no estimate of the time required for the infrared
observation is given here.  Because the majority of stars are rather
bright a very substantial fraction (30--50\%) of the stars can be
observed during twilight while the remainder can be observed during
bright time.  It is evident that a high degree of homogeneity of
spectral data needs to be achieved across the full set of stars in
order to achieve the scientific goals of the project. In particular
the combination of spectra calls for excellent calibration across
instruments. To this end the ECF's Instrument Physical Modelling Group
which is involved in the calibration of instruments on HST and at ESO
is prepared to make major contributions thereby ensuring maximum
quality of the data and the resulting high level products.

A comparable `Nearby Stars' project (PI: R.E.~Luck) is being conducted
in the northern hemisphere (http://astrwww.cwru.edu/adam/) using the
McDonald 2.1-m Struve reflector and the Sandiford Echelle
Spectrograph.  

A dedicated survey would provide data useful for other applications in
astrophysics. Results would include a very fine spectral library
which, combined with e.g.\ Gaia information on the luminosity function
and kinematics, would be very powerful.

\subsection{Potential Overlap and Competition}
\label{sec:role-d}

{\bf Eddington and Kepler}

The arguments for undertaking the Eddington mission are unchanged from
the time that the ESA Astronomy Working Group placed it at the top of
the list of astronomy projects to be undertaken if financing can be
made available.  Essentially, Kepler is designed to study a single
low-latitude target field and thus a single population of stars, in a
limited mass range.  In the same way that the star formation history
of the Galaxy cannot be ascertained by observing a single field, the
acceptance of Kepler results as being of general applicability to the
complete Galaxy is questionable. This is exemplified by the absence of
planets in 47~Tuc, and is a question that can only be resolved by
further observations.

Without any further mission dedicated to terrestrial planet searches,
there will be no real understanding of the distribution of such
planets in general (e.g., in other parts of the Galaxy or in
clusters), and the number of targets discovered by Kepler may in any
case be very small. Eddington would observe stars in a variety of
environments, i.e.,\ field stars and stars in open clusters, and will
thus consolidate the search for planets across different stellar
parameters (including metallicity, age, mass, binary status, density
of surrounding stellar population, etc.). 

More generally, experiments need repeatability: single experiment
results always provoke more questions, and if the results are
controversial it is only reasonable that they are checked
independently.  Having just one mission aimed at the detection of
terrestrial planets appears risky and inadequate, in particular
considering the effort being made for future missions that aim at
studying these planets (Darwin/TPF).  The present working group
emphasises the value of aiming to launch such a mission at the
earliest opportunity, preferably before 2010. Should it turn out that
it is only possible to launch Eddington significantly later, ESA
should consider a larger-scale post-Kepler mission on a longer time
schedule.

\vspace{20pt}
{\bf OWL and Kepler/Eddington}

The OWL science case is currently not so much based on finding
Earth-like planets, but rather on imaging or spectroscopy of nearby
candidates detected by other means. To this extent OWL's primary
science case is as a follow-up instrument. On the other hand there are
no prospects other than Darwin/TPF for finding all nearby Earth-mass
planets (transits can only discover a small fraction), so its use for
Earth-mass planet searches in a survey-type mode would be valuable,
and should add to its contribution to the field. Such a survey might
be undertaken during the mirror-filling phase, although detailed
feasibility studies of this approach have not yet been made.

\vspace{20pt}
{\bf OWL/Dome~C versus Darwin/TPF}

As summarised in Section~\ref{sec:antarctic} and
Table~\ref{tab:angel}, appropriate Antarctic observations could at
least partially compete with some aspects of the Darwin/TPF missions.
However, some key spectral regions are accessible only from space,
e.g.\ O$_3$ at 9.6\,$\mu$m and CO$_2$ at 15\,$\mu$m.  Trade-offs would
depend on the number of objects visible, the S/N ultimately attainable
from Darwin, and the accessible spectral range and the resulting
scientific information that can be inferred for exo-planetology and
exobiology. Further investigations will be needed in order to
understand whether such Antarctic-based observations could be
considered as realistic in the medium term.

\subsection{Open Areas: Survey Mission Beyond Kepler/Eddington}
\label{sec:role-e}

The absence of specific survey missions beyond 2015 noted in
Section~\ref{sec:expected-direction} can only be justified once all
interesting targets in the sky to be studied in the future have been
discovered.  For small terrestrial planets, the Kepler mission is the
only approved discovery mission. Even with Eddington, statistics of
the occurrence of Earth-mass planets will still be poorly known, and
the need for a larger, more performant, space transit survey seems
already rather compelling.  A case could also be made for an all-sky
transit survey from space, in order to find the nearest transiting
terrestrial planets, facilitating follow-up with astrometry and
spectroscopy. The arguments presented in Section~\ref{sec:esa-themes},
`Understanding the Planetary Population of our Galaxy', also draw
attention to a lack of further search programmes for terrestrial
planets beyond Kepler.

\subsection{Other Considerations}
\label{sec:role-f}

\subsubsection{Fundamental Physical Data}

It is not evident that present knowledge of basic atomic data is
adequate to understand the high-resolution spectra of normal host
stars over the broad wavelength region discussed in
Section~\ref{sec:dedicated-survey}, and the spectral signatures of
planets and their atmospheres.  Preliminary investigations suggest
that there is a need for improved atomic and molecular lines
particularly in the near infrared. To illustrate the point and the
potential risk of inadequate atomic data we give two examples: (i) Li
in low mass post-AGB stars; expanded wavelength tables for Ce have
shown that the line at 670.8~nm previously identified with (slightly
redshifted) Li is coincident with a Ce~II line at 670.8099~nm.
Abundance analysis subsequently showed that for all practical purposes
no Li~is present in these stellar photospheres \citep{rwb+02}.  This
is an important finding for stellar evolution and nucleosynthesis
since an elaborate mechanism had to be evoked in order to explain the
presence of Li in these evolved stars; (ii) variability of the fine
structure constant: evidence from the absorption features in distant
quasars suggested a variability of the fine structure constant over
time. Originally, these studies had been hampered by limited accuracy
of laboratory wavelengths for the relevant species \citep{ptm+00}.
Most recently a further complication introduced by the sensitivity to
the abundance of heavy isotopes has been described for~Mg
\citep{amo04}, where uncertainties in the isotope ratios might imitate
a variation of the fine structure constant at the observed level.
Nearer to the issues of exo-planets, the debate about the nature of
Jupiter's interior centres on the (unknown) equation of state of H/He
mixtures at high pressures.

Regardless of the exact details these examples illustrate that the
lack of accurate and reliable atomic data can lead to serious
misinterpretation of astrophysical data. When dealing with fundamental
issues such as the properties of extra-solar planets and their
suitability for supporting life such pitfalls can be rather
embarrassing and damaging to the credibility of the field.  In order
to assess the current situation properly, further consultation will be
needed between the astronomical community and the relevant atomic and
molecular physics community involved in the laboratory work. Links
between these communities are also being investigated in the context
of the Virtual Observatory.

\subsubsection{Fundamental Planetary Data}
\label{sec:fundamental}

The solar system research community has the best knowledge of the only
planetary system studied in detail, with extensive data archives 
like the Planetary Data System (PDS).  For extra-solar planet research
full use of that information must be made without being limited by it.
Most of the planetary systems found so far are quite different from
ours.  Items of particular relevance might be: (i) distribution of
minor bodies and dust, relevant for zodiacal light; (ii) mass radius
relation for various kinds of planets; (iii) reflective properties of
different planetary surfaces (gaseous/solid, rock/ice) in particular
integrated over the sphere (and thus the influence of weather); (iv)
spectral signatures of planetary atmospheres; (v) mass-loss from
atmospheres; documented by planetary probes for Mars, indication for
mass loss has also been found in HST observations of the transit of
HD~209458; (vi) signatures of volcanism, like SO$_2$ (cf.\ Earth and~Io).

Scientific results on solar system bodies and extra-solar planets are
presented at various conferences of both communities, which in part
have already established `cross-discipline' sessions to support the
exchange of knowledge between the two fields (e.g., the EGU and DPS).
ESO and ESA might consider fostering the exchange of information and
collaboration by organising joint cross-disciplinary workshops,
including ESA's Earth observation community. Such dedicated workshops
could aim at the specific needs of extra-solar planet research and
serve to define the needs for further observations of solar system
bodies (e.g., atmospheres of planets and moons). In addition, building
a data base collecting, for example, the spectra on planets, moons and
minor bodies in our solar system as a reference for extra-solar planets
could be discussed within both communities at such dedicated meetings.

\subsubsection{Amateur Networks}
\label{sec:amateur}

As described above, thousands of candidate planets will be discovered
by ongoing surveys both from the ground and in space. Follow-up of all
these candidates will be essential in order to verify their existence
and to derive additional information about their properties.
Observations of transits in particular will be valuable for candidates
found by radial velocity or astrometric measurements. Dedicated
(groups of) amateurs with state of the art equipment at reasonably
large ($>$60\,cm) telescopes could contribute in this area. The AAVSO
is already calling on its observers to engage in such activity. Very
recently, a dedicated campaign to observe possible transits of GJ~876b
in the system IL~Aqr was announced spanning a ten-hour period starting
21~Oct 2004 (http://www.aavso.org/news/ilaqr.shtml). ESO and ESA could
take a leading role here by establishing and coordinating a dedicated
Amateur Transit Network, an interesting opportunity for scientific
reasons as well as for its potential for public outreach.

Scientifically, the most attractive aspect is that a global network
(10--20 individual observatories) could provide very significant
amounts of observing time and excellent time coverage. The latter is
also essential for finding planetary signatures in microlensing
events. Such a network probably will be the most cost-effective and
efficient way to achieve a large increase in follow-up observations of
candidates.  The major challenge is to ensure that observations from
different sources can successfully be combined for analysis. To achieve
the required homogeneity, ESO/ESA could consider the use of common
hardware (CCD and filters) by the members of the network and possibly
software support. Relatively modest expenditures would be required to
equip each participating observatory with a high-quality photometric
system, very beneficial in terms of data quality.  Although the effort
required to coordinate such a network should not be underestimated, if
properly coordinated and supported, an ESO/ESA amateur transit network
could provide a cost-effective way to substantially increase the
capablility for follow-up observations of candidates.  Such an
increase seems necessary in order to keep up with the demand projected
to materialise over the next decade.

Sky \& Telescope (December 2004, p18) reported that only eight days
after the transiting system TrES--1 had been announced, a transit was
observed by a Belgian amateur as well as by two observers from
Slovenia.  The first transit system HD~209458 is fairly bright at
$V=7.6$~mag, but TrES--1 is only 11.8~mag.  The detector hardware
used for the discovery of TrES--1 is quite comparable to what advanced
amateurs have.  One main difference is the software which enables the
monitoring of thousands of stars, and therefore acceptable chances for
discovery.  In this context, one aspect of
Section~\ref{sec:ground-transits} is recalled: \cite{hm05} and
\cite{ass+05} describe how repeated observations of transits may lead
to the discovery of Earth-mass planets: the gravitational influence of
the small planet will induce pertubations in the orbit of the much
larger planet causing the transit; over a period of time this will
lead to a shift of the observed transit times on the order of minutes.
The effect is more pronounced for long-period massive planets, with
low-mass companions in orbital resonance.

Involvement of amateurs in extra-solar planet research also opens a
very attractive field in terms of public outreach and eduction. The
general public clearly is very interested in the topic of extra-solar
planets. The subject therefore is well suited to provide information
and eductational material explaining the prospects and limitations of
searches for extra-solar planets to the public and decision makers.
Given the very substantial investments required to realise the next
generation of extremely large telescopes, astronomy should make a
dedicated effort to take advantage of this overlap of public and
research interest. While an outreach effort on extra-solar planets will
be worthwhile in its own right, we expect that an outreach effort
centered around an amateur (public) involvement in science will be
much more effective thanks to its novelty and emotional appeal.  In
order to get some feedback from the amateur community directly one of
us (FK) gave a presentation at the 8~November 2004 meeting of the
astronomy club `Max Valier' in Bolzano, Italy. The club has more than
100~members and is well-equipped with an 80\,cm telescope and modern
CCDs.  Their new observatory at Gummer was built with funds of the
local government and is now run as a public observatory
(http://www.sternwarte.it) with an extensive program of tours and
observing sessions for schools and the general public. Further details
of this meeting, and a more detailed proposal for follow-up, are 
available.

\clearpage
\section{Recommendations}

\vspace{-10pt}

Europe has already established leadership in major areas of the
exo-planet field, including radial velocity (HARPS), transit searches
(COROT), and astrometry (Gaia).

The first goal of future actions should be to take full advantage of
this situation, with an offensive policy to optimize the scientific
return of instruments already built or foreseen in the near future
[the ongoing or planned ESO instrument programmes (HARPS, UVES, CRIRES,
OmegaTranS, PRIMA, GENIE, Planet Finder, etc.) are not considered 
further here].

The second goal is to prepare new initiatives. Suggested directions are:

\vspace{10pt}
{\bf 1. ESA }

1.1. Eddington: provide a clearer message to the community about the
plans and schedule for an Eddington-type exo-planetary mission
[Section~\ref{sec:space-transits}].

1.2. Gaia: recognise that the highest accuracy will supply the
most comprehensive information on low-mass planets in the Solar
neighbourhood (down to about 10\,$M_\oplus$), thus assisting the
identification of targets for Darwin/OWL/PRIMA, and the largest 
number of detections, scaling as the third power of the accuracy
[Section~\ref{sec:space-astrometry}].

1.3. Darwin: maintain the research and development plan for this key
project in the domain of exo-planet science at high priority,
consistent with the current 2015 target launch date, cf.\ TPF-C launch
planned for 2014 [Section~\ref{sec:space-2015}].

1.4. JWST: support the `Astrobiology and JWST' panel recommendations,
to ensure that JWST can follow up low-mass transits discoveries
[Section~\ref{sec:other-space}].

1.5. Encourage the community to submit mission proposals covering
the important themes in Section~\ref{sec:esa-themes}: e.g.\ a future
transit survey mission significantly more performant than currently
planned; an all-sky transit mission; a UV spectroscopic mission for
transit spectroscopy; astrometric detection of Earth-mass planets 
out to 20--30~pc, etc.\ [Section~\ref{sec:esa-themes}].

\vspace{10pt}
{\bf 2. ESO }

2.1. Support experiments to improve radial velocity mass detection
limits, e.g.\ based on experience from HARPS, down to those imposed by
stellar surface phenomena [Section~\ref{sec:limits}].

2.2. Characterise nearby potential planet host stars, e.g.\ 
$T_{\rm eff}$, log~$g$, [Fe/H], $M$, radius, etc.\ 
[Section~\ref{sec:role-c}].

2.3. Improve the capabilities of main-stream VLT instruments for high
cadence, high S/N transit spectroscopy in the visible and infrared
[Section~\ref{sec:ground-transits}].

2.4. Evaluate observation time for follow-up observations over the
next 10~years of transit candidates obtained by space missions (COROT,
Kepler, Eddington-like) and ground-based observations, including
high-resolution imaging and photometry, on small to very large
telescopes [Section~\ref{sec:follow-up}].

2.5. In addition to the use of OWL as a follow-up facility, consider
prospects for OWL searches for Earth-mass planets in the solar
neighbourhood, including during the filling phase [Section~\ref{sec:owl}].

2.6. Investigate the astrometric capabilities of OWL, possibly
leading to the inclusion of an astrometric facility in the
instrumentation plan [Section~\ref{sec:owl}].

\vspace{15pt}
{\bf 3. ESO--ESA -- Joint Initiatives}

3.1. Consider radial velocity follow-up of COROT (and possibly Kepler
or Eddington-type missions -- but note caveats of Kepler visibility)
transit candidates . This would involve hundreds of targets and
requires: (1) a major time allocation of La Silla 3.6-m telescope to
HARPS; (2) probably an instrument with a precision similar to HARPS
but on a larger telescope [Section~\ref{sec:follow-up}].

3.2. Consider facilities for radial velocity follow-up of the large
number of candidates (20--30\,000) from Gaia, requiring observing time
on relatively small, possibly robotic, telescopes for a few years of
full-time operations [Section~\ref{sec:follow-up}].

3.3. Consider the photometric (transit) monitoring of large number of
high-mass planet candidates which will come from space experiments
[Section~\ref{sec:follow-up}].

3.4. Support valid observing time requests for preparatory work to
space exo-planet missions, e.g., field characterisation for an 
Eddington-like mission [Section~\ref{sec:follow-up}].

3.5. Evaluate the prospects of implementing a small interferometric
array, along the lines of GENIE, at Dome~C/A, given that it could
radically change the capabilities of ELTs if (a second) OWL were
located there [Section~\ref{sec:antarctic}].

3.6. Consider coordination of amateur networks, along the lines of
AAVSO, for the follow-up of hot Jupiters detected from ground transit
surveys, from Gaia astrometry and photometry, and for surveys for
longer-period objects [Section~\ref{sec:amateur}].

3.7. Foster cooperation between the solar system research community
and the extra-solar planets community, e.g.\ by supporting or
establishing joint meetings of both communities addressing common
topics, such as formation, comparative planetology, emergence and
evolution of life on Earth and elsewhere, etc.\ 
[Section~\ref{sec:fundamental}].

3.8. Coordinate exo-planet public communication, with some common `code
of conduct', e.g.: (i)~no press release without a supporting scientic
paper; (ii)~claims should not be overstated, with support sought from
external organisations or universities; (iii)~retractions should be
posted in the same form to retain credibility.

\clearpage
\section*{Index to ESA and ESO facilities:}

Descriptions of each experiment or facility are given as follows:

\vspace{-10pt}
\subsubsection*{ESA:}
\vspace{-10pt}

COROT: see Section~\ref{sec:space-transits} 

Eddington: see Section~\ref{sec:space-transits} 

Gaia: see Section~\ref{sec:space-astrometry}

HST (ESA/NASA): see Section~\ref{sec:space-transits} 

JWST (ESA/NASA): see Section~\ref{sec:other-space}

\vspace{-10pt}
\subsubsection*{ESO:}
\vspace{-10pt}

ALMA: see Section~\ref{sec:alma}

CRIRES: see Section~\ref{sec:ground-transits}

GENIE: see Section~\ref{sec:genie}

HARPS: see Section~\ref{sec:ground-radvel}

NAOS-CONICA: see Section~\ref{sec:ground-direct}

OWL: see Section~\ref{sec:owl}

Planet Finder: see Section~\ref{sec:ground-direct}

PRIMA: see Section~\ref{sec:ground-direct}

VLTI: see Section~\ref{sec:ground-direct}

\vspace{-10pt}
\subsubsection*{NASA/Other:}
\vspace{-10pt}

FAME/AMEX/OBSS: see Section~\ref{sec:space-astrometry}

JASMINE (Japan): see Section~\ref{sec:space-astrometry}

Kepler: see Section~\ref{sec:space-transits} 

MPF/GEST: see Section~\ref{sec:space-microlensing}

SIM: see Section~\ref{sec:space-astrometry}

SOFIA: see Section~\ref{sec:other-space}

Spitzer (ex-SIRTF): see Section~\ref{sec:other-space}

\clearpage
\addcontentsline{toc}{section}{Appendices} 
\appendix
\section{Space Precursors: Interferometers, Coronographs and Apodizers}
\label{sec:space-precursors}

The McKee-Taylor Decadal Survey Committee \citep{mt00} qualified its
endorsement of the TPF mission with the condition that the abundance
of Earth-size planets be determined prior to the start of the TPF
mission. Many ideas for scientific and technological precursors for
TPF have been examined. The main contenders are summarised here
for completeness, although with the recent (April 2004) NASA
announcement on TPF strategy, it seems unlikely that any of these
concepts will be developed further:

{\bf Eclipse} (coronography) is a proposed NASA Discovery-class mission to
perform a direct imaging survey of nearby planetary systems, including
a complete survey for Jovian-sized planets orbiting 5\,AU from all
stars of spectral types A--K within 15\,pc of the Sun \citep{thb+03}.
Its optical design incorporates a telescope with an unobscured
aperture of 1.8~m, a coronographic camera for suppression of
diffracted light, and precision active optical correction for
suppression of scattered light, and imaging/spectroscopy.  A
three-year science mission would survey the nearby stars
accessible to TPF.  Eclipse may be resubmitted for NASA's
Discovery round in 2004. 

{\bf Jovian Planet Finder} (JPF) was a MIDEX proposal to directly
image Jupiter-like planets around some 40 nearby stars using a 1.5-m
optical imaging telescope and coronographic system, originally on the
International Space Station (ISS) \citep{cfi+02}.  Its sensitivity
results from super-smooth optical polishing, and should be sensitive
to Jovian planets at typical distances of 2--20\,AU from the parent
star, and imaging of their dusty disks -- potentially solar system
analogues.  A 3-yr mission lifetime was proposed.  
\ignore{ Some successor to JPF may be resubmitted for NASA's Discovery
  round in 2004, probably through a merging with ESPI (`EPIC',
  Clampin, private communication).}

{\bf Extra-Solar Planet Imager} (ESPI) is another proposed precursor to TPF
\citep{lgm+03}.  Originally proposed as a NASA Midex mission as a
$1.5\times1.5$~m$^2$ apodized square aperture telescope, reducing the
diffracted light from a bright central source, and making possible
observations down to 0.3~arcsec from the central star. Jupiter-like
planets could be detected around 160--175 stars out to 16\,pc, with
S/N~$>5$ in observations lasting up to 100~hours. Spectroscopic
follow-up of the brightest discoveries would be made.  The Extra-Solar
Planet Observatory (ExPO) is a similar concept proposed as a
Discovery-class mission \citep{gnp+03}.

{\bf Self-luminous Planet Finder} (SPF) is a further TPF precursor under
study by N.~Woolf and colleagues, aiming at the search for younger or
more massive giant planets in Jupiter/Saturn like orbits, where they
will be highly self-luminous and bright at wavelengths of
5--10\,$\mu$m, where neither local nor solar system zodiacal glow will
limit observations. SPF will demonstrate the key technologies of
passive cooling associated with interferometric nulling and truss
operation that are required for a TPF mission. SPF targets young
Jupiter-like planets both around nearby stars such as $\epsilon$~Eri,
and around A and early F~stars.

{\bf Fourier-Kelvin Stellar Interferometer} (FKSI) is a concept under
study at NASA GSFC \citep{ddk+03}. It is a space-based mid-infrared
imaging interferometer mission concept being developed as a precursor
for TPF.  It aims to provide 3~times the angular resolution of JWST
and to demonstrate the principles of interferometry in space. In its
minimum configuration, it uses two 0.5-m apertures on a 12.5-m
baseline, and predicts that some 7~known exo-planets will be directly
detectable, with low-resolution spectroscopy ($R\sim20$) possible in
favourable cases.

{\bf Optical Planet Discoverer} (OPD) is a concept midway between coronography
and Bracewell nulling \citep{mss+03}. 

{\bf Phase-Induced Amplitude Apodization} (PIAA, \cite{guy03}) is an
alternative to classical pupil apodization techniques (using an
amplitude pupil mask).  An achromatic apodized pupil is obtained by
reflection of an unapodized flat wavefront on two mirrors. By
carefully choosing the shape of these two mirrors, it is possible to
obtain a contrast better than $10^9$ at a distance smaller than
$2\lambda/d$ from the optical axis. The technique preserves both the
angular resolution and light-gathering capabilities of the unapodized
pupil, and claims to allow efficient detection of terrestrial
planets with a 1.5-m telescope in the visible.

\vspace{10pt}
Occulting masks are another approach to tackle in a conceptually
simple manner the basic problem of how to separate dim sources from
bright ones, and have been considered as precursor missions to Darwin/TPF.
Interest in this approach at NASA level currently appears limited. 

{\bf UMBRAS} (Umbral Missions Blocking Radiating Astronomical Sources)
refers to a class of missions \citep{slk+03}, currently designed
around a 4-m telescope and a 10-m occulter, with earlier concepts
including a 5--8~m screen (CORVET), or as NOME (Nexus Occulting
Mission Extension) a modification to Nexus, itself foreseen as a
test of technologies for JWST.

{\bf BOSS} (Big Occulting Steerable Satellite \citep{cs00}) consists
of a large occulting mask, typically a $70\times70$~m$^2$ transparent
square with a 35~m radius, and a radially-dependent, circular
transmission function inscribed, supported by a framework of
inflatable or deployable struts. The mask is used by appropriately
aligning it with a ground- or space-based observing telescope. In
combination with JWST, for example, both would be in a Lissajous-type
orbit around the Sun-Earth Lagrange point L2, with the mask steered to
observe a selected object using a combination of solar sailing and ion
or chemical propulsion. All but about $4\times10^{-5}$ of the light at
1\,$\mu$m would be blocked in the region of interest around a star
selected for exo-planet observations. Their predictions suggest that
planets separated by as little as 0.1--0.2~arcsec from their parent
star could be seen down to a relative intensity of $1\times10^{-9}$
for a magnitude~8 star.  Their simulations indicate that for systems
mimicking our solar system, Earth and Venus would be visible for stars
out to 5\,pc, with Jupiter and Saturn remaining visible out to about
20\,pc.

\ignore{
\begin{figure}[bth]
\begin{center}
\centerline{\hfill
\epsfig{file=fig-boss-a.eps,width=0.35\textwidth}\hfill
\epsfig{file=fig-boss-b.eps,width=0.35\textwidth}\hfill
}
\end{center}
\vspace{-20pt}
\caption[]{\footnotesize
  Left: a $2\times2$ arcsec$^2$ image of the log intensity of 
  our solar system at 3\,pc from BOSS observed at 1\,$\mu$m by an 8-m
  space telescope in a 3000~s exposure. The Sun is located at the
  centre, with the central 0.2~arcsec square, corresponding to BOSS,
  cut out of the image. Both Venus (above) and Earth (left) are
  observable.  Right: as viewed from 10\,pc.  Venus and Earth are now
  occulted, and Jupiter (upper right) and Saturn (lower left) are
  visible.}
\label{fig:boss}
\end{figure}
} 

\clearpage
\section{Beyond 2025: Life Finder and Planet Imager}
\label{sec:beyond-2025}

Within NASA's Origins Program HST, Spitzer and others are referred to as
`precursor missions', with SIM and JWST as `First Generation Missions'
leading to the `Second Generation Mission' TPF which will begin to
examine the existence of life beyond our Solar System. Once habitable
planets are identified, a `Life Finder' type of mission would expand
on the TPF principles to detect the chemicals that reveal biological
activities. And once a planet with life is found, `Planet Imager'
would be needed to observe it. These `Third Generation Missions', Life
Finder and Planet Imager, are currently just visions because the
required technology is not on the immediate horizon. Short descriptions 
are included here for completeness.

{\footnotesize
{\bf Life Finder: }
Taking pictures of the nearest planetary system (Darwin/TPF) is
considered to be a reasonable goal on a 10-year timescale, with
low-quality spectra a realistic by-product. Life Finder, which would
only be considered after Darwin/TPF results are available, and once
oxygen or ozone has been discovered in the atmosphere, would aim to
produce confirmatory evidence of the presence of life, searching for
an atmosphere significantly out of chemical equilibrium, for example
through its oxygen (20\% abundance on Earth) and methane ($10^{-6}$
abundance on Earth).  Some pointers to the technology requirements and
complexity of Life Finder have been described in the `Path to Life
Finder' \citep{woo01}.  Given that Darwin/TPF will take low-resolution
low-S/N spectra, a large area high angular resolution telescope will
be needed for detailed spectral study in order to confirm the presence
of life.  Recalling that the target objects will be as faint as the
Hubble Deep Field galaxies, buried in the glare of their parent star
some 0.05--0.1~arcsec away, the light collecting area of Life Finder
will have to be substantially larger than TPF's 50~m$^2$: a useful
target is 500--5000~m$^2$.  One of the primary technical challenges
will be to produce such a collecting area at affordable cost and mass.

The required development of new low mass and better wavefront optics,
coronography versus nulling, pointing control by solar radiation
pressure, sunshield, vibration damping, and space assembly, were
addressed by \cite{woo01}. According to their study, a `mini-Life
Finder' might be a $50\times10$~m$^2$ telescope, made with 12~segments
of $8.3\times5$~m$^2$, made of 5~kg~m$^{-2}$ glass, piezo-electric
controlled adaptive optics, and a total mass (optics and structure) of
about 10~tons.  Cooling would be by an attached sunshade also used for
solar pressure pointing, in a `sun orbiting fall away' orbit to avoid
the generation of thruster heat needed to maintain the L2-type orbit.
There are still unsolved complexities underlying the actual science
case for Life Finder: {\it if\/} the goal is to detect the 7.6\,$\mu$m
methane feature --- which is not definitively the relevant goal; see,
for example discussions of the use of the `vegetation signature' in
\cite{agl+02} --- the required collecting area accelerates from a
plausible 220~m$^2$ (four or five 8-m telescopes) for a planet at
3.5\,pc, to a mighty 4000~m$^2$ (eighty 8-m telescopes) even at only
15\,pc. A new proposal to study Life Finder has recently been submitted
to NASA by Shao, Traub, Danchi \& Woolf (N.~Woolf, private
communication).  Various reports on related studies can be found under
NASA's Institute for Advanced Concepts (NIAC) www pages
(http://www.niac.usra.edu/) including `Very large optics for the study
of Extra-Solar Terrestrial Planets' (N.~Woolf); and `A structureless
extremely large yet very lightweight swarm array space telescope'
(I.~Bekey). The former includes an outline technology development plan
for Life Finder, with costs simply stated as $\gg$\$2~billion.

{\bf Planet Imager:}
TPF aims to obtain images of planetary systems in which the planets
appear as point sources.  Resolving the surface of a planet is, at
best, a far future goal requiring huge technology development that is
not yet even in planning. Much longer baselines will be required, from
tens to hundreds of km in extent. Formation flying of these systems
will require technology development well beyond even the daunting
technologies of Darwin/TPF -- complex control systems, ranging and
metrology, wavefront sensing, optical control and on-board computing.
Having accepted that we are now peering into a much more distant and
uncertain future, we can examine some of the ideas which are being
discussed.

Life Finder studies \citep{woo01} have been used to evaluate the
requirements for Planet Imager which, they consider, would require
some 50--100 Life Finder telescopes used together in an
interferometric array. Their conclusions were that {\it `the
  scientific benefit from this monstrously difficult task does not
  seem commensurate with the difficulty'}. This echoes the conclusions
of \cite{bs96} who undertook a partial design of a separated
spacecraft interferometer which could achieve visible light images
with $10\times10$ resolution elements across an Earth-like planet at
10\,pc.  This called for 15--25 telescopes of 10-m aperture, spread
over 200~km baselines. Reaching $100\times100$ resolution elements
would require 150--200 spacecraft distributed over 2000~km baselines,
and an observation time of 10~years per planet.  These authors noted
that the resources they identified would dwarf those of the Apollo
Program or the Space Station, concluding that it was {\it `difficult
  to see how such a program could be justified'}. The effects of
planetary rotation on the time variability of the spectral features
observed by an imager, complicates the imaging task although may be
tractable, while more erratic time variability (climatic, cloud
coverage, etc.) will greatly exacerbate any imaging attempts.

{\bf Hypertelescopes/OVLA: }

Parallel to the Planet Imager studies, in Europe the LISE
group (Laboratoire d'Interf\'erom\'etrie Stellaire et
Exo-plan\'etaire) carries out research in the area of high-resolution
astronomical imaging, including imaging extra-solar planets. The group
is studying several complementary projects for `hypertelescopes' on
Earth and in space \citep{rbg+02, grl+03}.  The steps needed
to reach this goal are set out as requiring: (1) a hypertelescope on
Earth -- the OVLA (Optical Very Large Array); (2) a 100-m precursor
geostationary version in space; (3) a km-scale version in a higher
orbital location; (4) a 100~km version, including dozens of mirrors of
typically 3~m aperture.  Labeyrie et al.\ proposed the mission
`Epicurus', an extra-solar Earth imager, to ESA in 1999 in response to
the F2/F3 call for mission proposals.
 
Their basic `hypertelescope' design involves a dilute array of smaller
apertures (an imaging interferometer) having a `densified' exit pupil,
meaning that the exit pupil has sub-pupils having a larger relative
size than the corresponding sub-apertures in the entrance pupil (see
Fig.~1 of \cite{pla+00}). Their applicability extends to observing
methods highly sensitive to the exit pupil shape, such as phase-mask
coronography.

In the most recent published studies \citep{rbg+02} the hypertelescope
is combined with such a coronograph to yield attenuations at levels of
$10^{-8}$. Simulations of 37~telescopes of 60~cm aperture distributed
over a baseline of 80~m in the infrared, observing the 389 Hipparcos M5--F0
stars out to 25\,pc (with simulated contributions from zodiacal and
exo-zodiacal background) yields 10-hour snapshot images in which an
Earth-like planet is potentially detectable around 73\% of the stars.
Gains of a factor 20--30 with respect to a simple Bracewell nulling
interferometer are reported.

In space, the plans call for a flotilla of dozens or hundreds of small
elements, deployed in the form of a large dilute mosaic mirror.
Pointing is achievable by globally rotating the array, which is slowly
steerable with small solar sails attached to each element. A
`moth-eye' version allows full sky coverage with fixed elements, using
several moving focal stations \citep{lab99a, lab99c}. The
geostationary precursor hypertelescope could be a version of TPF. An
exo-Earth discoverer would require a 100--1000~m hypertelescope with
coronograph, while an exo-Earth imager would require a 150~km
hypertelescope with coronograph. In the approach of \cite{lab99a} a
30-min exposure using a hypertelescope comprising 150 3-m diameter
mirrors in space with separations up to 150~km, would be sufficient to
detect `green' spots similar to the Earth's Amazon basin on a planet
at 10~light-years, although these vegetation features are more
prominent in the infrared \citep{agl+02}.  \ignore{A 2-month ESO
  investigation into the detection of exo-planets with hypertelescopes
  and advance coronography may commence before the end of 2004.}

} 

\clearpage
\section{ESO 1997 Working Group on Extra-Solar Planets}

The ESO 1997 Working Group on the Detection of Extra-Solar Planets
consisted of 16 members, and met on four occasions during 1996 and
1997.  The assignment was to {\it `advise ESO [...] on how to help
  designing a competitive strategy in this field that is predicted to
  expand dramatically in the next years, and to become one of the
  leading fields of astronomy in the next century.'}  The 1997 
Working Group and its task were somewhat comparable to the current
effort, with the notable difference that this time the Working Group
was asked to provide recommendations to both ESA and ESO, thereby
encompassing both space- and ground-based astronomy.

This appendix looks briefly at the findings and recommendations of their
final report, published on 10~September 1997.  This may provide some
understanding of why some recommendations were successfully
implemented while others were not, and what lessons can be learned
from the past.

In the introduction of their report the 1997 Working Group states:
{\it `Only by allocating a major fraction of time on some of its
  telescopes and developing new technology --- and doing it now --
  will ESO fully exploit its potential in this field and be truly
  competitive.'} They then laid out their plan to achieve this goal.
They identified six areas in which ESO could play a critical role,
namely: radial velocity searches, narrow-angle astrometry,
microlensing, direct detection, transits and timing of eclipses. The
recommendations, planned timeframe, and status and achievements of
each of these is summarised in Table~10.

\vspace{10pt} {\bf Radial Velocity Searches:} The main recommendation
was to devote a major fraction of the observing time at the CAT and
the 3.6\,m to monitor the radial velocities of about 1000~stars over
the next 5--10~years. They called for the development of a dedicated
spectrograph providing an accuray of about 1\,m/s. Such a survey was
considered indispensable to provide targets for VLTI which was assumed
to become operational by 2002. They also advocated high-fidelity
calibration of the iodine cell for UVES and the timely development of
CRIRES to provide a survey using twilight observations in order to
obtain complimentary information in the IR.

\vspace{10pt} {\bf Narrow-Angle Astrometry:} The Working Group
considered astrometry a very attractive method for the study of
planets because it allows for determination of orbits and planet
masses directly. It was considered complimentary to other search
methods exploring different regions of parameter space. Achieving an
accuracy of about 10\,micro-arcsec was considered feasible.
Realisation of the technique was regarded as technically very
challenging, requiring knowledge of the length of the baseline to
within 50~$\mu$m and the knowledge of the delay between the two stars
with 5\,nm precision over a 100\,m baseline.

\vspace{10pt} {\bf Microlensing:} Microlensing was identified as a
method which is in principle able to detect Earth-mass planets. Since
searches for Jovian planets were already in the development phase in
1997 the Working Group suggested that ESO should focus on detecting
Earth-mass planets. They suggested that a dedicated 2.5\,m telescope
should monitor the bulge with a large (1$^\circ$) field detector (16k by
16k CCD) during a 120~night season. It would observe a few million
uncrowded stars achieving 1\,\% photometric accuracy reaching $V=20$
in a 4-minute exposure. A high sampling rate is crucial to obtain the
characteristic of the short (5~hr) planet event on the microlensing
light curve. No specific time frame was given but the Working Group
called for an aggressive ESO-based campaign.

When VST/OmegaCam becomes operational in 2006, the technical
capabilities for the above programme will be available. Microlensing
searches for Earth-mass planets, though, are not part of the major
science goals for VST/OmegaCam.

\vspace{10pt} {\bf Direct Detection:} Direct detection was considered
essential for deriving many physical properties of extra-solar plants
such as size, temperature, chemical composition etc. The Working Group
stressed that it would be extremely challening to achieve the required
contrast levels in particular from the ground.  They identified a very
powerful adaptive optics system as a key ingredient in combination
with coronographic instrumentation and sophisticated data processing.
They pointed out that spectroscopic signatures of planetary
atmospheres should be detectable with high-resolution spectroscopy in
the NIR (CRIRES) via their time-dependent Doppler shift. Other
spectral features will be unique to the planet and therefore appear as
`alien' features in the stellar spectrum.  They called for further
studies of instrumental requirements, and an early realization of an
instrument like CRIRES.

\vspace{10pt} {\bf Transits and Timing Eclipses:} The Working Group
noted the potential of this method for detecting planets down to
Uranus size, along with planets with rings and moons of giant planets.
They mention the possibility of obtaining spectra of planets'
atmospheres during transits.  They also noted that timing of eclipses
in binary stars was a simple method for detecting giant planets in
these systems.

\begin{landscape}
\begin{table}
\label{tab:eso-1997}
\footnotesize
\caption[]{ESO Working Group 1997: Recommendations and Outcome}
\begin{center}
\begin{tabular}{lll} \\ \hline
{\bfseries{Radial velocity searches}} & & \\
\hline
Recommendation & Planned Timeframe & Status and Achievements\\ \hline
CORALIE survey at ca 5 m/s & start end of 1997 & first light May 1998, 3 m/s, ongoing 200 nights/year\\
FEROS survey & end 1998 & not done, FEROS moved to 2.2\,m\\
Dedicated new spectrograph at 3.6\,m & late 1999 at 1\,m/s & HARPS: first light Feb 2003, operational at 1\,m/s \\
UVES ready for RV studies & mid 1999 & UVES first light, Sep 1999, iodine cell since 2000\\
CRIRES & mid 2000 & first light late 2005\\
\hline
{\bfseries{Narrow angle astrometry}} & & \\
\hline
Recommendation & Planned Timeframe & Status and Achievements\\ \hline
Develop astrometric capability for VLTI & 10 $\mu$arcsec by ca 2005 &
First AT installed Jan 2004\\
Astrometric program with auxiliary telescopes (AT)  & asap, before ca 2002 & 
ATs and PRIMA available by 2006 \\
Substantial time on ATs for 10-15 years & start 2002 & start 2006 ?\\
Design VLTI to achieve high astrometric precision & before procurement & done\\
\hline
{\bfseries{Microlensing}} & & \\
\hline
Recommendation & Planned Timeframe & Status and Achievements\\ 
\hline
Dedicate 120 nights on 2.5\,m telescope on Paranal & not specified/Sep 2001 & VLT Survey Telescope (VST) first light late 2005 ?\\
Develop 16k by 16k CCD & not specified/Sep 2001 & OmegaCam, ready for shipment mid 2005\\
Develop new photometric data processing techniques & not specified & under development\\
Real-time follow-up & not specified & \\
\hline
{\bfseries{Direct detection}} & & \\
\hline
Recommendation & Planned Timeframe & Status and Achievements\\ 
\hline
High-order AO/coronograph at VLT & not specified & Planet Finder: phase A 
ended 2004\\
High-resolution IR spectroscopy  CRIRES & mid 2000 & first light late 2005\\
VLTI measurements & not specified & by 2006 ? \\
\hline
{\bfseries{Transits and eclipse timing}} & & \\
\hline
Recommendation & Planned Timeframe & Status and Achievements\\ 
\hline
1\,m telescope/CCD & not specified & not implemented\\
100 deg$^2$ Schmidt telescope/CCD array & not specified & not implemented\\
1\,m telescope with GPS clock & not specified & not implemented\\

\end{tabular}
\end{center}
\end{table}
\end{landscape}

\end{document}